%% file: main.tex
\documentclass[sigconf,nonacm]{acmart}

\AtBeginDocument{%
  \providecommand\BibTeX{{%
    \normalfont B\kern-0.5em{\scshape i\kern-0.25em b}\kern-0.8em\TeX}}}

\setcopyright{acmcopyright}
\copyrightyear{2025}
\acmYear{2025}
\usepackage{graphicx}
\usepackage{float} 
\usepackage{subfigure}
\usepackage{multirow}
\usepackage{enumitem}
\usepackage{mdframed}
\usepackage{amsmath, amssymb}
\usepackage{makecell}
\usepackage{pifont}

\usepackage{multirow}
\usepackage{balance}
\usepackage{adjustbox} 
\usepackage{threeparttable}
\usepackage{xcolor}
\usepackage{color,colortbl}
\usepackage{tikz}
\usepackage{booktabs}
\usepackage{tikz-qtree}
\usetikzlibrary{fit,calc}
\usepackage{pgfplots}
\usepackage{varwidth}
\usepackage{url}
\usepackage{mathtools}
\usepackage{rotating} 

\usepackage{amsmath}
\usepackage{graphicx}
\usepackage{todonotes}
\usepackage{cleveref}

\newtheorem{example}{Example}
\usepackage[ruled,linesnumbered,vlined]{algorithm2e}
\usetikzlibrary{patterns}

\usepackage[noend]{algpseudocode}   
\newcommand{\tabincell}[2]{\begin{tabular}{@{}#1@{}}#2\end{tabular}}

\definecolor{linecolor}{rgb}{0.55, 0, 0} 
\definecolor{bgcolor}{rgb}{1, 0.95, 0.95} 
\newcommand{\Yes}{\ding{51}}
\newcommand{\No}{\ding{55}}
\usepackage[most]{tcolorbox}
\tcbset{highlightcode/.style=
    {enhanced, 
    colframe=gray!20, 
    colback=gray!20, 
    boxrule=0pt, 
    boxsep=0pt, 
    left=1mm, 
    right=0mm, 
    top=1mm, 
    bottom=1mm,
    overlay={
        \begin{scope}[shift={(frame.north west)}]
            \foreach \i in {1,2,3} {
                \node[anchor=east] at (-5mm,{-\i*15mm + 1.5mm}) {\footnotesize\the\numexpr\value{AlgoLine}+\i\relax};
            }
        \end{scope}
    },
}}
\usepackage{framed}

\definecolor{mycolor1}{RGB}{53,122,162}   
\definecolor{mycolor2}{RGB}{164,224,187} 

\newcommand{\change}[1]{\textcolor{black}{#1}}
\newcommand{\colorcell}[1]{%
    \cellcolor{mycolor1!#1!mycolor2}%
    \ifnum#1>60
      {\Large \textcolor{white}{#1}}%
    \else
      {\Large \textcolor{black}{#1}}%
    \fi
}

\newcommand\vldbdoi{10.14778/3773731.3773738}
\newcommand\vldbpages{5623 - 5637}
\newcommand\vldbvolume{18}
\newcommand\vldbissue{13}
\newcommand\vldbyear{2025}
\newcommand\vldbauthors{\authors}
\newcommand\vldbtitle{\shorttitle} 
\newcommand\vldbpagestyle{empty} 

\usepackage[most]{tcolorbox}
\usepackage{float}
\usepackage{xspace}
\tcbset{
  aibox/.style={
    width=474.18663pt,
    top=10pt,
    colback=white,
    colframe=black,
    colbacktitle=black,
    enhanced,
    center,
    attach boxed title to top left={yshift=-0.1in,xshift=0.15in},
    boxed title style={boxrule=0pt,colframe=white,},
  }
}
\newtcolorbox{AIbox}[2][]{aibox,title=#2,#1}

\begin{document}
\title{In-depth Analysis of Graph-based RAG in a Unified Framework}



\input{sections/author}

\input{sections/abstract}



\maketitle
\pagestyle{\vldbpagestyle}
\begingroup\small\noindent\raggedright\textbf{PVLDB Reference Format:}\\
\vldbauthors. \vldbtitle. PVLDB, \vldbvolume(\vldbissue): \vldbpages, \vldbyear.\\
\href{https://doi.org/\vldbdoi}{doi:\vldbdoi}
\endgroup
\begingroup
\renewcommand\thefootnote{}\footnote{\noindent
This work is licensed under the Creative Commons BY-NC-ND 4.0 International License. Visit \url{https://creativecommons.org/licenses/by-nc-nd/4.0/} to view a copy of this license. For any use beyond those covered by this license, obtain permission by emailing \href{mailto:info@vldb.org}{info@vldb.org}. Copyright is held by the owner/author(s). Publication rights licensed to the VLDB Endowment. \\
\raggedright Proceedings of the VLDB Endowment, Vol. \vldbvolume, No. \vldbissue\ %
ISSN 2150-8097. \\
\href{https://doi.org/\vldbdoi}{doi:\vldbdoi} \\
}\addtocounter{footnote}{-1}\endgroup

\ifdefempty{\vldbavailabilityurl}{}{
\vspace{.3cm}
\begingroup\small\noindent\raggedright\textbf{PVLDB Artifact Availability:}\\
The source code, data, and/or other artifacts have been made available at \url{https://github.com/JayLZhou/GraphRAG}.
\endgroup
}


\definecolor{1000w}{HTML}{AA0030} 

\definecolor{1w}{HTML}{6B238E}   
\definecolor{10w}{HTML}{00AA00}  

\definecolor{DarkCoral}{HTML}{CD5B45}
\definecolor{DarkSoftApricot}{HTML}{D1A655}

\newcommand{\zhou}[1]{\textcolor{magenta}{[zhou: #1]}}
\input{sections/1_Introduciton}
\input{sections/2_ProblemStatement}

\input{sections/3_GenericModel}

\input{sections/4_GraphBuild}
\input{sections/5_IndexBuild}

\input{sections/6_Retrieve}

\input{sections/6_Experiment}

\input{sections/7_LessonsOpportunities}
\input{sections/7_RelatedWork}
\input{sections/8_Conclution}
\section{Acknowledgments}

This work was supported in part by the 1+1+1 CUHK-CUHK(SZ)-GDSTC Joint Collaboration Fund under Grant 2025A0505000045, Guangdong Provincial Key Laboratory of Mathematical Foundations for Artificial Intelligence (2023B1212010 001), Shenzhen Research Institute of Big Data under Grant SIF20240002, and Huawei Collaboration Fund under Grant TC20240920019.
\clearpage

\balance
\bibliographystyle{ACM-Reference-Format}
\bibliography{ref}

\input{sections/10_Appdenix}

\end{document}

%% file: sections/author.tex
\settopmatter{authorsperrow=3}
\author{Yingli Zhou}
\affiliation{%
  \institution{The Chinese University of Hong Kong, Shenzhen, China}
\country{}
}
\email{yinglizhou@link.cuhk.edu.cn}

\author{Yaodong Su}
\affiliation{%
  \institution{The Chinese University of Hong Kong, Shenzhen, China}
\country{}
}
\email{224045002@link.cuhk.edu.cn}

\author{Youran Sun}
\affiliation{%
  \institution{The Chinese University of Hong Kong, Shenzhen, China}
\country{}
}
\email{youransun@link.cuhk.edu.cn}

\author{Shu Wang}
\affiliation{%
  \institution{The Chinese University of Hong Kong, Shenzhen, China}
\country{}
}
\email{shuwang3@link.cuhk.edu.cn}

\author{Taotao Wang}
\affiliation{%
  \institution{The Chinese University of Hong Kong, Shenzhen, China}
\country{}
}
\email{taotaowang@link.cuhk.edu.cn}

\author{Runyuan He}
\affiliation{%
  \institution{The Chinese University of Hong Kong, Shenzhen, China}
\country{}
}
\email{123090163@link.cuhk.edu.cn}

\author{Yongwei Zhang}
\affiliation{%
\institution{Huawei Cloud Computing Technologies Co., Ltd.}
\country{}
}
\email{zhangyongwei11@huawei.com}

\author{Sicong Liang}
\affiliation{%
\institution{Huawei Cloud Computing Technologies Co., Ltd.}
\country{}
}
\email{liangsicong2@huawei.com}

\author{Xilin Liu}
\affiliation{%
\institution{Huawei Cloud Computing Technologies Co., Ltd.}
\country{}
}
\email{liuxilin3@huawei.com}

\author{Yuchi Ma}
\affiliation{%
\institution{Huawei Cloud Computing Technologies Co., Ltd.}
\country{}
}
\email{mayuchi1@huawei.com}

\author{Yixiang Fang}
\authornote{Yixiang Fang is the corresponding author.}
\affiliation{%
  \institution{The Chinese University of Hong Kong, Shenzhen, China}
\country{}
}
\email{fangyixiang@cuhk.edu.cn}

%% file: sections/abstract.tex
\begin{abstract}
Graph-based Retrieval-Augmented Generation (RAG) has proven effective in integrating external knowledge into large language models (LLMs), improving their factual accuracy, adaptability, interpretability, and trustworthiness.
A number of graph-based RAG methods have been proposed in the literature. 
However, these methods have not been systematically and comprehensively compared under the same experimental settings.
In this paper, \change{we first summarize a unified framework to incorporate all graph-based RAG methods from a high-level perspective.}
We then extensively compare representative graph-based RAG methods over a range of questing-answering (QA) datasets -- from specific questions to abstract questions -- and examine the effectiveness of all methods, providing a thorough analysis of graph-based RAG approaches.
As a byproduct of our experimental analysis, we are also able to identify new variants of the graph-based RAG methods over specific QA and abstract QA tasks respectively, by combining existing techniques, which outperform the state-of-the-art methods.
Finally, based on these findings, we offer promising research opportunities.
We believe that a deeper understanding of the behavior of existing methods can provide new valuable insights for future research.
\end{abstract}

%% file: sections/1_Introduciton.tex
\section{Introduction}
\label{sec:intro}

The development of Large Language Models (LLMs) like GPT-4 ~\cite{achiam2023gpt}, Qwen2.5~\cite{yang2024qwen2}, and Llama 3.1~\cite{dubey2024llama} has sparked a revolution in the field of artificial intelligence~\cite{liu2024survey,huang2023survey,wang2024survey,zheng2024large,li2023large,nie2024survey,ghimire2024generative,wang2024large}.
%
\begin{figure}[]
    \centering
        \setlength{\abovecaptionskip}{-0.05cm}
    \setlength{\belowcaptionskip}{-0.3cm}
    \includegraphics[scale=0.57]{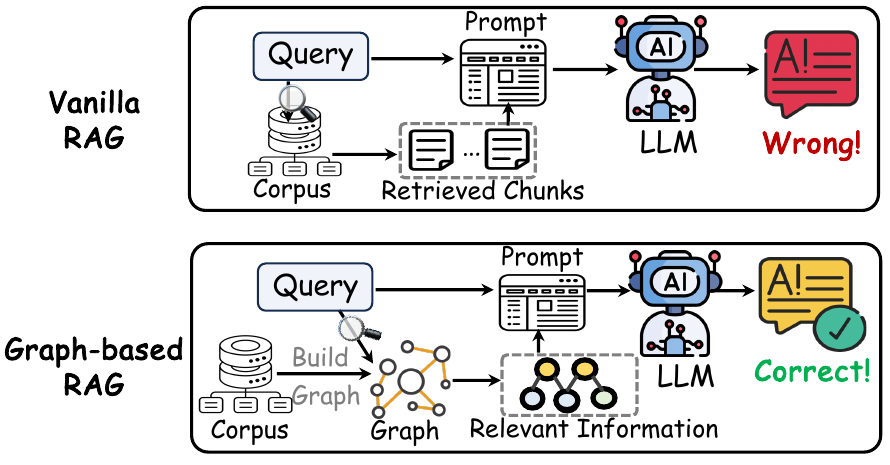}
    \caption{\change{Overview of vanilla RAG and graph-based RAG.}}
    \label{fig:intro}
\end{figure}
Despite their remarkable comprehension and generation capabilities, LLMs may still generate incorrect outputs due to a lack of domain-specific knowledge, real-time updated information, and proprietary knowledge, which are outside LLMs' pre-training corpus, known as ``hallucination''~\cite{peng2024graph}.

\begin{table*}[]
  \centering
  \caption{\change{Classification of existing representative graph-based RAG methods. Index Component indicates which graph elements (e.g., nodes, relationships, communities) are stored in the index. }}
  \renewcommand{\arraystretch}{1.25} 
  \label{tab:overall_table}
  \resizebox{\textwidth}{!}{
  \begin{tabular}{l|l|l|l|l|c|c}
     \toprule
     	 \textbf{Method} & \textbf{Graph Type} & \textbf{Index Component} &  
       \textbf{Retrieval Primitive} & \textbf{Retrieval Granularity} & 
       \textbf{Specific QA} & \textbf{Abstract QA} \\
     \midrule
     
     \texttt{RAPTOR}~\cite{sarthi2024raptor} & Tree & Tree node & Question vector & Tree node & \Yes & \Yes \\  

     \texttt{KGP}~\cite{wang2024knowledge} & Passage Graph & Entity & Question & Chunk & \Yes & \No \\  

    \texttt{HippoRAG}~\cite{gutierrez2024hipporag} & Knowledge Graph & Entity & Entities in question & Chunk & \Yes & \No \\  
    
     \texttt{G-retriever}~\cite{he2024g} & Knowledge Graph & Entity, Relationship & Question vector & Subgraph & \Yes & \No \\  
     \texttt{ToG}~\cite{sunthink} & Knowledge Graph & Entity, Relationship & Question & Subgraph & \Yes & \No \\  

     \texttt{DALK}~\cite{li2024dalk} & Knowledge Graph & Entity & Entities in question & Subgraph & \Yes & \No \\  
     \texttt{LGraphRAG}~\cite{edge2024local} & Textual Knowledge Graph  
     & Entity, Community & Question vector & Entity, Relationship, Chunk, Community &  \Yes & \No \\ 
     
     \texttt{GGraphRAG}~\cite{edge2024local} & Textual Knowledge Graph
     & Community & Question vector & Community  & \No & \Yes \\ 

     \texttt{FastGraphRAG}~\cite{fastgraphrag} & Textual Knowledge Graph & Entity & Entities in question & Entity, Relationship, Chunk & \Yes & \Yes \\  
     \texttt{LLightRAG}~\cite{guo2024lightrag} & Rich Knowledge Graph
     & Entity, Relationship & Low-level keywords in question & Entity, Relationship, Chunk & \Yes & \Yes  \\  

     \texttt{GLightRAG}~\cite{guo2024lightrag} & Rich Knowledge Graph  
     & Entity, Relationship  & High-level keywords in question & Entity, Relationship, Chunk & \Yes & \Yes \\  

     \texttt{HLightRAG}~\cite{guo2024lightrag} & Rich Knowledge Graph  
     & Entity, Relationship & Both high- and low-level keywords & Entity, Relationship, Chunk & \Yes & \Yes  \\

     \bottomrule
  \end{tabular}
  }
\end{table*}

To bridge this gap, the Retrieval Augmented Generation (RAG) technique~\cite{fan2024survey,gao2023retrieval,hu2024rag,huang2024survey,wu2024retrieval,yu2024evaluation,zhao2024retrieval} has been proposed, which supplements LLM with external knowledge to enhance its factual accuracy and trustworthiness. 
\change{Given a user query $Q$, the key idea of naive-based RAG \cite{lewis2020retrieval} (i.e., vanilla RAG) is to retrieve relevant chunks from the external corpus, and then feed them along with $Q$ as a prompt into LLM to generate answers.}
Consequently, RAG techniques have been widely applied in various fields, especially in domains where LLMs need to generate reliable outputs, such as healthcare ~\cite{liu2024survey,wang2024survey,zheng2024large}, finance \cite{nie2024survey,li2023large}, and education \cite{ghimire2024generative,wang2024large}.
Moreover, RAG has proven highly useful in many data management tasks, including NL2SQL~\cite{fan2024combining,li2024dawn}, data cleaning~\cite{li2024autodcworkflow,naeem2024retclean,fan2024cost,qian2024unidm}, knob tuning~\cite{lao2023gptuner,giannankouris2024lambda}, DBMS diagnosis~\cite{zhou2023d,zheng2024revolutionizing,singh2024panda}, and SQL rewrite~\cite{sun2024r,li2024llm}. 
\change{In turn, the database community has recently begun to actively explore how to build efficient and reliable RAG systems~\cite{jiang2024chameleon,agarwal2025cache}.}
Due to the important role of the RAG technique in LLM-based applications, numerous RAG methods have been proposed in the past year~\cite {hu2024rag,li2023survey}.
Among these methods, the state-of-the-art RAG approaches typically use the graph data as the external data (also called graph-based RAG), since they capture the rich semantic information and link relationships between entities.
\change{Unlike vanilla RAG, graph-based RAG methods retrieve relevant information related to the query $Q$—such as nodes, relationships, or subgraphs—from the graph, and then incorporate this information into the prompt along with $Q$ for the LLM to generate an answer. }
The overview of naive-based RAG and graph-based RAG is shown in Figure~\ref{fig:intro}.

\change{Recently, several mainstream database systems have started supporting graph-based RAG, including PostgreSQL \cite{postgresrag}, Neo4j \cite{neo4jrag}, and Databricks \cite{databricksrag}.
At the same time, graph-based RAG has become a core component in modern graph-native agentic systems, such as LangGraph \cite{LangGraph} and Chat2Graph \cite{chat2graph}.
}
Following the success of graph-based RAG, researchers from fields such as database, data mining, machine learning, and natural language processing have designed efficient and effective graph-based RAG methods~\cite{sarthi2024raptor,wu2024medical,wang2024knowledge,li2024dalk,gutierrez2024hipporag,edge2024local,guo2024lightrag,peng2024graph,huang2025ket,li2023gslb,zhang2025erarag,chen2025you,zhuang2025linearrag,xiang2025use}.
In Table \ref{tab:overall_table}, we summarize the key characteristics of 12 representative graph-based RAG methods based on the graph types they rely on, their index components, retrieval primitives and granularity, and the types of tasks they support.
After a careful literature review, we make the following observations. First, no prior work has proposed a unified framework to abstract the graph-based RAG solutions and identify key
performance factors. Second, existing works focus on evaluating
the overall performance, but not individual components. Third,
there is no existing comprehensive comparison between all these
methods in terms of accuracy and efficiency.

\textbf{Our work.} To address the above issues, in this paper, we conduct an in-depth study on graph-based RAG methods.
We first \change{summarize a novel unified framework with four stages,} namely  {\Large \ding{182}} {\textit{Graph building}},  {\Large \ding{183}} {\textit{Index construction}}, {{{\Large \ding{184}}} {\it Operator configuration}, and {{{\Large \ding{185}}} {\it Retrieval \& generation}, which captures the core ideas of all existing methods. 
Under this framework, we systematically compare 12 existing representative graph-based RAG methods.
We conduct comprehensive experiments on the widely used question-answering (QA) datasets, including the specific and abstract questions, which evaluate the effectiveness of these methods in handling diverse query types and provide an in-depth analysis.

In summary, our principal contributions are as follows.

\begin{itemize}
    \item Summarize a novel unified framework with four stages for graph-based RAG solutions from a high-level perspective (Sections \ref{sec:generic} $\sim$ \ref{sec:retrieve_and_generation}).
     
    \item Conduct extensive experiments from different angles using various benchmarks, providing a thorough analysis of graph-based RAG methods. Based on our analysis, we identify new variants of graph-based RAG methods, by combining existing techniques, which outperform the state-of-the-art methods (Section \ref{sec:experiments}).
    
    \item  Summarize lessons learned and propose practical research opportunities that can facilitate future studies (Section \ref{sec:lessons_opp}).
\end{itemize}

The rest of the paper is organized as follows.
In Section \ref{sec:pre}, we present the preliminaries and introduce a novel unified framework for graph-based RAG solutions in Section \ref{sec:generic}.
In Sections \ref{sec:graph_build} through \ref{sec:retrieve_and_generation}, we compare the graph-based RAG methods under our unified framework.
The comprehensive experimental results and analysis are reported in Section \ref{sec:experiments}. 
Section \ref{sec:related}  reviews related work while Section \ref{sec:conclusions} summarizes the paper.

%% file: sections/2_ProblemStatement.tex
\begin{figure*}[]
    \centering
         \setlength{\abovecaptionskip}{-0.01em}
        \setlength{\belowcaptionskip}{-1em}
    \includegraphics[scale=0.7]{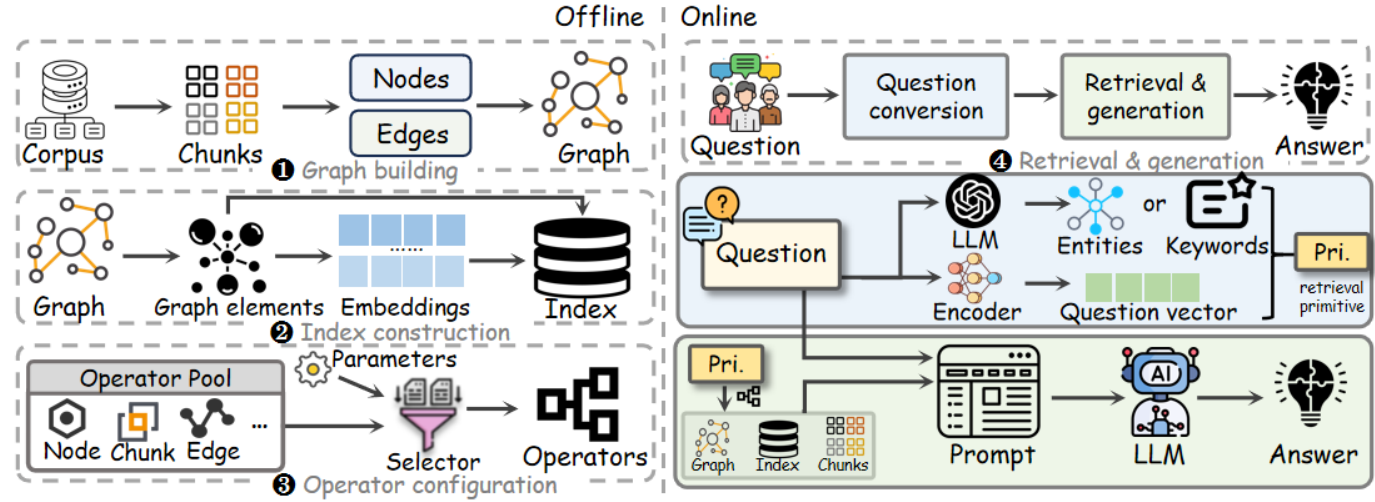}
    \caption{\change{Workflow of graph-based RAG methods under our unified framework.}}
    \label{fig:overview}
\end{figure*}

\section{Preliminaries}
\label{sec:pre}

In this section, we review some key concepts of LLM and the general workflow of graph-based RAG methods.

\subsection{Large Language Models (LLMs)}
We introduce some fundamental concepts of LLMs, including LLM
prompting and retrieval augmented generation (RAG).

{\bf LLM Prompting.} After instruction tuning on large corpus of human interaction scenarios, LLM is capable of following human instructions to complete different tasks \cite{dong2022survey,ouyang2022training}. Specifically, given the task input, we construct a prompt that encapsulates a comprehensive task description. The LLM processes this prompt to fulfill the task and generate the corresponding output.
Note that pre-training on trillions of bytes of data enables LLM to generalize to diverse tasks by simply adjusting the prompt \cite{ouyang2022training}.

{\bf Retrieval Augmented Generation.} During completing tasks
with prompting, LLMs often generate erroneous or meaningless responses, i.e., the hallucination problem ~\cite{huang2023survey}. 
To mitigate the problem, retrieval augmented generation (RAG) is utilized as an advanced LLM prompting technique by using the knowledge within the external corpus, typically including two major steps~\cite{gao2023retrieval}: {\it (1) retrieval}: given
a user question $Q$, using the index to retrieve the most relevant (i.e., top-$k$) chunks to $Q$, where the large corpus is first split into smaller chunks, and (2) {\it generation}: guiding LLM to generate answers with the retrieved chunks along with $Q$ as a prompt.

\subsection{Graph-based RAG}

Unlike vanilla RAG, graph-based RAG methods employ graph structures built from external corpus to enhance contextual understanding in LLMs and generate more informed and accurate responses~\cite{peng2024graph}.
Typically, graph-based RAG methods are composed of three major stages: {\it (1) graph building}: given a large corpus \(\mathcal{D}\) with \(d\) chunks, for each chunk, an LLM extracts nodes and edges, which are then combined to construct a graph \(\mathcal{G}\); {\it (2) retrieval}: given a user question $Q$, using the index to retrieve the most relevant information (e.g., nodes or subgraphs) from $\mathcal{G}$, and (3) {\it generation}: guiding LLM to generate answers by incorporating the retrieved information into the prompt along with $Q$.
%


%% file: sections/3_GenericModel.tex
\section{A UNIFIED FRAMEWORK}
\label{sec:generic}

In this section, we develop a novel unified framework, consisting of four stages: {\Large \ding{182}} {\textit{Graph building}},  {\Large \ding{183}} {\textit{Index construction}}, {{{\Large \ding{184}}} {\it Operator configuration}, and {{{\Large \ding{185}}} {\it Retrieval \& generation}, which can cover all existing graph-based RAG methods, as shown in Algorithm \ref{alg:framework}.

\begin{algorithm}[h]
  \caption{{A unified framework for graph-based RAG}}
  \label{alg:framework}
  \small
   \SetKwInOut{Input}{input}\SetKwInOut{Output}{output}
    \Input{Corpus $\mathcal{D}$, user question $Q$, and parameters $\mathcal{P}$}
    \Output{The answers for user question $Q$}
        $\mathcal{C} \gets$ split $\mathcal{D}$ into multiple chunks\;
        \tcp{\textcolor{teal}{(1) Graph building.}}
        $\mathcal{G} \gets${\tt  GraphBuilding}($\mathcal{C}$)\;
        
        \tcp{\textcolor{teal}{(2) Index construction.}}
        $\mathcal{I} \gets$ {\tt IndexConstruction}($\mathcal{G}$, $\mathcal{C}$)\;
        
        \tcp{\textcolor{teal}{(3) Operator configuration.}}
        $\mathcal{O} \gets$ {\tt OperatorConfiguration}( $\mathcal{P}$);

        \tcp{\textcolor{teal}{(4) Retrieve relevant information and generate response.}}
        $\mathcal{R} \gets$ {\tt Retrieval\&generation}($\mathcal{G}$, $\mathcal{I}$, $\mathcal{O}$, $Q$, $\mathcal{C}$);
    
    
    \Return{$\mathcal{R}$;}

\end{algorithm}

Specifically, given the large corpus $\mathcal{D}$, we first split it into multiple chunks $\mathcal{C}$ (line 1).
We then sequentially execute operations in the following four stages (lines 2-5):
(1) Build the graph $\mathcal{G}$ for input chunks $\mathcal{C}$ (Section \ref{sec:graph_build});
(2) Construct the index based on the graph $\mathcal{G}$ from the previous stage (Section \ref{sec:index_build});
(3) Configure the retriever operators for subsequent retrieving stages (Section \ref{sec:retrieve_and_generation}), and   
(4) For the input user question $Q$, retrieve relevant information from $\mathcal{G}$ using the selected operators and feed them along with the question $Q$ into the LLM to generate the answer.
\change{
The workflow of graph-based RAG methods under our framework is shown in Figure~\ref{fig:overview}.
We note that graph-based RAG methods differ across the four stages. Specifically, as shown in Table~\ref{tab:overall_table}, different methods construct distinct types of graphs in Stage {\Large \ding{182}}, build different indices in Stage {\Large \ding{183}}, and retrieve information at varying levels of granularity. Consequently, they employ different operators for retrieval, leading to variations in both Stage {\Large \ding{184}} and Stage {\Large \ding{185}}.
Note that the query is first converted into a retrieval primitive, which serves as the basis for retrieval in each method (see Section~\ref{sec:retri_gene}).
}

%% file: sections/4_GraphBuild.tex

\section{Graph building}
\label{sec:graph_build}

The {\it graph building} stage aims to transfer the input corpus into a graph, serving as a fundamental component in graph-based RAG methods. 
\change{Before building a graph, the corpus is first split into smaller chunks. Then, an LLM or other tools are used to construct nodes and edges based on these chunks, as shown in Figure \ref{fig:overview}{\Large \ding{182}}.
We note that this preprocessing step is essential for all RAG methods. Further details are provided in our technical report~\cite{appendix}.}
There are five types of graphs, each with a corresponding construction method; we present a brief description of each graph type and its construction method below:

\noindent {\Large \ding{182}} {\textit{Passage Graph.}} 
In the passage graph (PG), each chunk represents a node, and edges are built by the entity linking tools~\cite{wang2024knowledge}.
If two chunks contain a number of the same entities larger than a threshold, we link an edge for these two nodes. 

\noindent {\Large \ding{183}} {\textit{Tree.}} 
The tree is constructed in a progressive manner, where each chunk represents the leaf node in the tree. 
Then, it uses an LLM to generate higher-level nodes.
Specifically, at the \(i\)-th layer, the nodes of \((i+1)\)-th layer are created by clustering nodes from the \(i\)-th layer that does not yet have parent nodes. 
For each cluster with more than two nodes, the LLM generates a virtual parent node with a high-level summary of its child node descriptions.


\begin{table*}[h]
\centering
\footnotesize
\caption{\textbf{\change{Comparison of different types of graph.}}}
\begin{tabular}{lccccccccccc}
\toprule
\textbf{Graph} 
& \makecell[c]{\textbf{Entity}\\\textbf{Name}} 
& \makecell[c]{\textbf{Entity}\\\textbf{Type}} 
& \makecell[c]{\textbf{Entity}\\\textbf{Description}} 
& \makecell[c]{\textbf{Relationship}\\\textbf{Name}} 
& \makecell[c]{\textbf{Relationship}\\\textbf{Keyword}} 
& \makecell[c]{\textbf{Relationship}\\\textbf{Description}} 
& \makecell[c]{\textbf{Edge}\\\textbf{Weight}} 
& \makecell[c]{\textbf{Token}\\\textbf{Consuming}} 
& \makecell[c]{\textbf{Graph}\\\textbf{Size}} 
& \makecell[c]{\textbf{Information}\\\textbf{Richness}} 
& \makecell[c]{\textbf{Construction}\\\textbf{Time}} \\
\midrule
Tree & \No & \No & \No & \No & \No & \No & \No & \ding{72} & \ding{72} & \ding{72}\ding{72} & \ding{72} \\
PG   & \No & \No & \No & \No & \No & \No & \No & N/A & \ding{72}\ding{72}\ding{72}\ding{72} & \ding{72} & \ding{72}\ding{72}\ding{72}\ding{72} \\
KG   & \Yes & \No & \No & \Yes & \No & \No & \Yes & \ding{72}\ding{72} & \ding{72}\ding{72} & \ding{72}\ding{72}\ding{72} & \ding{72}\ding{72} \\
TKG  & \Yes & \Yes & \Yes & \No & \No & \Yes & \Yes & \ding{72}\ding{72}\ding{72} & \ding{72}\ding{72}\ding{72} & \ding{72}\ding{72}\ding{72}\ding{72} & \ding{72}\ding{72}\ding{72} \\
RKG  & \Yes & \Yes & \Yes & \No & \Yes & \Yes & \Yes & \ding{72}\ding{72}\ding{72} & \ding{72}\ding{72}\ding{72} & \ding{72}\ding{72}\ding{72}\ding{72}\ding{72} & \ding{72}\ding{72}\ding{72} \\
\bottomrule
\end{tabular}
\label{tab:compare_graph}
\end{table*}
\begin{figure*}[]
    \centering
    \includegraphics[width=1.0\linewidth]{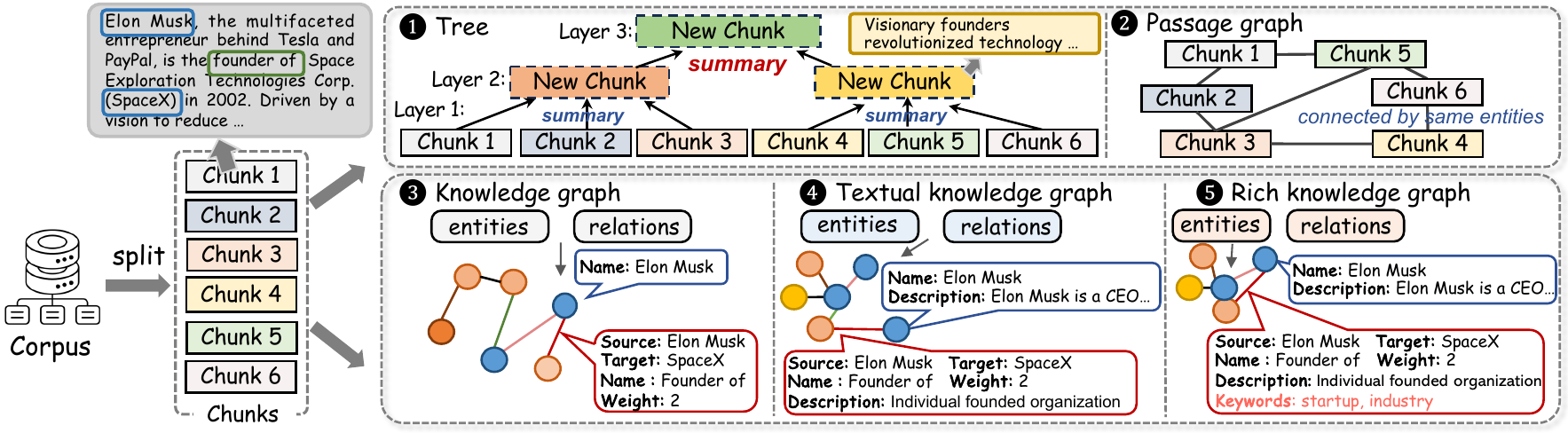}
    \caption{\change{Examples of five types of graphs.}}
    \label{fig:graph_example}
\end{figure*}

\noindent {\Large \ding{184}} {\textit{Knowledge Graph.}} The knowledge graph (KG) is constructed by extracting entities and relationships from each chunk, where each entity represents an object and the relationship denotes the semantic relation between two entities.

\noindent {\Large \ding{185}} {\textit{Textual Knowledge Graph.}} A textual knowledge graph (TKG) is a specialized KG (following the same construction step as KG), with the key difference being that in a TKG, each entity and relationship is assigned a brief textual description.

\noindent {\Large \ding{186}} {\textit{Rich Knowledge Graph.}}  The rich knowledge graph (RKG)  is an extended version of TKG, containing more information, including textual descriptions for entities and relationships, as well as keywords for relationships.
\change{We summarize the key characteristics of each graph type in Table~\ref{tab:compare_graph}, considering their contained attributes (e.g., inclusion of entity names and descriptions), the time and number of tokens required for construction, the resulting graph size, and the richness of information contained within each graph.
In addition, we provide a case study of how the five types of graphs are constructed from the external corpus in Figure \ref{fig:graph_example}.
The detailed descriptions are shown in our technical report \cite{appendix}.
}

%% file: sections/5_IndexBuild.tex
\section{Index Construction}
\label{sec:index_build}
To support efficient online querying, existing graph-based RAG methods typically include an index-construction stage, \change{which involves storing graph elements — such as entities and relationships — in a vector database and computing community reports to enable efficient online retrieval, as shown in Figure~\ref{fig:overview}{\Large \ding{183}}.}
Generally, there are three types of indices, {\Large \ding{182}} \textit{Node index},  {\Large \ding{183}} \textit{Relationship index}, and  {\Large \ding{184}} \textit{Community index}, where for the first two types, we use the well-known text-encoder models, such as BERT~\cite{devlin2018bert} or ColBert~\cite{khattab2020colbert} to generate embeddings for nodes or relationships in the graph.

\noindent {\Large \ding{182}} {\textit{Node index}}  stores the graph nodes in the vector database.
For \texttt{RAPTOR}, \texttt{G-retriever}, \texttt{DALK}, \texttt{FastGraphRAG}, \texttt{LGraphRAG}, \texttt{LLightRAG}, and \texttt{HLightRAG}, all nodes in the graph are directly stored in the vector database. For each node in KG, its embedding vector is generated by encoding its entity name, while for nodes in Tree, TKG, and RKG, the embedding vectors are generated by encoding their associated textual descriptions.
In \texttt{KGP}, it stores the TF-IDF matrix~\cite{han2022data}, which represents the term-weight distribution across different nodes (i.e., chunks) in the index.

\noindent {\Large \ding{183}} {\textit{Relationship index}} stores the relationships of the graph in a vector database, where for each relationship, its embedding vector is generated by encoding a description that combines its associated context (e.g., description) and the names of its linked entities.

\noindent {\Large \ding{184}} {\textit{Community index}} stores the community reports for each community, where communities are generated by the clustering algorithm and the LLM produces the reports.
Specifically, Leiden~\cite{traag2019louvain} algorithm is utilized by {\tt LGraphRAG} and {\tt GGraphRAG}. 

\change{\textbf{Remark.} {\it Relationship index} tends to have a larger size, whereas the {\it Community index} is more compact incurs the highest construction cost in terms of tokens and time, and provides the detailed comparisons of these indices in our technical report \cite{appendix}.}

%% file: sections/6_Retrieve.tex
\section{Retrieval and Generation}
\label{sec:retrieve_and_generation}

In this section, we explore the key steps in graph-based RAG methods, i.e., selecting operators,  and using them to retrieve relevant information to question $Q$.

\subsection{Retrieval operators}
\change{In this subsection, we demonstrate that the retrieval stage of various graph-based RAG methods can be abstracted into a modular sequence of operators. Different methods select and compose these operators in distinct ways, enabling flexible and extensible retrieval pipelines.
By systematically analyzing existing graph-based RAG implementations, we identify a comprehensive set of 19 retrieval operators, and based on the granularity of retrieval, we classify the operators into five categories.
We note that most operators are derived from designs described in the original papers—though often unnamed—so we assign them meaningful and consistent names. For the remaining operators, which are not explicitly defined in the literature, we extract and summarize them based on source code analysis.
Importantly, by selecting and arranging these operators in different sequences, all existing (and potentially future) graph-based RAG methods can be implemented.}
%


$\bullet$ \textit{Node type.} This type of operator focuses on retrieving ``important'' nodes for a given question, and based on the selection policy, there are seven different operators to retrieve nodes.
\noindent {\Large \ding{182}} {\tt VDB} leverages the vector database to retrieve nodes by computing the vector similarity with the query vector. 
\noindent {\Large \ding{183}} {\texttt{RelNode}} extracts nodes from the provided relationships.
\noindent {\Large \ding{184}} {\texttt{PPR}} uses the Personalized PageRank (PPR) algorithm~\cite{haveliwala2002topic} to identify the top-$k$ similar nodes to the question, where the restart probability of each node is based on its similarity to the entities in the given question.
\noindent {\Large \ding{185}} {\texttt{Agent}} utilizes the capabilities of LLMs to select nodes from a list of candidate nodes.
\noindent {\Large \ding{186}} {\texttt{Onehop}} selects the one-hop neighbor entities of the given entities.
\noindent {\Large \ding{187}} {\texttt{Link}} selects the top-1 most similar entity for each entity in the given set from the vector database.
\noindent {\Large \ding{188}} {\texttt{TF-IDF}} retrieves the top-$k$ relevant entities by ranking them based on term frequency and inverse document frequency from the TF-IDF matrix.

\begin{table*}[]
\centering
\small
\setlength{\tabcolsep}{7.5pt} 
\caption{Operators utilized in graph-based RAG methods; ``N/A'' means that this type of operator is not used.}
\begin{tabular}{l|c|c|c|c|c}
\toprule
\textbf{Method} & \textbf{Node} & \textbf{Relationship} & \textbf{Chunk} & \textbf{Subgraph} & \textbf{Community} \\ 
\midrule
{\tt RAPTOR}         & {\tt VDB}  & N/A             & N/A                & N/A            & N/A    \\ 
{\tt KGP}            & {\tt TF-IDF} & N/A            & N/A                & N/A            & N/A    \\ 
{\tt HippoRAG}       & {\tt Link} + {\tt PPR}  & {\tt Aggregator}  & {\tt Aggregator}  & N/A  & N/A    \\ 
{\tt G-retriever}    & {\tt VDB}  & {\tt VDB}             & N/A                & {\tt Steiner}        & N/A    \\ 
{\tt ToG}            & {\tt Link} + {\tt Onehop} + {\tt Agent}  & {\tt Onehop} + {\tt Agent}  & N/A  & N/A  & N/A    \\ 
{\tt DALK}           & {\tt Link} + {\tt Onehop} + {\tt Agent} & N/A & N/A  & {\tt KhopPath} + {\tt AgentPath} & N/A    \\
{\tt FastGraphRAG}   & {\tt Link} + {\tt VDB} + {\tt PPR}  & {\tt Aggregator}  & {\tt Aggregator}  & N/A  & N/A    \\ 

{\tt LGraphRAG}      & {\tt VDB}  & {\tt Onehop}          & {\tt Occurrence}         & N/A            & {\tt Entity} \\ 
{\tt RGraphRAG}      & N/A  & N/A             & N/A                & N/A            & {\tt Layer}  \\ 
{\tt LLightRAG}      & {\tt VDB}  & {\tt Onehop}          & {\tt Occurrence}         & N/A            & N/A    \\ 
{\tt GLightRAG}      & {\tt FromRel} & {\tt VDB}           & {\tt FromRel}           & N/A            & N/A    \\ 
{\tt HLightRAG}      & {\tt VDB} + {\tt FromRel}  & {\tt Onehop} + {\tt VDB}  & {\tt Occurrence} + {\tt FromRel} & N/A    & N/A    \\

\bottomrule
\end{tabular}
\label{tab:operator}
\end{table*}

$\bullet$ \textit{Relationship type.} These operators are designed to retrieve relationships from the graph that are most relevant to the user question. 
There are four operators:{\Large \ding{182}} {\texttt{VDB}},  {\Large \ding{183}} {\texttt{Onehop}}, {\Large \ding{184}} {\texttt{Aggregator}}, and {\Large \ding{185}} {\texttt{Agent}}.
Specifically, the \texttt{VDB} operator also uses the vector database to retrieve relevant relationships. 
The {\texttt{Onehop}} operator selects relationships linked by one-hop neighbors of the given selected entities. 
The {\texttt{Aggregator}} operator builds upon the \texttt{PPR} operator in the node operator. 
Given the PPR scores of entities, the most relevant relationships are determined by leveraging entity-relationship interactions. Specifically, the score of each relationship is obtained by summing the scores of the two entities it connects. Thus, the top-$k$ relevant relationships can be selected.
The key difference for the \texttt{Agent} operator is that, instead of using a candidate entity list, it uses a candidate relationship list, allowing the LLM to select the most relevant relationships based on the question.

$\bullet$ \textit{Chunk type.} The operators in this type aim to retrieve the most relevant chunks to the given question.
There are three operators: {\Large \ding{182}} {\texttt{Aggregator}},  {\Large \ding{183}} {\texttt{FromRel}}, and {\Large \ding{184}}  {\texttt{Occurrence}}.
\change{Specifically, {\tt Aggregator} uses the relationship score vector from the {\tt Link} operator and a relationship–chunk interaction matrix to aggregate chunk scores via matrix multiplication, selecting the top-$k$ chunks with the highest scores.
For \texttt{FromRel}: Given a set of relationships, all chunks containing at least one of them are retrieved.
The \texttt{Occurrence} selects the top-$k$ chunks based on the given relationships.
Specifically, for each relationship, we identify its two associated entities. If both entities appear in the same chunk, the chunk's score is incremented by 1. After processing all relationships, the top-$k$ chunks with the highest scores are selected.}

$\bullet$ \textit{Subgraph type.} There are three operators to retrieve the relevant subgraphs from the graph $\mathcal{G}$:
The {{\Large \ding{182}} \texttt{KhopPath}} operator aims to identify $k$-hop paths in $\mathcal{G}$ by iteratively finding such paths where the start and end points belong to the given entity set. After identifying a path, the entities within it are removed from the entity set, and this process repeats until the entity set is empty.
Note that if two paths can be merged, they are combined into one path.
For example, if we have two paths \(A \to B \to C\) and \(A \to B \to C \to D\), we can merge them into a single path \(A \to B \to C \to D\).
The {\Large \ding{183}} {\texttt{Steiner}} operator first identifies the relevant entities and relationships, then uses these entities as seed nodes to construct a Steiner tree \cite{han2022data}.
The {\Large \ding{184}} {\texttt{AgentPath}} operator aims to identify the most relevant $k$-hop paths to a given question, by using LLM to filter out the irrelevant paths.

$\bullet$ \textit{Community type.} 
\change{Only the {\tt LGraphRAG} and {\tt GGraphRAG} using the community operators, which includes two detailed operators, {\Large \ding{182}} {\texttt{Entity}},  and {\Large \ding{183}} {\texttt Layer}.
This operator first identifies communities that contain the specified entities, with each community maintaining an associated entity list. It then ranks the selected communities based on relevance scores assigned by the LLM, returning the top-$k$ highest-scoring ones.
Each community is associated with a layer attribute, and the {\tt Layer} operator retrieves all communities at or below the specified layer.}

%


\subsection{Operator configuration}

Under our unified framework, any existing graph-based RAG method can be implemented by leveraging the operator pool along with specific method parameter $\mathcal{P}$, \change{as shown in Figure \ref{fig:overview}{\Large \ding{184}}.} 
\change{Instead, $\mathcal{P}$ acts as a control module that configures the retrieval pipeline for a given graph-based RAG method by determining: (1) which atomic operators should be used in the method; and (2) the execution order of these operators within the retrieval process.}

In Table \ref{tab:operator}, we present how the existing graph-based RAG methods utilize our provided operators to assemble their retrieval stages.
\change{For example, {\tt LLightRAG} first applies the {\tt VDB} operator to retrieve relevant nodes, then uses the {\tt Onehop} operator to retrieve relevant relationships, and finally employs the {\tt Occurrence} operator to obtain the relevant chunks. 
In the above example, we can set $\mathcal P=\langle$ {\tt VDB}, {\tt Onehop}, {\tt Occurrence}$ \rangle$.
Essentially, the parameter $\mathcal{P}$ represents the retrieval configuration to distinguish the retrieval stage for each specific graph-based RAG method.
}
Due to this independent and modular decomposition of all graph-based RAG methods, we not only gain a deeper understanding of how these approaches work but also gain the flexibility to combine these operators to create new methods. 
Besides, new operators can be easily created, for example, we can create a new operator \texttt{VDB} within the community type, which allows us to retrieve the most relevant communities by using vector search to compare the semantic similarity between the question and the communities.
In our later experimental results (see Exp.5 in Section~\ref{sec:exp:abs_qa}), thanks to our modular design, we can design a new state-of-the-art RAG method by first creating two new operators and combining them with the existing operators.

\subsection{Retrieval \& generation}
\label{sec:retri_gene}
In the {\it Retrieval \& generation} stage, the graph-based RAG methods first go through a {\it Question conversion} stage (see the second subfigure on the right side of Figure~\ref{fig:overview}), which aims to transfer the user input question $Q$ into the retrieval primitive $\mathcal{D}$, where $\mathcal{D}$ denotes the atomic retrieval unit, such as entities or keywords in $Q$, and the embedding vector of $Q$.

In the  {\it Question conversion} stage, \texttt{DALK}, \texttt{HippoRAG}, and \texttt{ToG} extract entities from the question; \texttt{KGP} directly uses the original question as the retrieval primitive.
The three versions of \texttt{LightRAG} extract keywords from the question as the retrieval primitive, and the remaining methods use the embedding vector of $Q$.

Based on the retrieval primitive $\mathcal{D}$ and the selected operators, the most relevant information to $Q$ is retrieved and combined with $Q$ to form the final prompt for LLM response generation.
Generally, there are two types of answer generation paradigms: {\ding{182} {\tt Directly} and {\ding{183} {\tt Map-Reduce}}. 
\change{The former directly utilizes the LLM to generate the
answer. The {\tt Map-Reduce} strategy in {\tt GGraphRAG} prompts the LLM to generate partial answers and confidence scores from each retrieved community. These (answer, score) pairs are ranked, and the top ones are appended to form the final prompt for answer generation.
An example is shown in~\cite{appendix}.}

%% file: sections/6_Experiment.tex
\section{Experiments}
\label{sec:experiments}

We now present the experimental results. Section \ref{sec:exp:setup} discusses
the setup. We discuss the results for specific QA and abstract QA tasks in Sections \ref{sec:exp:speci} and \ref{sec:exp:abs_qa}, respectively.

\input{sections/6_Experiment/setup}

\input{sections/6_Experiment/specific}

\input{sections/6_Experiment/abstract}

%% file: sections/6_Experiment/setup.tex
\subsection{Setup}
\label{sec:exp:setup}
  \input{sections/6_Experiment/experiment_plot}

\definecolor{sota}{RGB}{191, 0, 64}
\definecolor{second}{RGB}{191, 129, 64}
\newcommand{\fst}[1]{\textbf{\textcolor{sota}{#1}}}
\newcommand{\snd}[1]{\textbf{\textcolor{second}{#1}}}
\begin{table}[]
    \centering
    \footnotesize
    \caption{Datasets used in our experiments; The underlined number of chunks denotes that the dataset is pre-split into chunks by the expert annotator.}
    \begin{tabular}{c|rrrc}
        \toprule
        Dataset & \# of Tokens & \# of Questions & \# of Chunks & QA Type\\
        \midrule
        MultihopQA & 1,434,889 & 2,556 & 609  & Specific QA \\
         Quality & 1,522,566 & 4,609 &265 & Specific QA \\
         PopQA  & 2,630,554 & 1,172  & \underline{33,595} & Specific QA\\
        MusiqueQA & 3,280,174 & 3,000 & \underline{29,898} & Specific QA \\ 
        HotpotQA   &8,495,056 & 3,702 &\underline{66,581} & Specific QA\\
        ALCE & 13,490,670 & 948 & \underline{89,562}& Specific QA \\
    \midrule
        Mix & 611,602 & 125 & 61& Abstract QA  \\
        MultihopSum & 1,434,889 & 125 & 609& Abstract QA  \\
        Agriculture &1,949,584 & 125 &12 & Abstract QA  \\
         CS & 2,047,923& 125 & 10& Abstract QA \\
       Legal &4,774,255 & 125 &94 & Abstract QA   \\
        \bottomrule
    \end{tabular}
    \label{tab:dataset}
\end{table}

\begin{table*}[]
\centering
\caption{Comparison of methods on different datasets, where \fst{Purple} denotes the best result, and \snd{Orange} denotes the best result excluding the best one; For the three largest datasets, we replace the clustering method in \texttt{RAPTOR} from Gaussian Mixture to K-means, as the former fails to finish within two days; The results of this version (i.e., K-means) are marked with $^{\dagger}$.}
\small
\begin{tabular}{l|r|r|r|r|r|r|r|r|r|r|r|r}
\toprule
\multirow{2}{*}{Method} & \multicolumn{2}{c|}{MultihopQA}  & Quality & \multicolumn{2}{c|}{PopQA}  & \multicolumn{2}{c|}{MusiqueQA} & \multicolumn{2}{c|}{HotpotQA}  & \multicolumn{3}{c}{ALCE}  \\ \cline{2-13}
 & Accuracy &  Recall & Accuracy  & Accuracy  & Recall & Accuracy & Recall & Accuracy & Recall & STRREC & STREM & STRHIT \\ \midrule
{\tt ZeroShot} & 49.022 & 34.256  & 37.058 & 28.592 & 8.263 & 1.833 & 5.072   & 35.467 & 42.407  & 15.454 & 3.692 & 30.696  \\ 
{\tt VanillaRAG}  & 50.626 & 36.918  & 39.141  & \snd{60.829} & \snd{27.058} & 17.233 & 27.874 & \snd{50.783} & 57.745  & \snd{34.283} & \fst{11.181} & \snd{63.608}  \\ \midrule
{\tt G-retriever} & 42.019 & 43.116 & 31.807  & 17.084 & 6.075 & 2.733 & 11.662 & --- & ---   & 9.754 & 2.215 & 19.726  \\ 
{\tt ToG} & 41.941 & 38.435 & 34.888 & 47.677 & 23.727  & 9.367 & 20.536 & ---  & --- & 13.975 & 3.059 & 29.114\\ 
{\tt KGP} & 48.161 & 36.272  & 33.955  & 57.255 & 24.635  & \snd{17.333} & 27.572  & --- & ---   & 27.692  & 8.755 & 51.899  \\ 
{\tt DALK}   & 53.952  & 47.232 & 34.251 & 45.604  & 19.159 & 11.367 & 22.484  & 33.252  & 47.232  & 21.408 & 4.114  & 44.937\\ 
{\tt LLightRAG} & 44.053 & 35.528 & 34.780& 38.885 & 16.764   & 9.667 & 19.810 & 34.144 & 41.811   & 21.937 & 5.591 & 43.776  \\ 
{\tt GLightRAG} & 48.474 & 38.365  & 33.413 & 20.944 & 8.146  & 7.267 & 17.204 & 25.581 & 33.297 & 17.859 & 3.587 & 37.131   \\ 
{\tt HLightRAG} & 50.313 & 41.613 & 34.368 & 41.244 & 18.071 & 11.000 & 21.143    & 35.647 & 43.334 & 25.578 & 6.540 & 50.422  \\ 
{\tt FastGraphRAG} & 52.895 & 44.278 & 37.275   & 53.324 & 22.433  & 13.633 & 24.470  & 43.193 & 51.007  & 30.190 & 8.544 & 56.962\\ 
{\tt HippoRAG} & 53.760 & \snd{47.671} & \snd{48.297}  & 59.900 & 24.946   & 17.000 & \snd{28.117} & 50.324 & \snd{58.860}  & 23.357 & 6.962 & 43.671 \\ 
{\tt LGraphRAG} & \snd{55.360} & \fst{50.429} & 37.036 & 45.461 & 18.657    & 12.467 & 23.996 & 33.063 & 42.691 & 28.448 & 8.544 & 54.747    \\ 
{\tt RAPTOR}  & \fst{56.064} & 44.832  & \fst{56.997} & \fst{62.545} & \fst{27.304}    &\fst{24.133$^\dagger$} & \fst{35.595$^{\dagger}$} & \fst{55.321$^{\dagger}$} & \fst{62.424$^{\dagger}$} & \fst{35.255$^\dagger$} & \snd{11.076$^\dagger$} & \fst{65.401$^\dagger$} \\ 
\bottomrule
\end{tabular}
\label{tab:methods_comparison}
\end{table*}

\noindent \begin{tikzpicture}
\filldraw (0,0) -- (-0.15,0.08) -- (-0.15,-0.08) -- cycle ; 
\end{tikzpicture} \textbf{\underline{Workflow of our evaluation.}} We present the first open-source testbed for graph-based RAG methods, which (1) collects and re-implements 12 representative methods within a unified framework (as depicted in Section \ref{sec:generic}).
(2) supports a fine-grained comparison over the building blocks of the retrieval stage with up to 100+ variants, and (3) provides a comprehensive evaluation over 11 datasets with various metrics in different scenarios.
We summarize the workflow of our empirical study in \cite{appendix}.

\noindent \begin{tikzpicture}
\filldraw (0,0) -- (-0.15,0.08) -- (-0.15,-0.08) -- cycle ; 
\end{tikzpicture} \textbf{\underline{Benchmark Dataset.}} 
\change{
We employ 11 widely used real-world datasets \cite{gutierrez2024hipporag,edge2024local,guo2024lightrag,huang2024survey} to evaluate the performance of each graph-based RAG method. These datasets span various corpus domains and cover diverse task types.
}

\begin{itemize}
    \item \textit{Specific.}
This group focuses on detail-oriented questions referencing specific entities (e.g., ``Who won the 2024 U.S. presidential election?''). We divide them into two types based on complexity: {\it Simple} (answerable from one or two chunks without reasoning): Quality~\cite{pang2021quality}, PopQA~\cite{mallen2022not}, HotpotQA~\cite{yang2018hotpotqa}) and {\it Complex} (requiring multi-hop reasoning and synthesis): MultihopQA~\cite{tang2024multihop}, MusiqueQA~\cite{trivedi2022musique}, ALCE~\cite{gao2023enabling}).

      \item \textit{Abstract.} Unlike the previous groups, the questions in this category are not centered on specific factual queries. Instead, they involve abstract, conceptual inquiries that encompass broader topics, summaries, or overarching themes.
      An example of an abstract question is: ``How does artificial intelligence influence modern education?''.
      The abstract question requires a high-level understanding of the dataset contents, including five datasets: Mix~\cite{qian2024memorag}, MultihopSum~\cite{tang2024multihop}, Agriculture~\cite{qian2024memorag}, CS~\cite{qian2024memorag}, and Legal~\cite{qian2024memorag}.
\end{itemize}
Their statistics, including the numbers of tokens, and questions, and the question-answering (QA) types are reported in Table~\ref{tab:dataset}.
For specific (both complex and simple) QA datasets, we use the questions provided by each dataset.
\change{While for abstract QA datasets, We follow the question generation method introduced in~\cite{edge2024local}, using GPT-4o to generate 125 questions per dataset with controlled difficulty, which is also aligned with existing works~\cite{edge2024local, guo2024lightrag,fastgraphrag,peng2024graph}.
The details and prompt template used for question generation are provided in our technical report~\cite{appendix}.}
Note that MultihopQA and MultihopSum originate from the same source, but differ in the types of questions they include—the former focuses on complex QA tasks, while the latter on abstract QA tasks.

\noindent \begin{tikzpicture}
\filldraw (0,0) -- (-0.15,0.08) -- (-0.15,-0.08) -- cycle ; 
\end{tikzpicture} \textbf{\underline{Evaluation Metric.}} 
For the specific QA tasks, we use Accuracy and Recall to evaluate performance on the first five datasets based on whether gold answers are included in the generations instead of strictly requiring exact matching, following ~\cite{schick2024toolformer,mallen2022not}.
For the ALCE dataset, answers are typically full sentences rather than specific options or words. Following existing works~\cite{gao2023enabling,sarthi2024raptor}, we use string recall (STRREC), string exact matching (STREM), and string hit (STRHIT) as evaluation metrics.
\change{For abstract QA tasks, we adopt four evaluation metrics following prior works~\cite{edge2024local, guo2024lightrag}: Comprehensiveness, Diversity, Empowerment, and Overall, which assess answer quality from different perspectives.
We employ a head-to-head comparison strategy using GPT-4o as the evaluator. Specifically, for each pair of answers, the LLM is prompted to judge which one is better with respect to a given metric, rather than assigning explicit scores.
This comparative approach is motivated by the strong performance of LLMs as evaluators of natural language generation, often matching or exceeding human judgments~\cite{wang2023chatgpt, zheng2023judging}.
We provide detailed descriptions of the evaluation protocol and example case studies for all four metrics in our technical report~\cite{appendix}.
}

\noindent  \begin{tikzpicture}
\filldraw (0,0) -- (-0.15,0.08) -- (-0.15,-0.08) -- cycle ; 
\end{tikzpicture} \textbf{\underline{Implementation.}} We implement all the algorithms in Python with our proposed unified framework and try our best to ensure a native and effective implementation.
All experiments are run on 350 Ascend 910B-3  NPUs~\cite{huawei}. 
\change{Besides, {\tt Zeroshot} \cite{brown2020language}, and vanilla RAG (denoted by {\tt VanillaRAG})~\cite{lewis2020retrieval} are also included in our study,
} which typically represent the model's inherent capability and the performance improvement brought by basic RAG, respectively.
If a method cannot finish in two days, we mark its result as \textbf{N/A} in the figures and ``---'' in the tables.

\noindent  \begin{tikzpicture}
\filldraw (0,0) -- (-0.15,0.08) -- (-0.15,-0.08) -- cycle ; 
\end{tikzpicture} \textbf{\underline{Hyperparameter Settings.}}
\change{In our experiments, we use Llama-3-8B~\cite{dubey2024llama} as the default LLM, not only because it is the most widely adopted model in recent RAG studies~\cite{zhang2024raglab}, but also due to its strong capabilities in language understanding and reasoning~\cite{dubey2024llama}, as well as its practical efficiency for deployment.}
For LLM, we set the maximum token length to 8,096, and use greedy decoding to generate one sample for the deterministic output.
For each method requiring top-$k$ selection (e.g., chunks or entities), we set $k=4$ to accommodate the token length limitation. 
We use one of the most advanced text-encoding models, BGE-M3~\cite{multim3}, as the embedding model across all methods to generate embeddings for vector search.
If an expert annotator pre-splits the dataset into chunks, we use those as they preserve human insight. Otherwise, following existing works~\cite{edge2024local,guo2024lightrag}, we divide the corpus into 1,200-token chunks.
\change{For other hyperparameters of each method, we adopt the original code settings when available; otherwise, we reproduce them based on the configurations described in the corresponding papers.
}

%% file: sections/6_Experiment/experiment_plot.tex
\definecolor{p1}{RGB}{174,223,172} 
\definecolor{p2}{RGB}{224,175,107}  
\definecolor{p3}{RGB}{138,170,214}  
\definecolor{p4}{RGB}{187, 18, 42} 
\definecolor{p5}{RGB}{216,174,174} 
\definecolor{p6}{RGB}{163,137,214} 
\definecolor{p7}{RGB}{226,145,153} 
\definecolor{p8}{RGB}{205,205,205} 
\definecolor{p9}{RGB}{255,0,127} 
\definecolor{p12}{RGB}{192,192,192} 

\definecolor{t1}{RGB}{148, 190, 146} 
\definecolor{t2}{RGB}{190, 149, 91} 
\definecolor{t3}{RGB}{0, 153, 255}
\definecolor{t4}{RGB}{255, 0, 51}
\pgfplotscreateplotcyclelist{mycolor}{
{fill=p4, draw=p4},
{fill=t1, draw=t1},
{fill=p3, draw=p3},
{fill=t2, draw=t2},
{fill=p5, draw=p5},
}

\pgfplotstableread[row sep=\\,col sep=&]{
	datasets &kcl & sctl & psctl & kcca & actual  \\ 
  2 & 1 & 1  & 972800    & 1  & 1 \\  
     4 & 1 & 1 & 972800  & 1 & 1 \\  
     6 & 1 & 1  &972800   &1   & 1\\  
}\runtime

\pgfplotstableread[row sep=\\,col sep=&]{
	datasets &kcl & sctl & psctl & kcca & actual  \\ 
     2 & 1 & 972800  & 972800    & 972800  & 3354 \\  
     4 & 1 & 972800 & 972800  & 972800 & 1303 \\  
     6 & 1 & 972800  &972800   &972800   & 6046\\  
}\memory
\pgfplotsset{
    discard if/.style 2 args={
        x filter/.code={
            \ifdim\thisrow{#1} pt=#2pt
                \def\pgfmathresult{inf}
            \fi
        }
    },
    discard if not/.style 2 args={
        x filter/.code={
            \ifdim\thisrow{#1} pt=#2pt
            \else
                \def\pgfmathresult{inf}
            \fi
        }
    }
}

\pgfplotstableread[row sep=\\,col sep=&]{
	datasets & Prompt & Comple \\
	2 & 3670881    & 717667   \\
	4 &3715554 & 740347 \\
	6 & 64318900   & 11720116   \\
	8 & 60562965   & 11456822  \\
       10 & 145568727   & 27809276  \\
       12 & 147531844   & 26254673\\
}\indexGraph

\pgfplotstableread[row sep=\\,col sep=&]{
	datasets & Prompt & Comple \\
	2 & 50390    & 15165   \\
	4 &3670881 & 717667 \\
	6 & 2485781   & 717811   \\
	8 & 1762600   & 526361  \\
       10 & 3389597   & 1003018  \\
}\indexToken


\pgfplotstableread[row sep=\\,col sep=&]{
	datasets & Prompt & Comple \\
	2 & 3670881    & 717667   \\
	4 &3715554 & 740347 \\
	6 & 64318900   & 11720116   \\
}\graphRAGVariants

\pgfplotstableread[row sep=\\,col sep=&]{
	datasets & Prompt & Comple \\
	2 & 3670881    & 717667   \\
	4 &3715554 & 740347 \\
	6 & 64318900   & 11720116   \\
	8 & 60562965   & 11456822  \\
       10 & 145568727   & 27809276  \\
       12 & 147531844   & 26254673\\
        14 & 147531844   & 26254673\\
     16 & 147531844   & 26254673\\
}\chunk

\pgfplotstableread[row sep=\\,col sep=&]{
	datasets & Prompt & Comple \\
	2 & 50.783  & 44.254 \\
	4 & 50.324  & 55.267 \\
	6 & 55.321  & 48.946 \\
	8 & 43.193  & 32.118 \\
    10 & 33.063  & 30.686 \\
    12 & 34.144  & 29.876 \\
    14 & 25.581  & 23.285 \\
    16 & 35.647  & 27.796 \\
}\HotpotqaACC

\pgfplotstableread[row sep=\\,col sep=&]{
	datasets & Prompt & Comple \\
	2 & 57.745  & 52.511 \\
	4 & 58.860  & 62.862 \\
	6 & 66.424  & 57.197 \\
	8 & 51.007  & 40.966 \\
    10 & 42.691  & 38.824 \\
    12 & 41.811  & 37.258 \\
    14 & 33.297  & 30.888 \\
    16 & 43.334  & 36.010 \\
}\HotpotqaRecall

\pgfplotstableread[row sep=\\,col sep=&]{
	datasets & Prompt & Comple \\
	2 & 60.829  & 57.041 \\
	4 & 59.900  & 60.472 \\
	6 & 62.545  & 58.899 \\
	8 & 53.324  & 46.748 \\
    10 & 45.461  & 40.315 \\
    12 & 38.885  & 36.312 \\
    14 & 20.944  & 18.227 \\
    16 & 41.244  & 31.594 \\
}\PopqaACC

\pgfplotstableread[row sep=\\,col sep=&]{
	datasets & Prompt & Comple \\
	2 & 27.058  & 25.877 \\
	4 & 24.946  & 25.041 \\
	6 & 27.304  & 25.712 \\
	8 & 22.433  & 19.996 \\
    10 & 18.657  & 15.623 \\
    12 & 16.764  & 14.947 \\
    14 & 8.146  & 6.643 \\
    16 & 18.071  & 12.812 \\
}\PopqaRecall

\pgfplotstableread[row sep=\\,col sep=&]{
	datasets & Prompt & Comple \\
	2 & 34.283  & 29.334 \\
	4 & 23.357  & 21.633 \\
	6 & 35.255  & 34.044 \\
	8 & 30.190  & 27.258 \\
    10 & 28.448  & 27.785 \\
    12 & 21.937  & 20.594 \\
    14 & 17.859  & 17.686 \\
    16 & 25.578  & 22.475 \\
}\ALCEStrRecall

\pgfplotstableread[row sep=\\,col sep=&]{
	datasets & Prompt & Comple \\
	2 & 11.181  & 8.228 \\
	4 & 6.962  & 5.696 \\
	6 & 11.076  & 10.971 \\
	8 & 8.544  & 7.490 \\
    10 & 8.544  & 8.017 \\
    12 & 5.591  & 5.169 \\
    14 & 3.587  & 4.852 \\
    16 & 6.540  & 6.329 \\
}\ALCEStrEM

\pgfplotstableread[row sep=\\,col sep=&]{
	datasets & Prompt & Comple \\
	2 & 63.608  & 56.329 \\
	4 & 43.671  & 41.561 \\
	6 & 65.401  & 62.342 \\
	8 & 56.962  & 53.376 \\
    10 & 54.747  & 52.954 \\
    12 & 43.776  & 40.506 \\
    14 & 37.131  & 34.599 \\
    16 & 50.422  & 43.776 \\
}\ALCEStrHit

\pgfplotstableread[row sep=\\,col sep=&]{
	datasets & Prompt & Comple \\
	2 & 652.0  & 4991.8 \\
	4 & 726.4  & 3933.9 \\
	6 & 598.8  & 6133.9 \\
	8 & 6029.9  & 6113.5 \\
    10 & 4806.2  & 5414.2 \\
    12 & 3041.1  & 4819.2 \\
    14 & 3223.3  & 5213.4 \\
    16 & 4817.2  & 6091.0 \\
}\HotpotqaToken

\pgfplotstableread[row sep=\\,col sep=&]{
	datasets & Prompt & Comple \\
	2 & 644.1  & 4991.6 \\
	4 & 751.3  & 3912.7 \\
	6 & 1231.4  & 5939.1 \\
	8 & 6190.5  & 6204.4 \\
    10 & 4161.1  & 5137.7 \\
    12 & 2383.1  & 4766.8 \\
    14 & 2826.4  & 4922.9 \\
    16 & 4710.7  & 7036.2 \\
}\PopqaToken

\pgfplotstableread[row sep=\\,col sep=&]{
	datasets & Prompt & Comple \\
	2 & 92.4  & 76.7 \\
	4 & 4440.3  & 94.3 \\
	6 & 867.6  & 117.9 \\
	8 & 8159.5  & 2336.9 \\
    10 & 12581.8  & 1510.9 \\
    12 & 2026.4  & 837.2 \\
    14 & 4816.9  & 237.5 \\
    16 & 5509.9 & 237.4 \\
    18 & 6151.5  & 341.0 \\
    20 & 5981.5  & 157.4 \\
    22 & 3739.1  & 138.6 \\
    24 & 6003.2  & 110.5 \\
    26 & 4005.5  & 135.2 \\
}\QualityGen

\pgfplotstableread[row sep=\\,col sep=&]{
	datasets & Prompt & Comple \\
	2 & 81.4  & 188.9 \\
	4 & 3487.7  & 135.7 \\
	6 & 1077.4  & 172.6 \\
	8 & 12385.1  & 4474.5 \\
    10 & 12034.8  & 1837.4 \\
    12 & 3688.6  & 1002.9 \\
    14 & 5545.8  & 228.4 \\
    16 & 5668.2  & 283.3 \\
    18 & 6844.4  & 318.8 \\
    20 & 5668.3  & 206.6 \\
    22 & 3102.0  & 159.1 \\
    24 & 6012.1  & 142.9 \\
    26 & 3028.2  & 181.8 \\
}\MultihopGen

\pgfplotstableread[row sep=\\,col sep=&]{
	datasets & Prompt & Comple \\
	2 & 38.7  & 59.4 \\
	4 & 620.9  & 46.3 \\
	6 & 4332.1  & 66.1 \\
	8 & 10520.3  & 2877.7 \\
    10 & 6917.8  & 1126.3 \\
    12 & 2030.3  & 949.8 \\
    14 & 2721.0  & 200.6 \\
    16 & 3661.6  & 224.5 \\
    18 & 5730.5  & 328.2 \\
    20 & 7255.9  & 93.7 \\
    22 & 784.1 & 76.9 \\
    24 & 5087.9  & 74.9 \\
    26 & 1344.5  & 74.0 \\
}\PopQAGen

\pgfplotstableread[row sep=\\,col sep=&]{
	datasets & Prompt & Comple \\
	2 & 39.9  & 97.9 \\
	4 & 665.4  & 78.7 \\
	6 & 3172.8  & 87.7 \\
	8 & 9524.2  & 2883.4 \\
    10 & 6347.9  & 1207.9 \\
    12 & 2570.5  & 940.4 \\
    14 & 3046.1  & 221.5 \\
    16 & 3382.2  & 238.6 \\
    18 & 5357.3  & 337.9 \\
    20 & 5922.9  & 120.6 \\
    22 & 716.8  & 111.6 \\
    24 & 4556.4  & 119.3 \\
    26 & 1649.8  & 93.1 \\
}\MusiqueGen

\pgfplotstableread[row sep=\\,col sep=&]{
	datasets & Prompt & Comple \\
	2 & 34.1  & 143.1 \\
	4 & 793.7  & 55.4 \\
	6 & 5040.2  & 55.8 \\
	8 & 9065.2  & 2318.0 \\
    10 & 8169.9  & 1156.7 \\
    12 & 3328.7  & 743.2 \\
    14 & 4266.2  & 161.8 \\
    16 & 3831.0  & 197.1 \\
    18 & 5968.2  & 264.1 \\
    20 & 5918.9  & 91.9 \\
    22 & 751.7  & 106.5 \\
    24 & 5349.8  & 91.3 \\
    26 & 710.9  & 82.8 \\
}\ALCEGen

\pgfplotstableread[row sep=\\,col sep=&]{
	datasets & Prompt & Comple \\
	2 & 43.4  & 81.5 \\
	4 & 580.1  & 58.6 \\
	6 & 1  & 1 \\
	8 & 1  & 1 \\
    10 & 1  & 1 \\
    12 & 3101.9  & 887.8 \\
    14 & 2881.4  & 192.8 \\
    16 & 2820.0  & 208.8 \\
    18 & 4612.3  & 309.5 \\
    20 & 5931.8  & 98.11 \\
    22 & 640.6  & 85.8 \\
    24 & 4716.9  & 89.3 \\
    26 & 688.8  & 68.8 \\
}\HotpotGen

%% file: sections/6_Experiment/specific.tex
\subsection{Evaluation for specific QA}
\label{sec:exp:speci}
In this section, we evaluate the performance of different methods on specific QA tasks.

\begin{tikzpicture}
\filldraw (0,0) -- (-0.15,0.08) -- (-0.15,-0.08) -- cycle ; 
\end{tikzpicture}  \textbf{Exp.1. Overall performance.} We report the metric values of all algorithms on specific QA tasks in Table \ref{tab:methods_comparison}. We can make the following observations and analyses: 
(1) Generally, the RAG technique significantly enhances LLM performance across all datasets, and the graph-based RAG methods (e.g., \texttt{HippoRAG} and \texttt{RAPTOR}) typically exhibit higher accuracy than \texttt{VanillaRAG}. However, if the retrieved elements are not relevant to the given question, RAG may degrade the LLM’s accuracy.
For example, on the Quality dataset, compared to \texttt{Zeroshot}, \texttt{RAPTOR} improves accuracy by 53.80\%, while \texttt{G-retriever} decreases it by 14.17\%. This is mainly because, for simple QA tasks, providing only entities and relationships from a subgraph is insufficient to answer such questions effectively.

(2) For specific QA tasks, retaining the original text chunks is crucial for accurate question answering, as the questions and answers in these datasets are derived from the text corpus.
This may explain why \texttt{G-retriever}, \texttt{ToG}, and \texttt{DALK}, which rely solely on graph structure information, perform poorly on most datasets. However, on MultihopQA, which requires multi-hop reasoning, \texttt{DALK} effectively retrieves relevant reasoning paths, achieving accuracy and recall improvements of 6.57\% and 27.94\% over \texttt{VanillaRAG}, respectively.

(3) If the dataset is pre-split into chunks by the expert annotator, \texttt{VanillaRAG} often performs better compared to datasets where chunks are split based on the token size, and we further investigate this phenomenon later in our technical report~\cite{appendix}.

\begin{figure*}[]
\centering
\setlength{\abovecaptionskip}{-0.01cm}
\setlength{\belowcaptionskip}{-0.3cm}
\subfigure[MultihopQA]{
           \begin{tikzpicture}[scale=0.47]
    \begin{axis} [
          ybar=0.5pt,
      width=0.5\textwidth,
      height=0.27\textwidth,
      axis on top, 
        bar width=40, bar shift=0,
        xtick={data},
          xticklabels={
        Tree, KG, TKG, RKG
      },
        xmin=350,
        xmax=750,
      ytick={1000000,  10000000, 100000000},
      ymode=log,
      ymin = 1000000,
        cycle list name=mycolor,
          ticklabel style={font=\Huge},
         every axis plot/.append style={line width=1.6pt},
      every axis/.append style={line width=1.6pt},
        ]
\addplot+ table[x=T,y=K] {graph.multihopqa};
\pgfplotsinvokeforeach{500,600,700}{        
  \addplot+[discard if not={T}{#1}] table[x=T,y=K] {graph.multihopqa};
}
\end{axis}
\end{tikzpicture}
    }
\subfigure[Quality]{
               \begin{tikzpicture}[scale=0.47]
    \begin{axis} [
          ybar=0.5pt,
      width=0.5\textwidth,
      height=0.27\textwidth,
      axis on top, 
        bar width=40, bar shift=0,
        xtick={data},
          xticklabels={
        Tree, KG, TKG, RKG
      },
        xmin=350,
        xmax=750,
      ytick={1000000,  10000000, 100000000},
      ymode=log,
      ymin = 1000000,
        cycle list name=mycolor,
          ticklabel style={font=\Huge},
         every axis plot/.append style={line width=1.6pt},
      every axis/.append style={line width=1.6pt},
        ]
\addplot+ table[x=T,y=K] {graph.quality};
\pgfplotsinvokeforeach{500,600,700}{        
  \addplot+[discard if not={T}{#1}] table[x=T,y=K] {graph.quality};
}
\end{axis}
\end{tikzpicture}
    }
 \subfigure[PopQA]{
                   \begin{tikzpicture}[scale=0.47]
    \begin{axis} [
          ybar=0.5pt,
      width=0.5\textwidth,
      height=0.27\textwidth,
      axis on top, 
        bar width=40, bar shift=0,
        xtick={data},
          xticklabels={
        Tree, KG, TKG, RKG
      },
        xmin=350,
        xmax=750,
      ytick={1000000,  10000000, 100000000},
      ymode=log,
      ymin = 1000000,
        cycle list name=mycolor,
          ticklabel style={font=\Huge},
         every axis plot/.append style={line width=1.6pt},
      every axis/.append style={line width=1.6pt},
        ]
\addplot+ table[x=T,y=K] {graph.popqa};
\pgfplotsinvokeforeach{500,600,700}{        
  \addplot+[discard if not={T}{#1}] table[x=T,y=K] {graph.popqa};
}
\end{axis}
\end{tikzpicture}
    }
        \subfigure[MusiqueQA]{
    	               \begin{tikzpicture}[scale=0.47]
    \begin{axis} [
          ybar=0.5pt,
      width=0.5\textwidth,
      height=0.27\textwidth,
      axis on top, 
        bar width=40, bar shift=0,
        xtick={data},
          xticklabels={
        Tree, KG, TKG, RKG
      },
        xmin=350,
        xmax=750,
      ytick={1000000,  10000000, 100000000},
      ymode=log,
      ymin = 1000000,
        cycle list name=mycolor,
          ticklabel style={font=\Huge},
         every axis plot/.append style={line width=1.6pt},
      every axis/.append style={line width=1.6pt},
        ]
\addplot+ table[x=T,y=K] {graph.musiqueqa};
\pgfplotsinvokeforeach{500,600,700}{        
  \addplot+[discard if not={T}{#1}] table[x=T,y=K] {graph.musiqueqa};
}
\end{axis}
\end{tikzpicture}
    }
    
    \caption{\change{Token cost of graph building on specific QA datasets.}}
    \label{fig:graph_build_spe}
\end{figure*}

(4) \texttt{RAPTOR} often achieves the best performance among most datasets, especially for simple questions.
For complex questions, \texttt{RAPTOR} also performs exceptionally well. This is mainly because, for such questions, high-level summarized information is crucial for understanding the underlying relationships across multiple chunks.
Hence, as we shall see, \texttt{LGraphRAG} is expected to achieve similar results, as it also incorporates high-level information (i.e., a summarized report of the most relevant community for a given question). However, we only observe this effect on the MultihopQA dataset. For the other two complex QA datasets, \texttt{LGraphRAG} even underperforms compared to \texttt{VanillaRAG}.
Meanwhile, \texttt{RAPTOR} still achieves the best performance on these two datasets. We hypothesize that this discrepancy arises from differences in how high-level information is retrieved (See Table \ref{tab:operator}). 
%

(5) For the three largest datasets, the K-means~\cite{han2022data}-based \texttt{RAPTOR} (denoted as \texttt{RAPTOR-K}) also demonstrates remarkable performance. This suggests that the clustering method used in \texttt{RAPTOR} merely impacts overall performance. This may be because different clustering methods share the same key idea: grouping similar items into the same cluster.
Therefore, they may generate similar chunk clusters. 
%
We note that  \texttt{RAPTOR-K} achieves comparable or even better performance than \texttt{RAPTOR}, and the detailed results are shown in technical report \cite{appendix}.
%
If {\tt RAPTOR} does not finish constructing the graph within two days, we use \texttt{RAPTOR-K} instead.

\change{\textbf{Remark.} We note that not all graph-based RAG methods consistently outperform the baseline {\tt VanillaRAG} on every question.
By carefully analyzing the failure case of the top-performing methods, we examine why {\tt HippoRAG},  {\tt RAPTOR}, and {\tt LGraphRAG} sometimes fall short in specific QA tasks.
Please see the failure cases in  \cite{appendix}.
}
\begin{figure}[]
    \centering
     \setlength{\abovecaptionskip}{-0.01cm}
  \setlength{\belowcaptionskip}{-0.25cm}
       	\begin{tikzpicture}[scale=0.52]
        		\begin{axis}[
                        grid = major,
        			ybar=0.11pt,
        			bar width=0.65cm,
        			width=0.9\textwidth,
    				height=0.23\textwidth,
        			xlabel={\huge \bf dataset}, 
        			xtick={2, 4, 6, 8, 10, 12},	xticklabels={MultihopQA, Quality, PopQA, MusiqueQA, HotpotQA, ALCE},
                     legend style={at={(0.5,1.30)}, anchor=north,legend columns=-1,draw=none},
                           legend image code/.code={
            \draw [#1] (0cm,-0.263cm) rectangle (0.4cm,0.15cm); },
        			xmin=0.8,xmax=13.2,
    					ymin=100000,ymax=200000000,
                    ymode = log,    
                log origin=infty,
        			tick align=inside,
        			ticklabel style={font=\Large},
        			every axis plot/.append style={line width = 1.6pt},
        			every axis/.append style={line width = 1.6pt},
        			]
        			\addplot[fill=p8] table[x=datasets,y=Prompt]{\indexGraph};
        			\addplot[fill=p4] table[x=datasets,y=Comple]{\indexGraph};
                \legend{\huge {\tt Prompt token},\huge {\tt Completion token}}
            		\end{axis}
        	\end{tikzpicture}
    	\caption{Token cost of index construction on specific QA.}
    \label{fig:spc:index}
\end{figure}

\begin{tikzpicture}
\filldraw (0,0) -- (-0.15,0.08) -- (-0.15,-0.08) -- cycle ; 
\end{tikzpicture}  \textbf{Exp.2. Token costs of graph and index building.} In this experiment, we first report the token costs of building four types of graphs across all datasets.  
\change{Notably, building PG incurs no token cost, as it does not rely on the LLM for graph construction. We only report the results on four datasets here, and leave the remaining results in our technical report \cite{appendix}.
As shown in Figure~\ref{fig:graph_build_spe}(a) to (d), we observe the following:
(1) Building trees consistently requires the least token cost, while TKG and RKG incur the highest token costs, with RKG slightly exceeding TKG. In some cases, RKG requires up to $40\times$ more tokens than trees.  
(2) KG falls between these extremes, requiring more tokens than trees but fewer than TKG and RKG.  
This trend aligns with the results in Table~\ref{tab:compare_graph}, where graphs with more attributes require higher token costs for construction.
(3) Recall that the token cost for an LLM call consists of two parts: the prompt token, which accounts for the tokens used in providing the input, and the completion part, which includes the tokens generated by the model as a response.
We can see that regardless of the graph type, the prompt part always incurs higher token costs than the completion part; the detailed results are shown in \cite{appendix}.}

We then examine the token costs of index building across all datasets.
Since only \texttt{LGraphRAG} and \texttt{GGraphRAG} require an LLM for index construction, we report only the token costs for generating community reports in Figure~\ref{fig:spc:index}.
We can see that the token cost for index construction is nearly the same as that for building TKG. This is mainly because it requires generating a report for each community, and the number of communities is typically large, especially in large datasets.  
For example, the HotpotQA dataset contains 57,384 communities, significantly increasing the overall token consumption.
That is to say, on large datasets, the two versions of \texttt{GraphRAG} often take more tokens than other methods in the offline stage.

\begin{table}[]
\footnotesize
\renewcommand\arraystretch{0.85}
\caption{{Time and token costs of all methods on specific QA.}}
\begin{tabular}{l|rr|rr|rr}
\toprule
\multirow{2}{*}{Method} &  \multicolumn{2}{c|}{MultihopQA} & \multicolumn{2}{c|}{PopQA} & \multicolumn{2}{c}{ALCE} \\ \cline{2-7}
~ & time & token &time & token &time & token   \\ \midrule
\texttt{ZeroShot} & 3.23 s & 270.3 &  1.17 s & 82.2  & 2.41 s & 177.2 \\ 
\texttt{VanillaRAG}   & 2.35 s & 3,623.4 & 1.41 s & 644.1  & 1.04 s & 849.1 \\ \midrule
\texttt{G-retriever}   &  6.87 s & 1,250.0 & 37.51 s & 3,684.5   & 101.16 s & 5,096.1 \\ 
\texttt{ToG}   &  69.74 s & 16,859.6  &42.02 s & 11,224.2  & 34.94 s & 11,383.2 \\ 
    \texttt{KGP}   & 38.86 s & 13,872.2  & 37.49 s & 6,738.9  & 105.09 s & 9,326.6 \\ 
\texttt{DALK}  & 28.03 s & 4,691.5 & 16.33 s & 2,496.5 & 17.04 s & 4,071.9  \\ 
\texttt{LLightRAG}  & 19.28 s & 5,774.1 & 10.71 s & 2,447.5 & 10.34 s & 4,427.9 \\ 
\texttt{GLightRAG}  & 18.37 s & 5,951.5 & 12.10 s & 3,255.6 & 13.02 s & 4,028.1 \\ 
\texttt{HLightRAG}  & 19.31 s & 7,163.2 & 17.71 s & 5,075.8  & 16.55 s & 6,232.3 \\ 
\texttt{FastGraphRAG}  &  7.17 s & 5,874.8 & 13.25 s & 6,157.0  & 25.82 s & 6,010.9 \\ 
\texttt{HippoRAG}  & 3.46 s & 3,261.1 & 2.32 s & 721.3  & 2.94 s & 858.2 \\ 
\texttt{LGraphRAG}  & 2.98 s & 6,154.9 & 1.72 s & 4,325.2 & 2.11 s & 5,441.1 \\ 
\texttt{RAPTOR}  & 3.18 s & 3,210.0 & 1.36 s & 1,188.3  & 1.54 s & 793.6 \\ 
\bottomrule
\end{tabular}
\label{tab:spc:time_token}

\end{table}

\begin{tikzpicture}
\filldraw (0,0) -- (-0.15,0.08) -- (-0.15,-0.08) -- cycle ; 
\end{tikzpicture}  \textbf{Exp.3. Evaluation of the generation costs.} In this experiment, we evaluate the time and token costs for each method in specific QA tasks.
Specifically, we report the average time and token costs for each query across all datasets in Table \ref{tab:spc:time_token}.
\change{We only report the results on three datasets here, and leave the remaining results in \cite{appendix}.}
%
It is not surprising that \texttt{ZeroShot} and \texttt{VanillaRAG} are the most cost-efficient methods in terms of both time and token consumption.
\change{In terms of token cost, {\tt RAPTOR} and {\tt HippoRAG} are generally more efficient than other graph-based RAG methods}, 
as they share a similar retrieval stage with \texttt{VanillaRAG}. The main difference lies in the chunk retrieval operators they use.
Besides, \texttt{KGP} and \texttt{ToG} are the most expensive methods, as they rely on the agents (i.e.,  different roles of the LLM) for information retrieval during prompt construction. 
The former utilizes the LLM to reason the next required information based on the original question and retrieved chunks, while the latter employs LLM to select relevant entities and relationships for answering the question.
On the other hand, the costs of \texttt{LLightRAG}, \texttt{GLightRAG}, and \texttt{HLightRAG} gradually increase, aligning with the fact that more information is incorporated into the prompt construction.
All three methods are more expensive than \texttt{LGraphRAG} in specific QA tasks, as they use LLM to extract keywords in advance.
Moreover, the time cost of all methods is proportional to the completion token cost.
We present the results in our technical report~\cite{appendix}, which explains why in some datasets, \texttt{VanillaRAG} is even faster than \texttt{ZeroShot}.

\begin{tikzpicture}
\filldraw (0,0) -- (-0.15,0.08) -- (-0.15,-0.08) -- cycle ; 
\end{tikzpicture}  \textbf{Exp.4. Detailed analysis for \texttt{RAPTOR} and \texttt{LGraphRAG}.}
\change{Due to space limitations, we highlight only the key insights derived from our analysis of \texttt{RAPTOR} and \texttt{LGraphRAG}.
First, \texttt{RAPTOR} reveals that a significantly higher proportion of retrieved high-level information (i.e., content from non-leaf nodes) appears in complex QA tasks compared to simple ones, suggesting that high-level information is crucial for multi-hop reasoning.
Second, we find that community reports serve as more effective high-level information than the chunk summaries used in {\tt RAPTOR}, and the similarity-driven community retrieval strategy proves more robust than the {\tt Entity} operator employed in {\tt LGraphRAG}.
Finally, for multi-hop reasoning tasks like MultihopQA, entity and relationship information provide valuable auxiliary signals that help the LLM connect relevant facts and guide the reasoning process.
The detailed analysis and respective experiment results are shown in \cite{appendix}.
}

\begin{table*}[]
\centering
\caption{\change{Comparison of methods on different datasets under different chunk sizes.}}
\small
\renewcommand{\arraystretch}{0.85} 
\setlength{\tabcolsep}{9pt} 
\begin{tabular}{l|l|r|r|r|r|r|r|r|r|r}
\toprule
\multirow{2}{*}{Method} & \multirow{2}{*}{Chunk Size} & \multicolumn{2}{c|}{MultihopQA}  & \multicolumn{2}{c|}{PopAll}  & \multicolumn{2}{c|}{HotpotAll}  & \multicolumn{3}{c}{ALCEAll}  \\ \cline{3-11}
& & Accuracy &  Recall  & Accuracy & Recall & Accuracy & Recall & STRREC & STREM & STRHIT \\ \midrule

\multirow{3}{*}{\tt VanillaRAG} & 600 & 54.421 & 42.740  & 57.255 & 24.171 & 49.190 & 56.935 &  30.174 & 8.333 & 57.911 \\
& 1,200 & 50.626 & 36.918 & 57.041 & 25.877 & 44.254 & 52.511 & 29.334 & 8.228  & 56.329 \\
& 2,400 & 50.665 & 37.172 & 47.677 & 19.122 & 27.553 & 34.293 & 26.350 & 7.490 & 51.371 \\ 
\midrule
\multirow{3}{*}{\tt DALK} & 600 & 55.986 & \snd{48.202} & 41.243 & 16.131 & 32.766 & 41.737 &  21.734 & 4.536 & 44.304  \\ 
& 1,200 & 53.952 & 47.232 & 42.602 & 17.024 & 30.416 & 39.544 & 21.327 & 4.430 & 43.987    \\  
& 2,400 & 53.208 & 46.829 & 45.318 & 18.651 & 28.633 & 37.826 & 20.350 & 4.430 & 41.456    \\  \midrule
\multirow{3}{*}{\tt HippoRAG} & 600  & 47.144 & 41.210 & \fst{62.401} & \snd{26.892} & 50.783 & 58.454 & 27.025 & 8.122 & 51.477   \\   
& 1,200 & 53.760 & 47.671 & 60.472 & 25.041 & \snd{55.267} & \snd{62.862} & 21.633  & 5.696  & 41.561    \\  
& 2,400 & 52.152 & 46.601 & 50.751 & 19.986 & 45.624 & 53.597 & 26.477 & 6.118 & 52.848    \\ \midrule 
\multirow{3}{*}{\tt LGraphRAG} & 600 & 55.282 & 46.267 & 53.181 & 26.292 & 41.194 & 49.801 & 33.692 & \snd{10.971} & \snd{62.447}   \\   
& 1,200 & {55.360} & \fst{50.429} & 39.814 & 17.998 & 30.686 & 38.824 & 27.785 & 8.017  & 52.954    \\  
& 2,400 & 54.930 & 44.588 & 43.317 & 20.185 & 37.061 & 45.366 & 28.398 & 7.806 & 54.008   \\  \midrule
\multirow{3}{*}{\tt RAPTOR} & 600 & \fst{56.729} & 46.358 & \snd{61.830} & \fst{28.176} & \fst{56.132} & \fst{63.584} &  \fst{35.111} & \fst{12.236} & \fst{63.186}  \\ 
& 1,200 & 56.064 & 44.832 & 47.963 & 21.399 & 31.983 & 39.864 & \snd{34.044} & \snd{10.971} & 62.342    \\  
& 2,400 & \snd{56.299} & 44.610 & 48.177 & 21.289 & 31.983 & 39.122 & 33.432 & 10.654 & 61.181    \\ 
\bottomrule

\end{tabular}
\label{tab:chunk_size}
\end{table*}
\begin{table}[]
\centering
\caption{Comparison of our newly designed methods on specific datasets with complex questions.}
\small
\renewcommand{\arraystretch}{0.85} 
\begin{tabular}{c|l|r|r|r}
\toprule
Dataset       & Metric              & \texttt{LGraphRAG} & \texttt{RAPTOR} & \texttt{VGraphRAG} \\ \midrule
\multirow{2}{*}{\textbf{MultihopQA}} & Accuracy           & 55.360 & \snd{56.064} & \fst{59.664} \\
                                     & Recall             & \snd{50.429} & 44.832 & \fst{50.893} \\ \midrule
\multirow{2}{*}{\textbf{MusiqueQA}}  & Accuracy           & 12.467 & \snd{24.133} & \fst{26.933} \\
                                     & Recall             & 23.996 & \snd{35.595} & \fst{40.026} \\ \midrule
\multirow{3}{*}{\textbf{ALCE}}  & STRREC           & 28.448 & \snd{35.255}  & \fst{41.023} \\ 
& STREM    & 8.544 & \snd{11.076} & \fst{15.401} \\                          
& STRHIT     & 54.747 & \snd{65.401} & \fst{71.835} \\                  
\bottomrule
\end{tabular}
\label{tab:new_sota_reduced}
\end{table}


\begin{tikzpicture}
\filldraw (0,0) -- (-0.15,0.08) -- (-0.15,-0.08) -- cycle ; 
\end{tikzpicture}\textbf{Exp.5. Effect of chunk size.}
We evaluate the impact of chunk size on all RAG methods for specific QA by splitting the corpus into chunks of 600, 1200, and 2400 tokens.
To this end, we construct three new datasets—PopAll, HotpotAll, and ALCEAll—by re-chunking the full corpora of PopQA, HotpotQA, and ALCE based on token length.
This re-chunking is necessary because these datasets are pre-split by expert annotators, which may not accurately reflect the effects of chunk sizes.
We only report the results of five methods in Table \ref{tab:chunk_size}, and present the remaining results in \cite{appendix}.
We can see that:
(1) For simple QA datasets (e.g., PopAll and HotpotAll), smaller chunk sizes generally yield better performance. This is because such questions often require information that is directly available in a single chunk or two. Smaller chunks provide more focused and precise context, improving answer accuracy.
(2) We note that performance on simple QA tasks is highly sensitive to chunk size, while for complex QA tasks, performance remains relatively stable across different chunk sizes. This is because complex questions typically require reasoning across multiple chunks, making them less dependent on individual chunk granularity.

\begin{tikzpicture}
\filldraw (0,0) -- (-0.15,0.08) -- (-0.15,-0.08) -- cycle ; 
\end{tikzpicture}
\textbf{Exp.6. New SOTA algorithm.} Based on the above analysis, we aim to develop a new state-of-the-art method for complex QA datasets, denoted as \texttt{VGraphRAG}.
Specifically, our algorithm first retrieves the top-$k$ entities and their corresponding relationships, this step is the same as \texttt{LGraphRAG}.
Next, we adopt the vector search-based retrieval strategy to select the most relevant communities and chunks.
Then, by combining the four elements above, we construct the final prompt of our method to effectively guide the LLM in generating accurate answers.
The results are also shown in Table \ref{tab:new_sota_reduced}, we can see that \texttt{VGraphRAG} performs best on all complex QA datasets.
For example, compared to \texttt{RAPTOR}, our new algorithm \texttt{VGraphRAG} improves Accuracy by 6.42\%  on the MultihopQA dataset and 11.6\%  on the MusiqueQA dataset, respectively.

\begin{table}[]
    \centering
\caption{{Effect of LLM backbones for specific QA task.}}
\renewcommand{\arraystretch}{0.85} 
\setlength{\tabcolsep}{4.2pt} 
    \footnotesize
    \begin{tabular}{llccccc}
        \toprule
\multirow{2}{*}{Method} &  \multirow{2}{*}{LLM backbone} & \multicolumn{2}{c}{MultihopQA} & \multicolumn{3}{c}{ALCEAll} \\
\cmidrule(lr){3-4} \cmidrule(lr){5-7}
     &   & Accuracy & Recall & STRREC & STREM & STRHIT\\
\midrule
\multirow{4}{*}{\tt ZeroShot} & Llama-3-8B         & 49.022 & 34.256 & 15.454 & 3.692 & 30.696 \\
                              &  Qwen-2.5-32B       & 45.070 & 33.332 & 30.512 & 10.127 & 56.118\\ 
                              &  Llama-3-70B       & 55.908 & 52.987 & 31.234 & 7.170 & 61.920 \\ 
                              &  GPT-4o-mini       & 59.546 & 48.322 & 34.965 & 10.232 & 66.245\\ \midrule   
\multirow{4}{*}{\tt VanillaRAG} & Llama-3-8B         & 50.626 & 36.918 & 29.334 & 8.228 & 56.329 \\
                              &  Qwen-2.5-32B       & 56.299 & 47.660 & 39.490 & 14.873 & 69.937\\ 
                              &  Llama-3-70B       & 56.768 & 49.127 & 34.961 & 9.810 & 68.038 \\ 
                              &  GPT-4o-mini       & 59.311 & 47.941 & 35.735 & 10.127 & 68.249\\ \midrule   

\multirow{4}{*}{\tt HLightRAG} & Llama-3-8B         & 50.313 & 41.613 & 22.475 & 6.329 & 43.776\\
                              &  Qwen-2.5-32B       & 53.678 & 51.403 & 34.168 & 10.971 & 63.819\\ 
                              &  Llama-3-70B       & 57.081 & 54.510 & 29.548 & 8.228 & 57.911\\ 
                              &  GPT-4o-mini       & 55.829 & 46.424 & \snd{41.334} & \snd{15.506} & 71.730\\ \midrule 
\multirow{4}{*}{\tt HippoRAG} & Llama-3-8B          & 53.760 & 47.671 & 21.633 & 5.696 & 41.561 \\
                              &  Qwen-2.5-32B       & 48.083 & 40.488 & 37.419 & 13.397 & 66.245\\ 
                              &  Llama-3-70B       & 57.277 & 57.736 & 32.904 & 9.916 & 32.534 \\ 
                              &  GPT-4o-mini       &\snd{67.723} & 55.482 & 39.274 & 12.447 & \snd{72.046}\\ \midrule 

\multirow{4}{*}{\tt RAPTOR} & Llama-3-8B         & 56.064 & 44.832 & 34.044 & 10.971 & 62.342 \\
                              &  Qwen-2.5-32B       & 60.485 & 56.359 & 39.267 & 13.924 & 70.359\\ 
                              &  Llama-3-70B       & 63.028 & \snd{61.042} & 37.286 & 12.236 & 68.671 \\ 
                              &  GPT-4o-mini       & 60.603 & 51.521 & 29.770 & 8.017 & 58.861\\ \midrule  
\multirow{4}{*}{\tt VGraphRAG} & Llama-3-8B         & 59.664 & 50.893 & 35.213 & 11.603 & 64.030 \\
                              &  Qwen-2.5-32B       & 57.277 & 55.151 & 39.234 & 14.557 & 69.831\\ 
                              &  Llama-3-70B       & 67.567 & \fst{68.445} & 37.576 & 12.447 & 69.198 \\ 
                              &  GPT-4o-mini       & \fst{68.193} & 56.564 & \fst{43.963} & \fst{18.038} & \fst{74.473}\\                               
        
        \bottomrule
    \end{tabular}
    \label{tab:backbone}
\end{table}

\begin{tikzpicture}
\filldraw (0,0) -- (-0.15,0.08) -- (-0.15,-0.08) -- cycle ; 
\end{tikzpicture}
\textbf{Exp.7. Effect of LLM backbone.} We evaluate the impact of different LLM backbones—Llama-3-8B~\cite{dubey2024llama}, Qwen-2.5-32B~\cite{yang2025qwen3}, Llama-3-70B~\cite{dubey2024llama}, and GPT-4o-mini—on the MultihopQA and ALCEAll datasets. The main results are shown in Table \ref{tab:backbone}, while the remaining parts are presented in \cite{appendix}.
We make the following observations:
(1) Stronger models generally yield better performance, especially in the {\tt Zeroshot} setting, which most directly reflects the inherent capabilities of the underlying LLM.
(2) The three variants of {\tt LightRAG}, {\tt LLightRAG}, {\tt GLightRAG}, and {\tt HLightRAG} as well as {\tt LGraphRAG}, achieve significant performance improvements when using more powerful LLMs. This can be attributed to their reliance on Rich Knowledge Graphs and Textual Knowledge Graphs, where stronger LLMs contribute to the construction of higher-quality graphs.
(3) {\tt HippoRAG} shows notably superior performance when using GPT-4o-mini compared to other LLM backbones. We attribute this to GPT-4o-mini's ability to extract more accurate entities from the question and to construct higher-quality knowledge graphs, thereby improving the retrieval of relevant chunks and the final answer accuracy.
(4) Regardless of the LLM backbone, our proposed method {\tt VGraphRAG} consistently achieves the best performance, demonstrating the advantages of our proposed unified framework.

%% file: sections/6_Experiment/abstract.tex
\begin{figure*}[]
\centering
\setlength{\abovecaptionskip}{-0.1cm}
\setlength{\belowcaptionskip}{-0.4cm}
\renewcommand{\arraystretch}{1.1} 

\subfigure[Comprehensiveness]{
    \resizebox{0.23\textwidth}{!}{ 
    \begin{tabular}{c*{5}{c}}
& VR & RA & GS & LR & FG \\ 
VR & \colorcell{50} & \colorcell{58} & \colorcell{30} & \colorcell{36} & \colorcell{93} \\ 
RA & \colorcell{42} & \colorcell{50} & \colorcell{39} & \colorcell{26} & \colorcell{82} \\ 
GS & \colorcell{70} & \colorcell{61} & \colorcell{50} & \colorcell{15} & \colorcell{89} \\ 
LR & \colorcell{64} & \colorcell{74} & \colorcell{85} & \colorcell{50} & \colorcell{98} \\ 
FG & \colorcell{7} & \colorcell{18} & \colorcell{11} & \colorcell{2} & \colorcell{50} \\ 
    \end{tabular}
    }
}
\subfigure[Diversity]{
    \resizebox{0.23\textwidth}{!}{ 
    \begin{tabular}{c*{5}{c}}
& VR & RA & GS & LR & FG \\ 
VR & \colorcell{50} & \colorcell{66} & \colorcell{58} & \colorcell{35} & \colorcell{90} \\ 
RA & \colorcell{34} & \colorcell{50} & \colorcell{54} & \colorcell{20} & \colorcell{76} \\ 
GS & \colorcell{42} & \colorcell{46} & \colorcell{50} & \colorcell{26} & \colorcell{86} \\ 
LR & \colorcell{65} & \colorcell{80} & \colorcell{74} & \colorcell{50} & \colorcell{96} \\ 
FG & \colorcell{10} & \colorcell{24} & \colorcell{14} & \colorcell{4} & \colorcell{50} \\ 
    \end{tabular}
    }
}
\subfigure[Empowerment]{
    \resizebox{0.23\textwidth}{!}{ 
    \begin{tabular}{c*{5}{c}}
& VR & RA & GS & LR & FG \\ 
VR & \colorcell{50} & \colorcell{60} & \colorcell{29} & \colorcell{28} & \colorcell{92} \\ 
RA & \colorcell{40} & \colorcell{50} & \colorcell{45} & \colorcell{22} & \colorcell{82} \\ 
GS & \colorcell{71} & \colorcell{54} & \colorcell{50} & \colorcell{12} & \colorcell{88} \\ 
LR & \colorcell{72} & \colorcell{78} & \colorcell{88} & \colorcell{50} & \colorcell{98} \\ 
FG & \colorcell{8} & \colorcell{18} & \colorcell{12} & \colorcell{2} & \colorcell{50} \\ 
    \end{tabular}
    }
}
\subfigure[Overall]{
    \resizebox{0.23\textwidth}{!}{ 
    \begin{tabular}{c*{5}{c}}
& VR & RA & GS & LR & FG \\ 
VR & \colorcell{50} & \colorcell{60} & \colorcell{19} & \colorcell{32} & \colorcell{93} \\ 
RA & \colorcell{40} & \colorcell{50} & \colorcell{44} & \colorcell{24} & \colorcell{82} \\ 
GS & \colorcell{81} & \colorcell{56} & \colorcell{50} & \colorcell{14} & \colorcell{88} \\ 
LR & \colorcell{68} & \colorcell{76} & \colorcell{86} & \colorcell{50} & \colorcell{98} \\ 
FG & \colorcell{7} & \colorcell{18} & \colorcell{12} & \colorcell{2} & \colorcell{50} \\ 
    \end{tabular}
    }
}
\caption{The abstract QA results on Mix dataset.}
\label{fig:winmap_mix}
\end{figure*}

\begin{figure*}[]
\centering
\setlength{\abovecaptionskip}{-0.1cm}
\setlength{\belowcaptionskip}{-0.4cm}
\renewcommand{\arraystretch}{1.1} 

\subfigure[Comprehensiveness]{
    \resizebox{0.23\textwidth}{!}{ 
    \begin{tabular}{c*{5}{c}}
& VR & RA & GS & LR & FG \\ 
VR & \colorcell{50} & \colorcell{50} & \colorcell{2} & \colorcell{46} & \colorcell{95} \\ 
RA & \colorcell{50} & \colorcell{50} & \colorcell{47} & \colorcell{48} & \colorcell{94} \\ 
GS & \colorcell{78} & \colorcell{53} & \colorcell{50} & \colorcell{79} & \colorcell{96} \\ 
LR & \colorcell{54} & \colorcell{52} & \colorcell{21} & \colorcell{50} & \colorcell{92} \\ 
FG & \colorcell{5} & \colorcell{6} & \colorcell{4} & \colorcell{8} & \colorcell{50} \\ 
    \end{tabular}
    }
}
\subfigure[Diversity]{
    \resizebox{0.23\textwidth}{!}{ 
    \begin{tabular}{c*{5}{c}}
& VR & RA & GS & LR & FG \\ 
VR & \colorcell{50} & \colorcell{64} & \colorcell{58} & \colorcell{64} & \colorcell{93} \\ 
RA & \colorcell{36} & \colorcell{50} & \colorcell{42} & \colorcell{49} & \colorcell{85} \\ 
GS & \colorcell{42} & \colorcell{55} & \colorcell{50} & \colorcell{52} & \colorcell{92} \\ 
LR & \colorcell{36} & \colorcell{51} & \colorcell{48} & \colorcell{50} & \colorcell{88} \\ 
FG & \colorcell{7} & \colorcell{15} & \colorcell{8} & \colorcell{12} & \colorcell{50} \\ 
    \end{tabular}
    }
}
\subfigure[Empowerment]{
    \resizebox{0.23\textwidth}{!}{ 
    \begin{tabular}{c*{5}{c}}
& VR & RA & GS & LR & FG \\ 
VR & \colorcell{50} & \colorcell{52} & \colorcell{36} & \colorcell{39} & \colorcell{95} \\ 
RA & \colorcell{48} & \colorcell{50} & \colorcell{45} & \colorcell{45} & \colorcell{93} \\ 
GS & \colorcell{64} & \colorcell{54} & \colorcell{50} & \colorcell{41} & \colorcell{97} \\ 
LR & \colorcell{61} & \colorcell{55} & \colorcell{59} & \colorcell{50} & \colorcell{95} \\ 
FG & \colorcell{5} & \colorcell{7} & \colorcell{3} & \colorcell{5} & \colorcell{50} \\ 
    \end{tabular}
    }
}
\subfigure[Overall]{
    \resizebox{0.23\textwidth}{!}{ 
    \begin{tabular}{c*{5}{c}}
& VR & RA & GS & LR & FG \\ 
VR & \colorcell{50} & \colorcell{52} & \colorcell{44} & \colorcell{46} & \colorcell{95} \\ 
RA & \colorcell{48} & \colorcell{50} & \colorcell{45} & \colorcell{47} & \colorcell{94} \\ 
GS & \colorcell{56} & \colorcell{55} & \colorcell{50} & \colorcell{52} & \colorcell{97} \\ 
LR & \colorcell{54} & \colorcell{53} & \colorcell{48} & \colorcell{50} & \colorcell{93} \\ 
FG & \colorcell{5} & \colorcell{6} & \colorcell{3} & \colorcell{7} & \colorcell{50} \\ 

    \end{tabular}
    }
}
\caption{The abstract QA results on MultihopSum dataset.}

\label{fig:winmap_multihop}
\end{figure*}
\begin{figure*}[]
\centering
\setlength{\abovecaptionskip}{-0.1cm}
\setlength{\belowcaptionskip}{-0.4cm}
\renewcommand{\arraystretch}{1.1} 

\subfigure[Comprehensiveness]{
    \resizebox{0.23\textwidth}{!}{ 
    \begin{tabular}{c*{5}{c}}
& VR & RA & GS & LR & FG \\ 
VR & \colorcell{50} & \colorcell{32} & \colorcell{39} & \colorcell{54} & \colorcell{85} \\ 
RA & \colorcell{68} & \colorcell{50} & \colorcell{19} & \colorcell{73} & \colorcell{94} \\ 
GS & \colorcell{61} & \colorcell{81} & \colorcell{50} & \colorcell{62} & \colorcell{89} \\ 
LR & \colorcell{46} & \colorcell{27} & \colorcell{38} & \colorcell{50} & \colorcell{78} \\ 
FG & \colorcell{15} & \colorcell{6} & \colorcell{11} & \colorcell{22} & \colorcell{50} \\ 
    \end{tabular}
    }
}
\subfigure[Diversity]{
    \resizebox{0.23\textwidth}{!}{ 
    \begin{tabular}{c*{5}{c}}
& VR & RA & GS & LR & FG \\ 
VR & \colorcell{50} & \colorcell{32} & \colorcell{45} & \colorcell{59} & \colorcell{77} \\ 
RA & \colorcell{68} & \colorcell{50} & \colorcell{16} & \colorcell{76} & \colorcell{90} \\ 
GS & \colorcell{55} & \colorcell{84} & \colorcell{50} & \colorcell{63} & \colorcell{82} \\ 
LR & \colorcell{41} & \colorcell{24} & \colorcell{37} & \colorcell{50} & \colorcell{71} \\ 
FG & \colorcell{23} & \colorcell{10} & \colorcell{18} & \colorcell{29} & \colorcell{50} \\ 
    \end{tabular}
    }
}
\subfigure[Empowerment]{
    \resizebox{0.23\textwidth}{!}{ 
    \begin{tabular}{c*{5}{c}}
& VR & RA & GS & LR & FG \\ 
VR & \colorcell{50} & \colorcell{24} & \colorcell{41} & \colorcell{52} & \colorcell{85} \\ 
RA & \colorcell{76} & \colorcell{50} & \colorcell{22} & \colorcell{76} & \colorcell{96} \\
GS & \colorcell{59} & \colorcell{78} & \colorcell{50} & \colorcell{58} & \colorcell{91} \\ 
LR & \colorcell{48} & \colorcell{24} & \colorcell{42} & \colorcell{50} & \colorcell{81} \\ 
FG & \colorcell{15} & \colorcell{4} & \colorcell{9} & \colorcell{19} & \colorcell{50} \\ 
    \end{tabular}
    }
}
\subfigure[Overall]{
    \resizebox{0.23\textwidth}{!}{ 
    \begin{tabular}{c*{5}{c}}
& VR & RA & GS & LR & FG \\ 
VR & \colorcell{50} & \colorcell{30} & \colorcell{38} & \colorcell{53} & \colorcell{85} \\ 
RA & \colorcell{70} & \colorcell{50} & \colorcell{16} & \colorcell{76} & \colorcell{95} \\ 
GS & \colorcell{62} & \colorcell{84} & \colorcell{50} & \colorcell{62} & \colorcell{90} \\ 
LR & \colorcell{47} & \colorcell{24} & \colorcell{38} & \colorcell{50} & \colorcell{79} \\ 
FG & \colorcell{15} & \colorcell{5} & \colorcell{10} & \colorcell{21} & \colorcell{50} \\ 

    \end{tabular}
    }
}
\caption{The abstract QA results on Agriculture dataset.}

\label{fig:winmap_agri}
\end{figure*}

\begin{figure*}[]
\centering
\small
\setlength{\abovecaptionskip}{-0.1cm}
\setlength{\belowcaptionskip}{-0.4cm}
\renewcommand{\arraystretch}{1.1} 

\subfigure[Comprehensiveness]{
    \resizebox{0.23\textwidth}{!}{ 
    \begin{tabular}{c*{5}{c}}
& VR & RA & GS & LR & FG \\ 
VR & \colorcell{50} & \colorcell{22} & \colorcell{25} & \colorcell{36} & \colorcell{80} \\ 
RA & \colorcell{78} & \colorcell{50} & \colorcell{55} & \colorcell{69} & \colorcell{99} \\ 
GS & \colorcell{75} & \colorcell{45} & \colorcell{50} & \colorcell{64} & \colorcell{97} \\ 
LR & \colorcell{64} & \colorcell{31} & \colorcell{36} & \colorcell{50} & \colorcell{95} \\ 
FG & \colorcell{20} & \colorcell{1} & \colorcell{3} & \colorcell{5} & \colorcell{50} \\ 
    \end{tabular}
    }
}
\subfigure[Diversity]{
    \resizebox{0.23\textwidth}{!}{ 
    \begin{tabular}{c*{5}{c}}
& VR & RA & GS & LR & FG \\ 
VR & \colorcell{50} & \colorcell{18} & \colorcell{25} & \colorcell{37} & \colorcell{75} \\ 
RA & \colorcell{82} & \colorcell{50} & \colorcell{51} & \colorcell{79} & \colorcell{99} \\ 
GS & \colorcell{75} & \colorcell{49} & \colorcell{50} & \colorcell{63} & \colorcell{91} \\ 
LR & \colorcell{63} & \colorcell{21} & \colorcell{37} & \colorcell{50} & \colorcell{93} \\ 
FG & \colorcell{25} & \colorcell{1} & \colorcell{9} & \colorcell{7} & \colorcell{50} \\ 
    \end{tabular}
    }
}
\subfigure[Empowerment]{
    \resizebox{0.23\textwidth}{!}{ 
    \begin{tabular}{c*{5}{c}}
& VR & RA & GS & LR & FG \\ 
VR & \colorcell{50} & \colorcell{15} & \colorcell{24} & \colorcell{29} & \colorcell{80} \\ 
RA & \colorcell{85} & \colorcell{50} & \colorcell{59} & \colorcell{72} & \colorcell{100} \\ 
GS & \colorcell{76} & \colorcell{41} & \colorcell{50} & \colorcell{60} & \colorcell{96} \\ 
LR & \colorcell{71} & \colorcell{28} & \colorcell{40} & \colorcell{50} & \colorcell{98} \\ 
FG & \colorcell{20} & \colorcell{0} & \colorcell{4} & \colorcell{2} & \colorcell{50} \\ 
    \end{tabular}
    }
}
\subfigure[Overall]{
    \resizebox{0.23\textwidth}{!}{ 
    \begin{tabular}{c*{5}{c}}
& VR & RA & GS & LR & FG \\ 
VR & \colorcell{50} & \colorcell{15} & \colorcell{22} & \colorcell{30} & \colorcell{79} \\ 
RA & \colorcell{85} & \colorcell{50} & \colorcell{54} & \colorcell{73} & \colorcell{67} \\ 
GS & \colorcell{78} & \colorcell{46} & \colorcell{50} & \colorcell{62} & \colorcell{97} \\ 
LR & \colorcell{70} & \colorcell{27} & \colorcell{38} & \colorcell{50} & \colorcell{97} \\ 
FG & \colorcell{21} & \colorcell{33} & \colorcell{3} & \colorcell{3} & \colorcell{50} \\ 

    \end{tabular}
    }
}
\caption{The abstract QA results on CS dataset.}
\label{fig:winmap}
\end{figure*}

\begin{figure*}[]
\centering
\setlength{\abovecaptionskip}{-0.1cm}
\setlength{\belowcaptionskip}{-0.4cm}
\renewcommand{\arraystretch}{1.1} 

\subfigure[Comprehensiveness]{
    \resizebox{0.23\textwidth}{!}{ 
    \begin{tabular}{c*{5}{c}}
& VR & RA & GS & LR & FG \\ 
VR & \colorcell{50} & \colorcell{26} & \colorcell{31} & \colorcell{41} & \colorcell{93} \\ 
RA & \colorcell{74} & \colorcell{50} & \colorcell{27} & \colorcell{67} & \colorcell{95} \\ 
GS & \colorcell{69} & \colorcell{73} & \colorcell{50} & \colorcell{62} & \colorcell{97} \\ 
LR & \colorcell{59} & \colorcell{33} & \colorcell{38} & \colorcell{50} & \colorcell{97} \\ 
FG & \colorcell{7} & \colorcell{5} & \colorcell{3} & \colorcell{3} & \colorcell{50} \\ 
    \end{tabular}
    }
}
\subfigure[Diversity]{
    \resizebox{0.23\textwidth}{!}{ 
    \begin{tabular}{c*{5}{c}}
& VR & RA & GS & LR & FG \\ 
VR & \colorcell{50} & \colorcell{36} & \colorcell{32} & \colorcell{45} & \colorcell{90} \\ 
RA & \colorcell{64} & \colorcell{50} & \colorcell{68} & \colorcell{68} & \colorcell{93} \\ 
GS & \colorcell{68} & \colorcell{33} & \colorcell{50} & \colorcell{66} & \colorcell{94} \\ 
LR & \colorcell{55} & \colorcell{32} & \colorcell{34} & \colorcell{50} & \colorcell{93} \\ 
FG & \colorcell{10} & \colorcell{7} & \colorcell{6} & \colorcell{7} & \colorcell{50} \\ 
    \end{tabular}
    }
}
\subfigure[Empowerment]{
    \resizebox{0.23\textwidth}{!}{ 
    \begin{tabular}{c*{5}{c}}
& VR & RA & GS & LR & FG \\ 
VR & \colorcell{50} & \colorcell{24} & \colorcell{29} & \colorcell{34} & \colorcell{95} \\ 
RA & \colorcell{76} & \colorcell{50} & \colorcell{31} & \colorcell{67} & \colorcell{96} \\ 
GS & \colorcell{71} & \colorcell{69} & \colorcell{50} & \colorcell{60} & \colorcell{96} \\ 
LR & \colorcell{66} & \colorcell{33} & \colorcell{40} & \colorcell{50} & \colorcell{97} \\ 
FG & \colorcell{5} & \colorcell{4} & \colorcell{4} & \colorcell{3} & \colorcell{50} \\ 
    \end{tabular}
    }
}
\subfigure[Overall]{
    \resizebox{0.23\textwidth}{!}{ 
    \begin{tabular}{c*{5}{c}}
& VR & RA & GS & LR & FG \\ 
VR & \colorcell{50} & \colorcell{26} & \colorcell{30} & \colorcell{37} & \colorcell{94} \\ 
RA & \colorcell{74} & \colorcell{50} & \colorcell{31} & \colorcell{67} & \colorcell{96} \\ 
GS & \colorcell{70} & \colorcell{69} & \colorcell{50} & \colorcell{62} & \colorcell{96} \\ 
LR & \colorcell{63} & \colorcell{33} & \colorcell{38} & \colorcell{50} & \colorcell{97} \\ 
FG & \colorcell{6} & \colorcell{4} & \colorcell{4} & \colorcell{3} & \colorcell{50} \\ 

    \end{tabular}
    }
}

\caption{The abstract QA results on Legal dataset.}

\label{fig:winmap_legal}
\end{figure*}

\subsection{Evaluation for abstract QA}
\label{sec:exp:abs_qa}

\begin{tikzpicture}
\filldraw (0,0) -- (-0.15,0.08) -- (-0.15,-0.08) -- cycle ; 
\end{tikzpicture}  \textbf{Exp.1. Overall Performance.} We evaluate the performance of methods that support abstract QA (see Table \ref{tab:overall_table}) by presenting head-to-head win rate percentages, comparing the performance of each row method against each column method.
Here, we denote VR, RA, GS, LR, and FG as \texttt{VanillaRAG}, \texttt{RAPTOR}, \texttt{GGraphRAG} with high-layer communities (i.e., two-layer for this original implementation), \texttt{HLightRAG} and \texttt{FastGraphRAG}, respectively.
The results are shown in Figure~\ref{fig:winmap_mix} to Figure~\ref{fig:winmap_legal}, and we can see that: (1) Graph-based RAG methods often outperform \texttt{VanillaRAG}, primarily because they effectively capture inter-connections among chunks.
\change{(2) Across all four metrics,  {\tt GGraphRAG} stands out across all metrics. It achieves the highest Comprehensiveness by leveraging community-level retrieval to reduce fragmented evidence and capture broader context. For {\it Diversity}, both {\tt RAPTOR} and {\tt GGraphRAG} perform well by aggregating content across clusters or communities, covering a wide range of subtopics. On {\it Empowerment}, {\tt GGraphRAG} and {\tt LightRAG} lead by integrating structured elements such as entities and relations, helping the LLM generate more grounded and actionable answers. Overall, {\tt GGraphRAG} consistently ranks first, with {\tt RAPTOR} typically second, demonstrating the value of high-level summaries and the effectiveness of {\tt Map-Reduce}.}
%
%

%

\begin{table*}[h]
\small
\renewcommand\arraystretch{0.83}
\caption{The average time and token costs on abstract QA datasets.}
\resizebox{0.98\textwidth}{!}{
\begin{tabular}{l|rr|rr|rr|rr|rr|rr}
\toprule
\multirow{2}{*}{Dataset} &  \multicolumn{2}{c|}{\tt VanillaRAG} & \multicolumn{2}{c|}{\tt RAPTOR} & \multicolumn{2}{c|}{\tt GGraphRAG} & \multicolumn{2}{c|}{\tt HLightRAG} & \multicolumn{2}{c|}{\tt FastGraphRAG} & \multicolumn{2}{c}{\tt CheapRAG} \\ \cline{2-13}
~ & time & token &time & token &time & token &time & token &time & token &time & token  \\ \hline
Mix & 18.7 s     &   4,114 & 35.5 s    &  4,921 & 72.2 s     &    10,922 &22.6 s    &   5,687 &20.9 s     &     4,779  &27.3 s   &     11,720 \\ \hline
MultihopSum  &  9.1 s & 1,680 & 32.7 s & 4,921 & 521.0 s & 353,889 & 33.7 s &5,329 & 34.4 s &5,839 & 54.1 s & 3,784    \\ \hline
Agriculture  &  17.4 s & 5,091 & 20.7 s & 3,753 & 712.3 s & 448,762 & 25.3 s & 4,364  & 28.8 s& 5,640 & 47.1 s & 10,544  \\ \hline
CS   &  17.8 s& 4,884 & 32.7 s& 4,921 & 442.0 s & 322,327 & 51.4 s & 4,908 & 28.2 s & 5,692  & 48.8 s & 17,699 
   \\ \hline
Legal   &  26.2 s & 2,943 & 59.8 s & 3,573 & 231.2 s & 129,969 & 31.1 s & 4,441 & 34.0 s & 5,411  & 34.8 s & 14,586 \\ 
\bottomrule
\end{tabular}
}
\label{tab:cost:abs}
\end{table*}
\begin{tikzpicture}
\filldraw (0,0) -- (-0.15,0.08) -- (-0.15,-0.08) -- cycle ; 
\end{tikzpicture}  \textbf{Exp.2. Evaluation of the generation costs.}
In this experiment, we present the time and token costs for each method in abstract QA tasks. 
As shown in Table~\ref{tab:cost:abs}, \texttt{GGraphRAG} is the most expensive method, as expected, while other graph-based methods exhibit comparable costs, although they are more expensive than \texttt{VanillaRAG}.
For example, on the MutihopSum dataset, \texttt{GGraphRAG} requires 57 $\times$ more time and 210 $\times$ more tokens per query compared to \texttt{VanillaRAG}.  
Specifically, each query in \texttt{GGraphRAG} takes around 9 minutes and consumes 300K tokens, making it impractical for real-world scenarios.
This is because, to answer an abstract question, \texttt{GGraphRAG} needs to analyze all retrieved communities, which is highly time- and token-consuming, especially when the number of communities is large (e.g., in the thousands).

\begin{figure*}[]
\centering
\setlength{\abovecaptionskip}{-0.1cm}
\setlength{\belowcaptionskip}{-0.3cm}
\renewcommand{\arraystretch}{1.25} 

\subfigure[Mix]{
    \resizebox{0.185\textwidth}{!}{ 
    \begin{tabular}{c*{5}{c}}
& Com. & Div. & Emp. & Overall  \\ 
VR & \colorcell{68} & \colorcell{56} & \colorcell{55} & \colorcell{62}  \\ 
RA & \colorcell{77} & \colorcell{56} & \colorcell{64} & \colorcell{70} \\ 
GS & \colorcell{73} & \colorcell{65} & \colorcell{66} & \colorcell{69} \\
LR & \colorcell{68}  & \colorcell{22} & \colorcell{34} & \colorcell{44} \\ 
FG & \colorcell{97} & \colorcell{81} & \colorcell{92} & \colorcell{93} \\ 
    \end{tabular}
    }
}
\subfigure[MultihopSum]{
    \resizebox{0.185\textwidth}{!}{ 
    \begin{tabular}{c*{5}{c}}
& Com. & Div. & Emp. & Overall  \\ 
VR & \colorcell{74} & \colorcell{47} & \colorcell{64} & \colorcell{68}  \\ 
RA & \colorcell{68} & \colorcell{48} & \colorcell{64} & \colorcell{67} \\ 
GS & \colorcell{72} & \colorcell{43} & \colorcell{65} & \colorcell{68} \\ 
LR & \colorcell{66}  & \colorcell{45} & \colorcell{56} & \colorcell{64} \\ 
FG & \colorcell{100} & \colorcell{86} & \colorcell{98} & \colorcell{99} \\ 
    \end{tabular}
    }
    }
\subfigure[Agriculture]{
    \resizebox{0.185\textwidth}{!}{ 
    \begin{tabular}{c*{5}{c}}
& Com. & Div. & Emp. & Overall  \\ 
VR & \colorcell{84} & \colorcell{54} & \colorcell{66} & \colorcell{75}  \\ 
RA & \colorcell{68} & \colorcell{42} & \colorcell{43} & \colorcell{54} \\ 
GS & \colorcell{81} & \colorcell{38} & \colorcell{62} & \colorcell{71} \\ 
LR & \colorcell{84}  & \colorcell{54} & \colorcell{62} & \colorcell{74} \\ 
FG & \colorcell{99} & \colorcell{79} & \colorcell{94} & \colorcell{94} \\ 
    \end{tabular}
    }
}
\subfigure[CS]{
    \resizebox{0.185\textwidth}{!}{ 
    \begin{tabular}{c*{5}{c}}
& Com. & Div. & Emp. & Overall  \\ 
VR & \colorcell{86} & \colorcell{47} & \colorcell{71} & \colorcell{78}  \\ 
RA & \colorcell{66} & \colorcell{19} & \colorcell{39} & \colorcell{49} \\ 
GS & \colorcell{66} & \colorcell{21} & \colorcell{48} & \colorcell{51} \\ 
LR & \colorcell{80}  & \colorcell{34} & \colorcell{54} & \colorcell{66} \\ 
FG & \colorcell{98} & \colorcell{79} & \colorcell{95} & \colorcell{96} \\ 
    \end{tabular}
    }
}
\subfigure[Legal]{
    \resizebox{0.185\textwidth}{!}{ 
    \begin{tabular}{c*{5}{c}}
& Com. & Div. & Emp. & Overall  \\ 
VR & \colorcell{82} & \colorcell{45} & \colorcell{62} & \colorcell{68}  \\ 
RA & \colorcell{63} & \colorcell{27} & \colorcell{40} & \colorcell{46} \\ 
GS & \colorcell{64} & \colorcell{26} & \colorcell{38} & \colorcell{46} \\ 
LR & \colorcell{76}  & \colorcell{42} & \colorcell{44} & \colorcell{59} \\ 
FG & \colorcell{97} & \colorcell{90} & \colorcell{94} & \colorcell{97} \\ 
    \end{tabular}
    }
}
\caption{Comparison of our newly designed method on abstract QA datasets.}

\label{fig:new_sota:abs}
\end{figure*}

\begin{figure*}[]
\centering
\setlength{\abovecaptionskip}{-0.1cm}
\setlength{\belowcaptionskip}{-0.3cm}
\renewcommand{\arraystretch}{1.1} 

\subfigure[chunk size=600]{
    \resizebox{0.185\textwidth}{!}{ 
    \begin{tabular}{c*{5}{c}}
&VR & RA & GS & LR & FG \\ 
{VR} & \colorcell{50} & \colorcell{27} & \colorcell{30} & \colorcell{78} & \colorcell{67} \\ 
{RA} & \colorcell{73} & \colorcell{50} & \colorcell{54} & \colorcell{90} & \colorcell{92} \\ 
{GS} & \colorcell{70} & \colorcell{46} & \colorcell{50} & \colorcell{86} & \colorcell{71} \\ 
{LR} & \colorcell{22} & \colorcell{10} & \colorcell{14} & \colorcell{50} & \colorcell{29} \\ 
{FG} & \colorcell{33} & \colorcell{8} & \colorcell{19} & \colorcell{71} & \colorcell{50} \\ 
    \end{tabular}
    }
}
\subfigure[chunksize=2400]{
    \resizebox{0.185\textwidth}{!}{ 
    \begin{tabular}{c*{5}{c}}
&VR & RA & GS & LR & FG \\ 
{VR} & \colorcell{50} & \colorcell{23} & \colorcell{37} & \colorcell{62} & \colorcell{68} \\ 
{RA} & \colorcell{77} & \colorcell{50} & \colorcell{66} & \colorcell{83} & \colorcell{90} \\ 
{GS} & \colorcell{63} & \colorcell{34} & \colorcell{50} & \colorcell{70} & \colorcell{72} \\ 
{LR} & \colorcell{38} & \colorcell{17} & \colorcell{30} & \colorcell{50} & \colorcell{54} \\ 
{FG} & \colorcell{32} & \colorcell{10} & \colorcell{28} & \colorcell{46} & \colorcell{50} \\ 
    \end{tabular}
    }
    }
\subfigure[Qwen-2.5-32B]{
    \resizebox{0.185\textwidth}{!}{ 
    \begin{tabular}{c*{5}{c}}
&VR & RA & GS & LR & FG \\ 
{VR} & \colorcell{50} & \colorcell{19} & \colorcell{10} & \colorcell{18} & \colorcell{68} \\ 
{RA} & \colorcell{81} & \colorcell{50} & \colorcell{24} & \colorcell{61} & \colorcell{88} \\ 
{GS} & \colorcell{90} & \colorcell{76} & \colorcell{50} & \colorcell{80} & \colorcell{93} \\ 
{LR} & \colorcell{82} & \colorcell{39} & \colorcell{20} & \colorcell{50} & \colorcell{91} \\ 
{FG} & \colorcell{32} & \colorcell{12} & \colorcell{7} & \colorcell{9} & \colorcell{50} \\ 
    \end{tabular}
    }
}
\subfigure[Llama-3-70B]{
    \resizebox{0.185\textwidth}{!}{ 
    \begin{tabular}{c*{5}{c}}
&VR & RA & GS & LR & FG \\ 
{VR} & \colorcell{50} & \colorcell{12} & \colorcell{6} & \colorcell{18} & \colorcell{78} \\ 
{RA} & \colorcell{88} & \colorcell{50} & \colorcell{30} & \colorcell{76} & \colorcell{97} \\ 
{GS} & \colorcell{94} & \colorcell{70} & \colorcell{50} & \colorcell{83} & \colorcell{99} \\ 
{LR} & \colorcell{82} & \colorcell{24} & \colorcell{17} & \colorcell{50} & \colorcell{95} \\ 
{FG} & \colorcell{22} & \colorcell{3} & \colorcell{1} & \colorcell{5} & \colorcell{50} \\ 
    \end{tabular}
    }
}
\subfigure[GPT-4o-mini]{
\resizebox{0.185\textwidth}{!}{ 
    \begin{tabular}{c*{5}{c}}
&VR & RA & GS & LR & FG \\ 
{VR} & \colorcell{50} & \colorcell{72} & \colorcell{4} & \colorcell{13} & \colorcell{60} \\ 
{RA} & \colorcell{28} & \colorcell{50} & \colorcell{4} & \colorcell{10} & \colorcell{37} \\ 
{GS} & \colorcell{96} & \colorcell{96} & \colorcell{50} & \colorcell{90} & \colorcell{95} \\ 
{LR} & \colorcell{87} & \colorcell{90} & \colorcell{10} & \colorcell{50} & \colorcell{87} \\ 
{FG} & \colorcell{40} & \colorcell{63} & \colorcell{5} & \colorcell{13} & \colorcell{50} \\ 
    \end{tabular}
    }
}
\caption{\change{Performance with different chunk sizes and LLM backbones on the MultihopSum dataset.}}

\label{fig:overall_add}
\end{figure*}

\begin{tikzpicture}
\filldraw (0,0) -- (-0.15,0.08) -- (-0.15,-0.08) -- cycle ; 
\end{tikzpicture} \textbf{Exp.3. New SOTA algorithm.} \change{While the \texttt{GGraphRAG} shows remarkable performance in abstract QA, its time and token costs are not acceptable in practice, since given a question $Q$, {\tt GGraphRAG} needs to use LLM to analyze all communities via {\tt Map-Reduce}. (See Section \ref{sec:retri_gene})
To alleviate this issue, we propose a cost-efficient variant of \texttt{GGraphRAG}, named \texttt{CheapRAG}.
Instead of applying the LLM to analyze all communities, \texttt{CheapRAG} first computes the vector similarity between each community and the query to filter out irrelevant ones. It then applies the LLM only to the most relevant communities, significantly reducing token costs compared to \texttt{GGraphRAG}.
Moreover, we observe that many top-performing methods, such as {\tt RAPTOR}, {\tt HLightRAG}, and {\tt GGraphRAG}, all leverage the original chunks.
This suggests that original chunks remain useful for certain questions.
Hence, {\tt CheapRAG} also incorporates original chunks into its retrieval process.
After retrieving the top-$k$ most relevant communities and chunks, \texttt{CheapRAG} adopts a \texttt{Map-Reduce} strategy: the LLM generates partial answers for each selected community and chunk independently, and then summarizes them into a final response.
As shown in Figure~\ref{fig:new_sota:abs} and Table~\ref{tab:cost:abs}, \texttt{CheapRAG} not only achieves better performance than \texttt{GGraphRAG} but also significantly reduces token costs (in most cases).
For example, on the MultihopSum dataset, \texttt{CheapRAG} reduces token costs by 100$\times$ compared to \texttt{GGraphRAG}, while achieving better answer quality.  
%
%
%
We leave improving the answer diversity of \texttt{CheapRAG} to future work.}
 
\change{\begin{tikzpicture}
\filldraw (0,0) -- (-0.15,0.08) -- (-0.15,-0.08) -- cycle ; 
\end{tikzpicture} \textbf{Exp.4. Effect of chunk size and LLM backbone.}
We also study the impact of chunk size and LLM backbone on abstract QA tasks, following the same experimental setup as in Section 7.2.
Due to space limitations, we report the results based on the ``Overall'' metric in Figure~\ref{fig:overall_add}, with additional details provided in~\cite{appendix}.
Our key observations are as follows:
(1) The performance of {\tt GGraphRAG} remains stable across different chunk sizes, likely due to its use of the {\tt Map-Reduce} strategy for final answer synthesis, which mitigates the influence of chunk granularity.
(2) In contrast, methods like {\tt FastGraphRAG} and {\tt VanillaRAG} show greater variance across chunk sizes, as their performance relies heavily on the granularity of individual chunks—smaller chunks tend to provide more precise information, directly impacting retrieval and generation quality.
(3) Regardless of chunk size, {\tt RAPTOR} and {\tt GGraphRAG} consistently achieve the best performance, reaffirming our earlier conclusion that high-level structural information is essential for abstract QA tasks.
(4) All methods still lag behind {\tt GGraphRAG}, further highlighting that community-level information is particularly beneficial for abstract QA tasks.
In addition, we evaluate our newly proposed method {\tt CheapRAG} against the baselines under varying chunk sizes and LLM backbones. As shown in~\cite{appendix}, {\tt CheapRAG} consistently achieves the best performance across all settings.
}

%% file: sections/7_LessonsOpportunities.tex
\section{Lessons and Opportunities}
\label{sec:lessons_opp}

We summarize the lessons (L) for practitioners and propose practical research opportunities (O) based on our observations.

\begin{figure}[]
	\centering
       \setlength{\abovecaptionskip}{-0.0005cm}
  \setlength{\belowcaptionskip}{-0.15cm}
	\includegraphics[width=0.99 \linewidth]{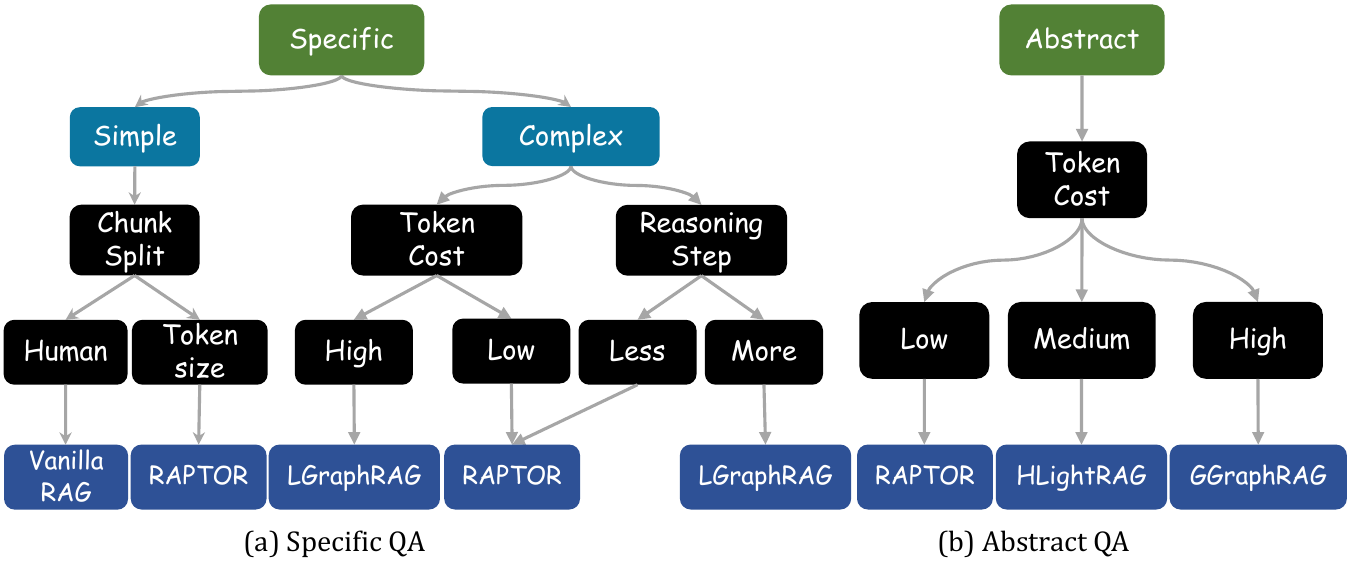}
	\caption{The taxonomy tree of RAG methods.}
	\label{fig:tree_rag}
\end{figure}


\noindent \begin{tikzpicture}
\filldraw (0,0) -- (-0.15,0.08) -- (-0.15,-0.08) -- cycle ; 
\end{tikzpicture} \textbf{\underline{L1.}} In \Cref{fig:tree_rag}, we depict a roadmap of the recommended RAG methods, highlighting which methods are best suited for different scenarios. 
\change{It is derived from all conducted experiments, which
is an overall conclusion for both graph-based RAG methods.}

\noindent \begin{tikzpicture}
\filldraw (0,0) -- (-0.15,0.08) -- (-0.15,-0.08) -- cycle ; 
\end{tikzpicture} \textbf{\underline{L2.}} Chunk quality is critical to the overall performance of RAG methods, and human experts typically produce more effective chunking than approaches based solely on token length.\change{(See results in Table \ref{tab:chunk_size} and Figure \ref{fig:overall_add}.)}

\noindent \begin{tikzpicture}
\filldraw (0,0) -- (-0.15,0.08) -- (-0.15,-0.08) -- cycle ; 
\end{tikzpicture} \textbf{\underline{L3.}} 
For complex questions in specific QA, high-level information is typically needed, as they capture the complex relationship among chunks, and the vector search-based retrieval strategy is better than the rule-based (e.g., {\tt Entity} operator) one.
\change{This lesson is supported by the results in Tables \ref{tab:methods_comparison} and \ref{tab:new_sota_reduced}.}

\noindent  \begin{tikzpicture}
\filldraw (0,0) -- (-0.15,0.08) -- (-0.15,-0.08) -- cycle ; 
\end{tikzpicture}  \textbf{\underline{L4.}} Community reports provide a more effective high-level structure than summarized chunk clusters for abstract QA tasks, as they better capture diversified topics and overarching themes within local modules of the corpus.
\change{(See is results in  Figures 6 $\sim$ 12).}


%

\noindent \noindent \begin{tikzpicture}
\filldraw (0,0) -- (-0.15,0.08) -- (-0.15,-0.08) -- cycle ; 
\end{tikzpicture}   \textbf{\underline{O1.}} All existing graph-based RAG methods (both specific QA and abstract QA) assume the setting of the external corpus is static. What if the external knowledge source evolves over time?
For example, Wikipedia articles are constantly evolving, with frequent updates to reflect new information. 
%
%
Can we design graph-based RAG methods that efficiently and effectively adapt to such dynamic changes in external knowledge sources?

\noindent \begin{tikzpicture}
\filldraw (0,0) -- (-0.15,0.08) -- (-0.15,-0.08) -- cycle ; 
\end{tikzpicture}  
\underline{\textbf{O2.}} 
The quality of a graph plays a key role in determining the effectiveness of graph-based RAG methods. 
However, evaluating graph quality before actually handling a question remains a critical challenge that needs to be addressed.
Existing graph construction methods consume a substantial number of tokens and often produce graphs with redundant entities or miss potential relationships, so designing a cost‑efficient yet effective construction method is a meaningful research direction.

\noindent \begin{tikzpicture}
\filldraw (0,0) -- (-0.15,0.08) -- (-0.15,-0.08) -- cycle; 
\end{tikzpicture}  \underline{\textbf{O3.}} In many domains, the corpus is private (e.g., finance, legal, and medical), and retrieving the relevant information from such corpus can reveal information about the knowledge source.
Designing a graph-based RAG method that incorporates local differential privacy is an interesting research problem.

\noindent \begin{tikzpicture}
\filldraw (0,0) -- (-0.15,0.08) -- (-0.15,-0.08) -- cycle ; 
\end{tikzpicture}   {\underline{\textbf{O4.}} 
\change{How to use LLMs and graph-based RAG methods to facilitate query optimization, such as generating efficient query plans and execution strategies, for database systems.}}
More details are in our technical report \cite{appendix}.



%
%

%% file: sections/7_RelatedWork.tex
\section{Related works}
\label{sec:related}

{
In this section, we mainly review the related works of existing RAG methods. We also present the applications of RAG in various areas, particularly in data management area.


$\bullet$ \textbf{RAG methods.} 
RAG has been proven to be very effective in migrating the ``hallucination'' of LLMs \cite{brown2020language,jeong2024adaptive,siriwardhana2023improving,chen2024auto,chen2023haipipe,chen2024automatic}.
%
Recently, most RAG approaches~\cite{wu2024medical,wang2024knowledge,li2024dalk,gutierrez2024hipporag,edge2024local,guo2024lightrag,sarthi2024raptor,peng2024graph} have adopted graph as the external knowledge to organize the information and relationships within documents, achieving improved overall retrieval performance, which is extensively reviewed in this paper. 
Nevertheless, there is a lack of a comprehensive work comparison between all
graph-based RAG methods in terms of accuracy and efficiency.
We note that there exists an empirical study \cite{han2025rag} that compares Microsoft's methods ({\tt LGraphRAG} and {\tt GGraphRAG}) with the standard {\tt VanillaRAG}, and a few survey papers on graph-based RAG systems~\cite{zhang2025survey,kamra2024enhancing}. However, our work differs significantly from these in both scope and depth.
First, our work focus on systematically comparing the different graph-based RAG methods,  and conduct a stage-wise comparison across a unified framework. This allows us to identify core design principles and enables the construction of a new state-of-the-art method through component recombination.
Second, unlike survey papers that offer high-level overviews, our work provides deep empirical analysis and practical insights grounded in extensive experiments.



$\bullet$ \textbf{RAG applications.}  Due to the wealth of developer experience captured in a vast array of database forum discussions, recent studies \cite{zhou2023d,lao2023gptuner,fan2024combining,zhou2024db,lillm,sun2024r,chen2024automatic,giannankouris2024lambda,singh2024panda} have begun leveraging RAGs to enhance database performance. For instance, GPTuner \cite{lao2023gptuner} proposes to enhance database knob tuning using RAG by leveraging domain knowledge to identify
important knobs and coarsely initialize their values for subsequent
refinement. Besides, D-Bot \cite{zhou2023d} proposes an LLM-based database
diagnosis system, which can retrieve relevant knowledge chunks
and tools, and use them to identify typical root causes accurately.
In addition, RAG-based SQL rewriting systems \cite{lillm,sun2024r,song2025quite} have recently attracted significant attention. 
The RAG-based data analysis systems have also been studied~\cite{liang2025revisiting,liu2025palimpchat,anderson2024design,lin2025twix,lin2024towards,patel2024lotus,liu2024declarative,chen2023seed}. 
For applications in other areas, we refer readers to recent RAG surveys \cite{zhang2024raglab,huang2023survey}. 

}


%% file: sections/8_Conclution.tex
\section{Conclusions}
\label{sec:conclusions}

In this paper, we provide an in-depth experimental evaluation and
comparison of existing graph-based Retrieval-Augmented Generation (RAG) methods. 
We first provide a novel unified framework, which can cover all
the existing graph-based RAG methods, using an abstraction of a few key operations. We then thoroughly analyze and compare different graph-based RAG methods under our framework.
We further systematically evaluate these methods from different angles using various datasets for both specific and abstract question-answering (QA) tasks, and also develop variations by combining existing techniques, which often outperform state-of-the-art methods.
From extensive experimental results and analysis, we have identified several important findings and analyzed the critical components that affect the performance.
In addition, we have summarized the lessons learned and proposed practical research opportunities that can facilitate future studies.

%% file: sections/10_Appdenix.tex
\clearpage

\appendix

\definecolor{c1}{RGB}{214,39,40} 
\definecolor{c2}{RGB}{31,119,180}  
\definecolor{c3}{RGB}{148,103,189} 
\definecolor{c4}{RGB}{140, 86, 75} 

\definecolor{c5}{RGB}{227,119,194} 
\definecolor{c6}{RGB}{188,189,34} 

\definecolor{c7}{RGB}{23,190,207} 
\definecolor{c8}{RGB}{255,127,14} %

\definecolor{c11}{RGB}{172, 196, 226} 
\definecolor{c12}{RGB}{208, 221, 238} 
\definecolor{c21}{RGB}{174,223,172} 
\definecolor{c22}{RGB}{138,170,214}  
\definecolor{c23}{RGB}{248,199,1} 
\definecolor{c24}{RGB}{255,0,127} 
\definecolor{c25}{RGB}{163,137,214} 

\pgfplotstableread[row sep=\\,col sep=&]{
chunkSize & Vanilla & HippoRAG & RAPTOR & FastGraphRAG & LGraphRAG & LLightRAG & HLightRAG & DALK\\
1 & 54.421 & 47.144 & 56.729 & 50.000 & 55.282 & 46.675 & 50.196  & 55.986 \\
3 & 50.626 & 53.760 & 56.064 & 50.861 & 55.360 & 44.053 & 50.313  & 53.952  \\
5 & 50.665 & 52.152 & 56.299 & 47.261 & 54.930 & 43.858 &  49.648 & 53.208 \\
}\multiAcc

\pgfplotstableread[row sep=\\,col sep=&]{
chunkSize & Vanilla & HippoRAG & RAPTOR & FastGraphRAG & LGraphRAG & LLightRAG & HLightRAG & DALK\\
1 & 42.740 & 41.210 & 46.359 & 44.736 & 46.267 & 39.190 &  40.626 & 48.202 \\
3 & 36.918 & 47.671 & 44.832 & 46.312 & 50.429 & 35.528 &  41.613 & 47.232 \\
5 & 37.172 & 46.601 & 44.610 & 46.251 & 44.588 & 36.878 &  41.775 & 46.829 \\
}\multiRec

\pgfplotstableread[row sep=\\,col sep=&]{
chunkSize & Vanilla & HippoRAG & RAPTOR & FastGraphRAG & LGraphRAG & LLightRAG & HLightRAG & DALK\\
1 & 57.255 & 62.402 & 61.830 & 49.392 & 53.181 & 38.671 & 34.668  & 41.243 \\
3 & 57.041 & 60.472 & 47.963 & 46.748 & 39.814 & 36.312 & 31.594  & 42.602  \\
5 & 47.677 & 50.751 & 48.177 & 32.809 & 43.317 & 35.811 & 30.093  & 45.318 \\
}\popallAcc

\pgfplotstableread[row sep=\\,col sep=&]{
chunkSize & Vanilla & HippoRAG & RAPTOR & FastGraphRAG & LGraphRAG & LLightRAG & HLightRAG & DALK\\
1 & 24.171 & 26.892 & 28.176 & 22.326 & 26.292 & 16.040 & 15.326  & 16.131  \\
3 & 25.877 & 25.041 & 21.399 & 19.996 & 17.998 & 14.947 & 12.812  & 17.024  \\
5 & 19.121 & 19.986 & 21.289 & 12.879 & 20.185 & 14.230 & 11.683  & 18.651  \\
}\popallRec

\pgfplotstableread[row sep=\\,col sep=&]{
chunkSize & Vanilla & HippoRAG & RAPTOR & FastGraphRAG & LGraphRAG & LLightRAG & HLightRAG & DALK\\
1 & 49.190 & 50.783 & 56.132 & 35.467 & 41.194 & 32.334 & 30.632  & 32.766 \\
3 & 44.254 & 55.267 & 31.929 & 32.118 & 30.686 & 29.876 & 27.796  & 30.416  \\
5 & 27.553 & 45.624 & 31.929 & 26.445 & 37.061 & 29.336 & 26.958  & 28.633  \\
}\hotallAcc

\pgfplotstableread[row sep=\\,col sep=&]{
chunkSize & Vanilla & HippoRAG & RAPTOR & FastGraphRAG & LGraphRAG & LLightRAG & HLightRAG & DALK\\
1 & 56.935 & 58.454 & 63.584 & 43.534 & 49.801 & 39.753 & 38.194  & 41.737 \\
3 & 52.511 & 62.862 & 39.864 & 40.966 & 38.824 & 37.258 & 36.010  & 39.544 \\
5 & 34.293 & 53.597 & 39.122 & 34.800 & 45.366 & 36.621 & 34.387  & 37.826 \\
}\hotallRec

\pgfplotstableread[row sep=\\,col sep=&]{
chunkSize & Vanilla & HippoRAG & RAPTOR & FastGraphRAG & LGraphRAG & LLightRAG & HLightRAG & DALK\\
1 & 30.174 & 27.025 & 35.111 & 30.116 & 33.692 & 20.084 & 23.833  & 21.734 \\
3 & 29.334 & 21.633 & 34.044 & 27.258 & 27.785 & 20.594 & 22.475  & 21.327 \\
5 & 26.350 & 26.477 & 33.432 & 22.020 & 28.398 & 19.576 & 20.816  & 20.350 \\
}\alceallStrRec

\pgfplotstableread[row sep=\\,col sep=&]{
chunkSize & Vanilla & HippoRAG & RAPTOR & FastGraphRAG & LGraphRAG & LLightRAG & HLightRAG & DALK\\
1 & 8.333 & 8.122 & 12.236 & 9.705 & 10.971 & 4.641 & 5.907 & 4.536 \\
3 & 8.228 & 5.696 & 10.971 & 7.490 & 8.017 & 5.169 & 6.329  & 4.430 \\
5 & 7.490 & 6.118 & 10.654 & 6.435 & 7.806 & 5.485 & 5.485  & 4.430 \\
}\alceallStrEM

\pgfplotstableread[row sep=\\,col sep=&]{
chunkSize & Vanilla & HippoRAG & RAPTOR & FastGraphRAG & LGraphRAG & LLightRAG & HLightRAG & DALK\\
1 & 57.911 & 51.477 & 63.186 & 55.907 & 62.447 & 40.717 & 47.363  & 44.304 \\
3 & 56.329 & 41.561 & 62.342 & 53.376 & 52.954 & 40.506 & 43.776  & 43.987 \\
5 & 51.371 & 52.848 & 61.181 & 42.300 & 54.008 & 38.397 & 40.823  & 41.456 \\
}\alceallStrHit

\pgfplotstableread[row sep=\\,col sep=&]{
chunkSize & Vanilla & RAPTOR & GGraphRAG & LightRAG & FastGraphRAG\\
9 & 97 & 94 & 94 & 99 & 99 \\
27 & 74 & 68 & 72 & 66 & 100 \\
45 & 97 & 89 & 94 & 99 & 99 \\
}\winwincom

\pgfplotstableread[row sep=\\,col sep=&]{
chunkSize & Vanilla & RAPTOR & GGraphRAG & LightRAG & FastGraphRAG\\
9 & 84 & 75 & 33 & 98 & 97 \\
27 & 47 & 48 & 43 & 45 & 86 \\
45 & 86 & 73 & 46 & 96 & 98 \\
}\winwindiv

\pgfplotstableread[row sep=\\,col sep=&]{
chunkSize & Vanilla & RAPTOR & GGraphRAG & LightRAG & FastGraphRAG\\
9 & 80 & 60 & 43 & 95 & 99 \\
27 & 64 & 64 & 65 & 56 & 98 \\
45 & 79 & 52 & 47 & 92 & 96 \\
}\winwinemp

\pgfplotstableread[row sep=\\,col sep=&]{
chunkSize & Vanilla & RAPTOR & GGraphRAG & LightRAG & FastGraphRAG\\
9 & 90 & 76 & 69 & 99 & 99 \\
27 & 68 & 67 & 68 & 64 & 99 \\
45 & 92 & 74 & 73 & 96 & 98 \\
}\winwinoverall

\section{ADDITIONAL EXPERIMENTS}

\subsection{Results on specific QA tasks}
\label{sec:exp:more}

In this Section, we present the additional results on the specific QA tasks. 
\begin{table}[]
\centering
\caption{Comparison \texttt{RAPTOR} and \texttt{RAPTOR-K}.}
\small

\textbf{Exp.1. Comparison {\tt RAPTOR-K} and {\tt RAPTOR}.} 
We compare {\tt RAPTOR-K} with {\tt RAPTOR} on the first three
datasets, and present results in Table \ref{tab:compare_raptor}. We observe that {\tt RAPTOR-K} achieves comparable or even better performance than RAPTOR.

\begin{figure*}[]
\centering
\setlength{\abovecaptionskip}{-0.01cm}
\setlength{\belowcaptionskip}{-0.3cm}

    \subfigure[HotpotQA]{
    	               %
    }
    \subfigure[The token costs on HotpotQA]{
    	%
    }
        \subfigure[The token costs on ALCE]{
    		%
    }
    \caption{Token cost of graph building on specific QA datasets.}
    \label{fig:graph_build_spe_2}
\end{figure*}
\begin{figure*}[h]
\centering
\setlength{\abovecaptionskip}{-0.01cm}
\subfigure[MultihopQA]{
    	%
    }
    \caption{Proportion of the token costs for prompt and completion in graph building stage across all datasets.}
    \label{fig:graph_build_spe_plus}
\end{figure*}
\begin{table*}[]

\renewcommand\arraystretch{0.85}
\caption{{The average time and token costs of all methods on specific QA datasets.}}
\begin{tabular}{l|rr|rr|rr|rr|rr|rr}
\toprule
\multirow{2}{*}{Method} &  \multicolumn{2}{c|}{MultihopQA} & \multicolumn{2}{c|}{Quality} & \multicolumn{2}{c|}{PopQA} & \multicolumn{2}{c|}{MusiqueQA} & \multicolumn{2}{c|}{HotpotQA} & \multicolumn{2}{c}{ALCE} \\ \cline{2-13}
~ & time & token &time & token &time & token &time & token &time & token &time & token  \\ \midrule
\texttt{ZeroShot} & 3.23 s & 270.3 & 1.47 s & 169.1 &  1.17 s & 82.2 & 1.73 s &    137.8  &1.51 s & 125.0 & 2.41 s & 177.2 \\ 
\texttt{VanillaRAG}   & 2.35 s & 3,623.4 & 2.12 s & 4,502.0 & 1.41 s & 644.1 & 1.31 s & 745.4 & 1.10 s & 652.0 & 1.04 s & 849.1 \\ \midrule
\texttt{G-retriever}   &  6.87 s & 1,250.0 & 5.18 s & 985.5 & 37.51 s & 3,684.5 & 31.21 s & 3,260.5 & --- & ---  & 101.16 s & 5,096.1 \\ 
\texttt{ToG}   &  69.74 s & 16,859.6 & 37.03 s & 10,496.4 &42.02 s & 11,224.2 & 53.55 s & 12,480.8 & --- & --- & 34.94 s & 11,383.2 \\ 
    \texttt{KGP}   & 38.86 s & 13,872.2 & 35.76 s & 14,092.7 & 37.49 s & 6,738.9 & 39.82 s & 7,555.8 & --- & --- & 105.09 s & 9,326.6 \\ 
\texttt{DALK}  & 28.03 s & 4,691.5 & 13.23 s & 2,863.6 & 16.33 s & 2,496.5 & 17.48 s & 3,510.9 & 21.33 s & 3,989.7 & 17.04 s & 4,071.9  \\
\bottomrule
\end{tabular}
\label{tab:app:spc:time_token}

\end{table*}

\begin{table}[]
\renewcommand\arraystretch{1.12}
\footnotesize
\caption{Proportion of retrieved nodes across tree layers.}
\centering
\begin{tabular}{c|r|r|r|r|r|r}
\toprule
Layer & MultihopQA & Quality & PopQA & MusiqueQA & HotpotQA & ALCE\\ \midrule
0 & 59.3\% & 76.8\% & 76.1\% & 69.3\%&89.7\% & 90.6\% \\
1 & 27.5\% & 18.7\% & 16.5\% & 28.1\% & 9.5\% &8.8\% \\ 
$\textgreater$ 1 & 13.2\% & 4.5\% & 7.4\% & 2.6\% & 0.8\% &0.6\% \\ 
\bottomrule
\end{tabular}
\label{tab:layer}
\end{table}

\begin{table}[]
\caption{Descriptions of the different variants of \texttt{LGraphRAG}.}
\label{tab:dec_variants}
\small
\resizebox{0.48\textwidth}{!}{
%
    }

    \caption{Token costs for prompt and completion tokens in the generation stage across all datasets.}
    \label{fig:compare:gen_token}
\end{figure*}

\begin{figure*}[]
\centering
\setlength{\abovecaptionskip}{-0.01cm}
\subfigure[HotpotQA (Accuracy)]{
             	%
    }
    \caption{Effect of chunk quality on the performance of specific QA tasks.}
    \label{fig:compare:all}
\end{figure*}

\begin{table*}[]
\centering
\caption{Comparison of our newly designed methods on specific datasets with complex questions.}
\small
\renewcommand{\arraystretch}{0.85} 
\resizebox{\textwidth}{!}{
\begin{tabular}{c|l|r|r|r|r|r|r|r|r}
\toprule
Dataset       & Metric             & \texttt{ZeroShot} & \texttt{VanillaRAG} & \texttt{LGraphRAG} & \texttt{RAPTOR} & \texttt{GraphRAG-ER} & \texttt{GraphRAG-CC} & \texttt{VGraphRAG-CC} & \texttt{VGraphRAG} \\ \midrule
\multirow{2}{*}{\textbf{MultihopQA}} & Accuracy          & 49.022 & 50.626 & 55.360 & \snd{56.064} & 52.739 & 52.113 & 55.203 & \fst{59.664} \\
                                     & Recall            & 34.526 & 36.918 & \snd{50.429} & 44.832 & 45.113 & 43.770 & 46.750 & \fst{50.893} \\ \midrule
\multirow{2}{*}{\textbf{MusiqueQA}}  & Accuracy          & 1.833  & 17.233 & 12.467 & \snd{24.133} & 11.200 & 13.767 & 22.400 & \fst{26.933} \\
                                     & Recall            & 5.072  & 27.874 & 23.996 & \snd{35.595} & 22.374 & 25.707 & 35.444 & \fst{40.026} \\ \midrule
\multirow{3}{*}{\textbf{ALCE}}  & STRREC          & 15.454  & 34.283 & 28.448 & 35.255 & 26.774 & 35.366 & \snd{37.820} & \fst{41.023} \\ 
& STREM   & 3.692  & 11.181 & 8.544 & 11.076 & 7.5949 & 11.920 & \snd{13.608} & \fst{15.401} \\                          
& STRHIT    & 30.696  & 63.608 & 54.747 & 65.401 &52.743 & 64.662 & \snd{68.460} & \fst{71.835} \\                  
\bottomrule
\end{tabular}
}
\label{tab:app:new_sota}
\end{table*}

\begin{table*}[]
\centering
\caption{\change{Comparison of methods on different datasets under different chunk sizes, where \fst{Purple} denotes the best result, and \snd{Orange} denotes the best result excluding the best one.}}
\small
\renewcommand{\arraystretch}{1.2} 
%
\textbf{Exp.2. Token costs of graph and index building.} 
We report the token costs of building four types
of graphs across HotpotQA and ALCE datasets in Figure \ref{fig:graph_build_spe_2}.
Recall that the token cost for an LLM call
consists of two parts: the prompt token, which accounts for the
tokens used in providing the input, and the completion part, which
includes the tokens generated by the model as a response. Here,
we report the token costs for prompt and completion on HotpotQA and ALCE in Figure  \ref{fig:graph_build_spe_2}(c) to (d), and show the results on other datasets in Figure \ref{fig:graph_build_spe_plus}.
We conclude that, regardless of the graph type, the prompt part
always incurs higher token costs than the completion part.

\begin{tikzpicture}
\filldraw (0,0) -- (-0.15,0.08) -- (-0.15,-0.08) -- cycle ; 
\end{tikzpicture}
\textbf{Exp.3. Evaluation of the generation costs.} 
In this experiment, we evaluate the time and token costs for each method in
specific QA tasks. Specifically, we report the average time and token costs for each query across all datasets in Table \ref{tab:app:spc:time_token}, the conclusions are consistent with those reported in our manuscript.
As shown in Figure~\ref{fig:compare:gen_token}, we present the average token costs for prompt tokens and completion tokens across all questions in all specific QA datasets.
We can observe that the running time of each method is highly proportional to the completion token costs, which aligns with the computational paradigm of the Transformer architecture.

\begin{tikzpicture}
\filldraw (0,0) -- (-0.15,0.08) -- (-0.15,-0.08) -- cycle ; 
\end{tikzpicture}
\textbf{Exp.4. Detailed analysis for {\tt RAPTOR} and {\tt LGraphRAG}.} 
Our first analysis about \texttt{RAPTOR} aims to explain why \texttt{RAPTOR} outperforms \texttt{VanillaRAG}.
Recall that in \texttt{RAPTOR}, for each question $Q$, it retrieves the top-$k$ items across the entire tree, meaning the retrieved items may originate from different layers.
That is, we report the proportion of retrieved items across different tree layers in Table \ref{tab:layer}.
As we shall see, for the MultihopQA and MusiqueQA datasets, the proportion of retrieved high-level information (i.e., items not from leaf nodes) is significantly higher than in other datasets. For datasets requiring multi-hop reasoning to answer questions, high-level information plays an essential role. This may explain why \texttt{RAPTOR} outperforms \texttt{VanillaRAG} on these two datasets.

We then conduct a detailed analysis of \texttt{LGraphRAG} on complex questions in specific QA datasets by modifying its retrieval methods or element types.
By doing this, we create three variants of \texttt{LGraphRAG}, and we present the detailed descriptions for each variant in Table \ref{tab:dec_variants}.
Here, \texttt{VGraphRAG-CC} introduces a new retrieval strategy. Unlike \texttt{LGraphRAG}, it uses vector search to retrieve the top-$k$ elements (i.e., chunks or communities) from the vector database.
Eventually, we evaluate their performance on the three complex QA datasets and present the results in Table \ref{tab:app:new_sota}.

\begin{figure*}[h]
    \centering
    \setlength{\abovecaptionskip}{-0.05cm}
    \setlength{\belowcaptionskip}{-0.1cm}
    \quad \ref{ALLleg} \\
\subfigure[MultihopQA (Accuracy)]{
\begin{tikzpicture}[scale=0.6]
      \begin{axis}[
      legend style = {
                    legend columns=4,
                    inner sep=0pt,
                    draw=none,
                },
      legend image post style={scale=1.1, line width=1pt},
        legend to name=ALLleg,
        xmin=0, xmax=6,
        ymin=42, ymax=58,
        xtick = {1,3,5},
        xticklabels = {600, 1200, 2400},
        mark size=3.5pt, 
        width=0.3\textwidth,
        height=0.3\textwidth,
        every axis plot/.append style={line width = 1.5pt},
        every axis/.append style={line width = 1.5pt},
      ]
      \addplot[c3, mark=triangle] table[x=chunkSize, y=Vanilla] {\multiAcc};
      \addplot[c6, mark=square, mark options={rotate=45}] table[x=chunkSize, y=HippoRAG] {\multiAcc};
      \addplot[c5, mark=o] table[x=chunkSize, y=RAPTOR] {\multiAcc};
      \addplot[c1, mark=triangle, mark options={rotate=180}] table[x=chunkSize, y=FastGraphRAG] {\multiAcc};
      \addplot[c4, mark=square] table[x=chunkSize, y=LGraphRAG] {\multiAcc};
      \addplot[c2, mark=diamond] table[x=chunkSize, y=LLightRAG] {\multiAcc};
      \addplot[c7, mark=x] table[x=chunkSize, y=HLightRAG] {\multiAcc};
      \addplot[c8, mark=triangle, mark options={rotate=270}] table[x=chunkSize, y=DALK] {\multiAcc};
      \legend{\tt{\small VanillaRAG},\tt{ \small HippoRAG}, \tt{\small RAPTOR}, \tt{\small FastGraphRAG} , \tt{\small LGraphRAG}, \tt{\small LLightRAG} ,\tt{ \small HLightRAG}, \tt{\small DALK}}
      \end{axis}
    \end{tikzpicture}
    }
\subfigure[MultihopQA (Recall)]{
\begin{tikzpicture}[scale=0.6]
      \begin{axis}[
        xmin=0, xmax=6,
        ymin=34, ymax=52,
        xtick = {1,3,5},
        xticklabels = {600, 1200, 2400},
        mark size=3.5pt, 
        width=0.3\textwidth,
        height=0.3\textwidth,
        every axis plot/.append style={line width = 1.5pt},
        every axis/.append style={line width = 1.5pt},
      ]
      
      \addplot[c3, mark=triangle] table[x=chunkSize, y=Vanilla] {\multiRec};
      \addplot[c6, mark=square, mark options={rotate=45}] table[x=chunkSize, y=HippoRAG] {\multiRec};
      \addplot[c5, mark=o] table[x=chunkSize, y=RAPTOR] {\multiRec};
      \addplot[c1, mark=triangle, mark options={rotate=180}] table[x=chunkSize, y=FastGraphRAG] {\multiRec};
      \addplot[c4, mark=square] table[x=chunkSize, y=LGraphRAG] {\multiRec};
      \addplot[c2, mark=diamond] table[x=chunkSize, y=LLightRAG] {\multiRec};
      \addplot[c7, mark=x] table[x=chunkSize, y=HLightRAG] {\multiRec};
      \addplot[c8, mark=triangle, mark options={rotate=270}] table[x=chunkSize, y=DALK] {\multiRec};
      \end{axis}
    \end{tikzpicture}
    }
\subfigure[PopAll (Accuracy)]{
\begin{tikzpicture}[scale=0.6]
      \begin{axis}[
        xmin=0, xmax=6,
        ymin=27, ymax=65,
        xtick = {1,3,5},
        xticklabels = {600, 1200, 2400},
        mark size=3.5pt, 
        width=0.3\textwidth,
        height=0.3\textwidth,
        every axis plot/.append style={line width = 1.5pt},
        every axis/.append style={line width = 1.5pt},
      ]
      
      \addplot[c3, mark=triangle] table[x=chunkSize, y=Vanilla] {\popallAcc};
      \addplot[c6, mark=square, mark options={rotate=45}] table[x=chunkSize, y=HippoRAG] {\popallAcc};
      \addplot[c5, mark=o] table[x=chunkSize, y=RAPTOR] {\popallAcc};
      \addplot[c1, mark=triangle, mark options={rotate=180}] table[x=chunkSize, y=FastGraphRAG] {\popallAcc};
      \addplot[c4, mark=square] table[x=chunkSize, y=LGraphRAG] {\popallAcc};
      \addplot[c2, mark=diamond] table[x=chunkSize, y=LLightRAG] {\popallAcc};
      \addplot[c7, mark=x] table[x=chunkSize, y=HLightRAG] {\popallAcc};
      \addplot[c8, mark=triangle, mark options={rotate=270}] table[x=chunkSize, y=DALK] {\popallAcc};
      \end{axis}
    \end{tikzpicture}
    }
\subfigure[PopAll (Recall)]{
\begin{tikzpicture}[scale=0.6]
      \begin{axis}[
        xmin=0, xmax=6,
        ymin=10, ymax=30,
        xtick = {1,3,5},
        xticklabels = {600, 1200, 2400},
        mark size=3.5pt, 
        width=0.3\textwidth,
        height=0.3\textwidth,
        every axis plot/.append style={line width = 1.5pt},
        every axis/.append style={line width =1.5pt},
      ]
      
      \addplot[c3, mark=triangle] table[x=chunkSize, y=Vanilla] {\popallRec};
      \addplot[c6, mark=square, mark options={rotate=45}] table[x=chunkSize, y=HippoRAG] {\popallRec};
      \addplot[c5, mark=o] table[x=chunkSize, y=RAPTOR] {\popallRec};
      \addplot[c1, mark=triangle, mark options={rotate=180}] table[x=chunkSize, y=FastGraphRAG] {\popallRec};
      \addplot[c4, mark=square] table[x=chunkSize, y=LGraphRAG] {\popallRec};
      \addplot[c2, mark=diamond] table[x=chunkSize, y=LLightRAG] {\popallRec};
      \addplot[c7, mark=x] table[x=chunkSize, y=HLightRAG] {\popallRec};
      \addplot[c8, mark=triangle, mark options={rotate=270}] table[x=chunkSize, y=DALK] {\popallRec};
      \end{axis}
    \end{tikzpicture}
    }
\subfigure[HotpotAll (Accuracy)]{
\begin{tikzpicture}[scale=0.6]
      \begin{axis}[
        xmin=0, xmax=6,
        ymin=23, ymax=62,
        xtick = {1,3,5},
        xticklabels = {600, 1200, 2400},
        mark size=3.5pt, 
        width=0.3\textwidth,
        height=0.3\textwidth,
        every axis plot/.append style={line width = 1.5pt},
        every axis/.append style={line width = 1.5pt},
      ]
      
      \addplot[c3, mark=triangle] table[x=chunkSize, y=Vanilla] {\hotallAcc};
      \addplot[c6, mark=square, mark options={rotate=45}] table[x=chunkSize, y=HippoRAG] {\hotallAcc};
      \addplot[c5, mark=o] table[x=chunkSize, y=RAPTOR] {\hotallAcc};
      \addplot[c1, mark=triangle, mark options={rotate=180}] table[x=chunkSize, y=FastGraphRAG] {\hotallAcc};
      \addplot[c4, mark=square] table[x=chunkSize, y=LGraphRAG] {\hotallAcc};
      \addplot[c2, mark=diamond] table[x=chunkSize, y=LLightRAG] {\hotallAcc};
      \addplot[c7, mark=x] table[x=chunkSize, y=HLightRAG] {\hotallAcc};
      \addplot[c8, mark=triangle, mark options={rotate=270}] table[x=chunkSize, y=DALK] {\hotallAcc};
      \end{axis}
    \end{tikzpicture}
    }
\subfigure[HotpotAll (Recall)]{
\begin{tikzpicture}[scale=0.6]
      \begin{axis}[
        xmin=0, xmax=6,
        ymin=30, ymax=68,
        xtick = {1,3,5},
        xticklabels = {600, 1200, 2400},
        mark size=3.5pt, 
        width=0.39\textwidth,
        height=0.3\textwidth,
        every axis plot/.append style={line width = 1.5pt},
        every axis/.append style={line width = 1.5pt},
      ]
      
      \addplot[c3, mark=triangle] table[x=chunkSize, y=Vanilla] {\hotallRec};
      \addplot[c6, mark=square, mark options={rotate=45}] table[x=chunkSize, y=HippoRAG] {\hotallRec};
      \addplot[c5, mark=o] table[x=chunkSize, y=RAPTOR] {\hotallRec};
      \addplot[c1, mark=triangle, mark options={rotate=180}] table[x=chunkSize, y=FastGraphRAG] {\hotallRec};
      \addplot[c4, mark=square] table[x=chunkSize, y=LGraphRAG] {\hotallRec};
      \addplot[c2, mark=diamond] table[x=chunkSize, y=LLightRAG] {\hotallRec};
      \addplot[c7, mark=x] table[x=chunkSize, y=HLightRAG] {\hotallRec};
      \addplot[c8, mark=triangle, mark options={rotate=270}] table[x=chunkSize, y=DALK] {\hotallRec};
      \end{axis}
    \end{tikzpicture}
    }
\subfigure[ALCEAll (STRREC)]{
\begin{tikzpicture}[scale=0.6]
      \begin{axis}[
        xmin=0, xmax=6,
        ymin=18, ymax=37,
        xtick = {1,3,5},
        xticklabels = {600, 1200, 2400},
        mark size=3.5pt, 
        width=0.39\textwidth,
        height=0.3\textwidth,
        every axis plot/.append style={line width = 1.5pt},
        every axis/.append style={line width = 1.5pt},
      ]
      
      \addplot[c3, mark=triangle] table[x=chunkSize, y=Vanilla] {\alceallStrRec};
      \addplot[c6, mark=square, mark options={rotate=45}] table[x=chunkSize, y=HippoRAG] {\alceallStrRec};
      \addplot[c5, mark=o] table[x=chunkSize, y=RAPTOR] {\alceallStrRec};
      \addplot[c1, mark=triangle, mark options={rotate=180}] table[x=chunkSize, y=FastGraphRAG] {\alceallStrRec};
      \addplot[c4, mark=square] table[x=chunkSize, y=LGraphRAG] {\alceallStrRec};
      \addplot[c2, mark=diamond] table[x=chunkSize, y=LLightRAG] {\alceallStrRec};
      \addplot[c7, mark=x] table[x=chunkSize, y=HLightRAG] {\alceallStrRec};
      \addplot[c8, mark=triangle, mark options={rotate=270}] table[x=chunkSize, y=DALK] {\alceallStrRec};
      \end{axis}
    \end{tikzpicture}
    }
\subfigure[ALCEAll (STREM)]{
\begin{tikzpicture}[scale=0.6]
      \begin{axis}[
        xmin=0, xmax=6,
        ymin=3, ymax=13,
        xtick = {1,3,5},
        xticklabels = {600, 1200, 2400},
        mark size=3.5pt, 
        width=0.39\textwidth,
        height=0.3\textwidth,
        every axis plot/.append style={line width = 1.5pt},
        every axis/.append style={line width = 1.5pt},
      ]
      
      \addplot[c3, mark=triangle] table[x=chunkSize, y=Vanilla] {\alceallStrEM};
      \addplot[c6, mark=square, mark options={rotate=45}] table[x=chunkSize, y=HippoRAG] {\alceallStrEM};
      \addplot[c5, mark=o] table[x=chunkSize, y=RAPTOR] {\alceallStrEM};
      \addplot[c1, mark=triangle, mark options={rotate=180}] table[x=chunkSize, y=FastGraphRAG] {\alceallStrEM};
      \addplot[c4, mark=square] table[x=chunkSize, y=LGraphRAG] {\alceallStrEM};
      \addplot[c2, mark=diamond] table[x=chunkSize, y=LLightRAG] {\alceallStrEM};
      \addplot[c7, mark=x] table[x=chunkSize, y=HLightRAG] {\alceallStrEM};
      \addplot[c8, mark=triangle, mark options={rotate=270}] table[x=chunkSize, y=DALK] {\alceallStrEM};
      \end{axis}
    \end{tikzpicture}
    }
\subfigure[ALCEAll (STRHIT)]{
\begin{tikzpicture}[scale=0.6]
      \begin{axis}[
        xmin=0, xmax=6,
        ymin=35, ymax=66,
        xtick = {1,3,5},
        xticklabels = {600, 1200, 2400},
        mark size=3.5pt, 
        width=0.39\textwidth,
        height=0.3\textwidth,
        every axis plot/.append style={line width = 1.5pt},
        every axis/.append style={line width = 1.5pt},
      ]
      
      \addplot[c3, mark=triangle] table[x=chunkSize, y=Vanilla] {\alceallStrHit};
      \addplot[c6, mark=square, mark options={rotate=45}] table[x=chunkSize, y=HippoRAG] {\alceallStrHit};
      \addplot[c5, mark=o] table[x=chunkSize, y=RAPTOR] {\alceallStrHit};
      \addplot[c1, mark=triangle, mark options={rotate=180}] table[x=chunkSize, y=FastGraphRAG] {\alceallStrHit};
      \addplot[c4, mark=square] table[x=chunkSize, y=LGraphRAG] {\alceallStrHit};
      \addplot[c2, mark=diamond] table[x=chunkSize, y=LLightRAG] {\alceallStrHit};
      \addplot[c7, mark=x] table[x=chunkSize, y=HLightRAG] {\alceallStrHit};
      \addplot[c8, mark=triangle, mark options={rotate=270}] table[x=chunkSize, y=DALK] {\alceallStrHit};
      \end{axis}
    \end{tikzpicture}
    }

    \caption{\change{Effect of chunk size.}}
    \label{fig:diff_chunk_size}
\end{figure*}

\begin{table*}[]
    \centering
\caption{\change{The specific QA performance comparison of graph-based RAG methods with different LLM backbones.}}
\renewcommand{\arraystretch}{0.91} 
\setlength{\tabcolsep}{14pt} 
    \small
    \begin{tabular}{llccccc}
        \toprule
\multirow{2}{*}{Method} &  \multirow{2}{*}{LLM backbone} & \multicolumn{2}{c}{MultihopQA} & \multicolumn{3}{c}{ALCEAll} \\
\cmidrule(lr){3-4} \cmidrule(lr){5-7}
     &   & Accuracy & Recall & STRREC & STREM & STRHIT\\
\midrule
\multirow{4}{*}{\tt ZeroShot} & Llama-3-8B         & 49.022 & 34.256 & 15.454 & 3.692 & 30.696 \\
                              &  Qwen-2.5-32B       & 45.070 & 33.332 & 30.512 & 10.127 & 56.118\\ 
                              &  Llama-3-70B       & 55.908 & 52.987 & 31.234 & 7.170 & 61.920 \\ 
                              &  GPT-4o-mini       & 59.546 & 48.322 & 34.965 & 10.232 & 66.245\\ \midrule   
\multirow{4}{*}{\tt VanillaRAG} & Llama-3-8B         & 50.626 & 36.918 & 29.334 & 8.228 & 56.329 \\
                              &  Qwen-2.5-32B       & 56.299 & 47.660 & 39.490 & 14.873 & 69.937\\ 
                              &  Llama-3-70B       & 56.768 & 49.127 & 34.961 & 9.810 & 68.038 \\ 
                              &  GPT-4o-mini       & 59.311 & 47.941 & 35.735 & 10.127 & 68.249\\ \midrule   
\multirow{4}{*}{\tt G-retriever} & Llama-3-8B         & 42.019 & 43.116 & 13.061 & 2.848 & 26.160\\
                              &  Qwen-2.5-32B       & 43.075 & 33.864 & 34.678 & 13.608 & 60.338\\ 
                              &  Llama-3-70B       & 50.430 & 50.144 & 24.575 & 5.907 & 49.051\\ 
                              &  GPT-4o-mini       & 56.534 & 46.427 & 31.681 & 7.068 & 62.764 \\ \midrule   
\multirow{4}{*}{\tt ToG} & Llama-3-8B         & 41.941 & 38.435 & 15.909 & 3.059 & 33.755 \\
                              &  Qwen-2.5-32B       & 34.390 & 35.566 & 19.167 & 4.430 & 39.662\\ 
                              &  Llama-3-70B       & 42.762 & 38.495 & 18.449 & 3.270 & 38.819\\ 
                              &  GPT-4o-mini       & 41.862 & 30.247 & 28.405 & 7.068 & 56.962\\ \midrule              
\multirow{4}{*}{\tt KGP} & Llama-3-8B         & 48.161 & 36.272 & 27.551 & 7.806 & 52.637\\
                              &  Qwen-2.5-32B       & 61.463 & 57.340 & 40.608 & 14.873 & 71.097\\ 
                              &  Llama-3-70B       & 55.008 & 47.878 & 35.420 & 10.549 & 68.379\\ 
                              &  GPT-4o-mini       & 63.146 & 55.789 & 38.015 & 12.764 & 69.620\\ \midrule   
\multirow{4}{*}{\tt DALK} & Llama-3-8B         & 53.952 & 47.232 & 21.327 & 4.430 & 43.987\\
                              &  Qwen-2.5-32B       & 32.003 & 20.158 & 16.653 & 4.430 & 33.333\\ 
                              &  Llama-3-70B       & 60.524 & 55.086 & 28.980 & 5.696 & 60.338\\ 
                              &  GPT-4o-mini       & 66.980 & 57.687 & 31.813 & 8.333 & 62.236\\ \midrule   
\multirow{4}{*}{\tt LLightRAG} & Llama-3-8B         & 44.053 & 35.528 & 20.594 & 5.169 & 40.506\\
                              &  Qwen-2.5-32B       & 48.552 & 45.387 & 31.549 & 9.916 & 59.177\\ 
                              &  Llama-3-70B       & 57.081 & 54.510 & 28.636 & 8.228 & 56.013\\ 
                              &  GPT-4o-mini       & 52.113 & 38.923 & 38.446 & 15.084 & 33.456\\ \midrule      
\multirow{4}{*}{\tt GLightRAG} & Llama-3-8B         & 48.474 & 38.365 & 17.686 & 4.852 & 34.599\\
                              &  Qwen-2.5-32B       & 52.582 & 48.236 & 27.961 & 9.599 & 53.165\\ 
                              &  Llama-3-70B       & 55.986 & 51.713 & 23.553 & 5.807 & 45.992\\ 
                              &  GPT-4o-mini       & 55.125 & 47.899 & 39.095 & \snd{15.506} & 67.933\\ \midrule 
\multirow{4}{*}{\tt HLightRAG} & Llama-3-8B         & 50.313 & 41.613 & 22.475 & 6.329 & 43.776\\
                              &  Qwen-2.5-32B       & 53.678 & 51.403 & 34.168 & 10.971 & 63.819\\ 
                              &  Llama-3-70B       & 57.081 & 54.510 & 29.548 & 8.228 & 57.911\\ 
                              &  GPT-4o-mini       & 55.829 & 46.424 & \snd{41.334} & \snd{15.506} & 71.730\\ \midrule 
\multirow{4}{*}{\tt FastGraphRAG} & Llama-3-8B         & 52.895 & 44.278 & 27.258 & 7.490 & 53.376 \\
                              &  Qwen-2.5-32B       & 46.088 & 50.370 & 31.387 & 10.021 & 59.388\\ 
                              &  Llama-3-70B       &  54.069  &55.787  & 35.658 & 12.236 & 65.612 \\ 
                              &  GPT-4o-mini       & 66.080 & 57.007 & 23.521 & 8.228 & 42.827\\ \midrule 
\multirow{4}{*}{\tt HippoRAG} & Llama-3-8B          & 53.760 & 47.671 & 21.633 & 5.696 & 41.561 \\
                              &  Qwen-2.5-32B       & 48.083 & 40.488 & 37.419 & 13.397 & 66.245\\ 
                              &  Llama-3-70B       & 57.277 & 57.736 & 32.904 & 9.916 & 32.534 \\ 
                              &  GPT-4o-mini       &\snd{67.723} & 55.482 & 39.274 & 12.447 & \snd{72.046}\\ \midrule 
\multirow{4}{*}{\tt LGraphRAG} & Llama-3-8B         & 55.360 &  50.429 & 27.785 & 8.017 & 52.954 \\
                              &  Qwen-2.5-32B       & 49.531 & 52.113 & 35.406 & 12.553 & 63.924\\ 
                              &  Llama-3-70B       & 58.060 & 55.390 & 34.256 & 10.232 & 66.561 \\ 
                              &  GPT-4o-mini       & 65.415 & 50.216 & 36.890 & 11.287 & 69.304\\ \midrule 
\multirow{4}{*}{\tt RAPTOR} & Llama-3-8B         & 56.064 & 44.832 & 34.044 & 10.971 & 62.342 \\
                              &  Qwen-2.5-32B       & 60.485 & 56.359 & 39.267 & 13.924 & 70.359\\ 
                              &  Llama-3-70B       & 63.028 & \snd{61.042} & 37.286 & 12.236 & 68.671 \\ 
                              &  GPT-4o-mini       & 60.603 & 51.521 & 29.770 & 8.017 & 58.861\\ \midrule  
\multirow{4}{*}{\tt VGraphRAG} & Llama-3-8B         & 59.664 & 50.893 & 35.213 & 11.603 & 64.030 \\
                              &  Qwen-2.5-32B       & 57.277 & 55.151 & 39.234 & 14.557 & 69.831\\ 
                              &  Llama-3-70B       & 67.567 & \fst{68.445} & 37.576 & 12.447 & 69.198 \\ 
                              &  GPT-4o-mini       & \fst{68.193} & 56.564 & \fst{43.963} & \fst{18.038} & \fst{74.473}\\                               
        
        \bottomrule
    \end{tabular}
    \label{tab:app:backbone}
\end{table*}

\begin{tikzpicture}
\filldraw (0,0) -- (-0.15,0.08) -- (-0.15,-0.08) -- cycle ; 
\end{tikzpicture}
\textbf{Exp.5. Effect of the chunk size.} Recall that our study includes some datasets that are pre-split by the export annotator. 
To investigate this impact, we re-split the corpus into multiple chunks based on token size for these datasets instead of using their original chunks.
Here, we create three new datasets from HotpotQA, PopQA, and ALCE, named HotpotAll, PopAll, and ALCEAll, respectively.

For each dataset, we use \texttt{Original} to denote its original version and \texttt{New chunk} to denote the version after re-splitting.
We report the results of graph-based RAG methods on both the original and new version datasets in Figure~\ref{fig:compare:all}, we can see that: (1) The performance of all methods declines, mainly because rule-based chunk splitting (i.e., by token size) fails to provide concise information as effectively as expert-annotated chunks.
(2) Graph-based methods, especially those relying on TKG and RKG, are more sensitive to chunk quality. This is because the graphs they construct encapsulate richer information, and coarse-grained chunk splitting introduces potential noise within each chunk. Such noise can lead to inaccurate extraction of entities or relationships and their corresponding descriptions, significantly degrading the performance of these methods.
(3) As for token costs, all methods that retrieve chunks incur a significant increase due to the larger chunk size in \texttt{New chunk} compared to \texttt{Original}, while other methods remain stable.  
These findings highlight that chunk segmentation quality is crucial for the overall performance of all RAG methods.

\change{We evaluate all 12 RAG methods under chunk sizes of 600, 1200, and 2400 tokens, as shown in Table~\ref{tab:app:chunk_size}, and further select the top-performing eight methods to analyze how their performance varies with chunk size, as shown in Figure~\ref{fig:diff_chunk_size}. We observe that the results across different settings remain largely consistent with the conclusions drawn using the default chunk size of 1200 in our submitted manuscript.
Specifically, we make the following observations and analysis: 
(1) For simple QA datasets (e.g., PopAll and HotpotAll), smaller chunk sizes generally yield better performance. This is because such questions often require information that is directly available in a single chunk or two. Smaller chunks provide more focused and precise context, improving answer accuracy.
(2) Across all chunk sizes, {\tt HippoRAG} and {\tt RAPTOR} perform best on simple QA tasks, while {\tt RAPTOR} and {\tt LGraphRAG} outperform others on complex QA datasets (e.g., MultihopQA, ALCEAll). These observations align with the results reported in our manuscript.
(3) The effect of chunk size varies by task type: performance on simple QA tasks is highly sensitive to chunk size, while for complex QA tasks, performance remains relatively stable across different chunk sizes. This is because complex questions typically require reasoning across multiple chunks, making them less dependent on individual chunk granularity.}

{
\begin{tikzpicture}
\filldraw (0,0) -- (-0.15,0.08) -- (-0.15,-0.08) -- cycle ; 
\end{tikzpicture}
\textbf{Exp.6. Effect of the LLM backbones.}  For each system, we evaluate multiple LLM backbones, including Llama-3-8B, Qwen-2.5-32B, Llama-3-70B, and GPT-4o-mini. These cover three open-source models ranging from small to large scales, as well as one proprietary yet strong GPT-family model.
In this experiment, we fix the chunk size at 1200 tokens.
As shown in Table~\ref{tab:app:backbone}, we evaluate each RAG method and the {\tt ZeroShot} baseline on the MultihopQA and ALCEAll datasets using different LLM backbones.
Based on the results, we can see that:
(1) Stronger models generally yield better performance, especially in the {\tt ZeroShot} setting, which most directly reflects the inherent capabilities of the underlying LLM.
(2) We observe that the three variants of {\tt LightRAG}, {\tt LLightRAG}, {\tt GLightRAG}, and {\tt HLightRAG} as well as {\tt LGraphRAG}, achieve significant performance improvements when using more powerful LLMs. This can be attributed to their reliance on Rich Knowledge Graphs and Textual Knowledge Graphs, where stronger LLMs contribute to the construction of higher-quality graphs.
(3) {\tt HippoRAG} shows notably superior performance when using GPT-4o-mini compared to other LLM backbones. We attribute this to GPT-4o-mini's ability to extract more accurate entities from the question and to construct higher-quality knowledge graphs, thereby improving the retrieval of relevant chunks and the final answer accuracy.
(4) Regardless of the LLM backbone, our proposed method {\tt VGraphRAG} consistently achieves the best performance, demonstrating the advantages of our proposed unified framework.
}

\begin{tikzpicture}
\filldraw (0,0) -- (-0.15,0.08) -- (-0.15,-0.08) -- cycle ; 
\end{tikzpicture}
\textbf{Exp.7. The size of graph.} 
For each dataset, we report the size of five types of graphs in Table~\ref{tab:graph_size}.
We observe that PG is typically denser than other types of graphs, as they connect nodes based on shared entity relationships, where each node represents a chunk in PG.  
In fact, the probability of two chunks sharing at least a few entities is quite high, leading to a high graph density (i.e., the ratio of edges to nodes), sometimes approaching a clique (fully connected graph).
In contrast, KG, TKG, and RKG are much sparser since they rely entirely on LLMs to extract nodes and edges. This sparsity is primarily due to the relatively short and incomplete outputs typically generated by LLMs, which miss considerable potential node-edge pairs.
Interestingly, the size or density of the constructed graph has not shown a strong correlation with the final performance of graph-based RAG methods.  
This observation motivates us to explore a method for evaluating the quality of the constructed graph before using it for LLM-based question answering.

\begin{table*}[]
\centering
\small
\caption{The size of each graph type across all datasets.}
\label{tab:graph_size}
\begin{tabular}{c|rr|rr|rr|rr|rr}
\toprule
& \multicolumn{2}{c}{Tree}
& \multicolumn{2}{c}{PG}
& \multicolumn{2}{c}{KG}
& \multicolumn{2}{c}{TKG}
& \multicolumn{2}{c}{RKG} \\
\textbf{Dataset} 
& \# of vertices & \# of edges
& \# of vertices & \# of edges
&  \# of vertices  & \# of edges
& \# of vertices &\# of edges
& \# of vertices & \# of edges \\
\midrule
MultihopQA & 2,053 & 2,052 & 1,658 & 564,446 & 35,953 & 37,173 & 12,737 & 10,063 & 18,227 & 12,441 \\
Quality & 1,862 & 1,861 & 1,518 & 717,468 & 28,882 & 30,580 & 10,884 & 8,992 & 13,836 & 9,044 \\
PopQA & 38,325 & 38,324 & 32,157 & 3,085,232 & 260,202 & 336,676 & 179,680 & 205,199 & 188,946 & 215,623 \\
MusiqueQA & 33,216 & 33,215 & 29,898 & 3,704,201 & 228,914 & 295,629 & 153,392 & 183,703 & 149,125 & 188,149 \\
HotpotQA & 73,891 & 73,890 & 66,559 & 13,886,807 & 511,705 & 725,521 & 291,873 & 401,693 & 324,284 & 436,362 \\
ALCE & 99,303 & 99,302 & 89,376 & 22,109,289 & 610,925 & 918,499 & 306,821 & 475,018 & 353,989 & 526,486 \\
Mix & 719 & 718 & 1,778 & 1,225,815 & 28,793 & 34,693 & 7,464 & 2,819 &7,701  & 3,336 \\
MultihopSum & 2,053 & 2,052 & N/A & N/A  & N/A  & N/A  & 12,737 & 10,063 & 18,227 & 12,441 \\
Agriculture &2,156  & 2,155 & N/A & N/A  & N/A  & N/A & 15,772 & 7,333 & 17,793 & 12,600 \\
CS & 2,244 & 2,243 & N/A & N/A  & N/A  & N/A & 10,175 & 6,560 & 12,340 & 8,692 \\
Legal & 5,380 & 5,379 & N/A & N/A  & N/A  & N/A & 15,034 & 10,920 & 16,565 & 17,633 \\

\bottomrule
\end{tabular}
\end{table*}

\begin{figure}[h]
    \centering
    \includegraphics[width=1.0\linewidth]{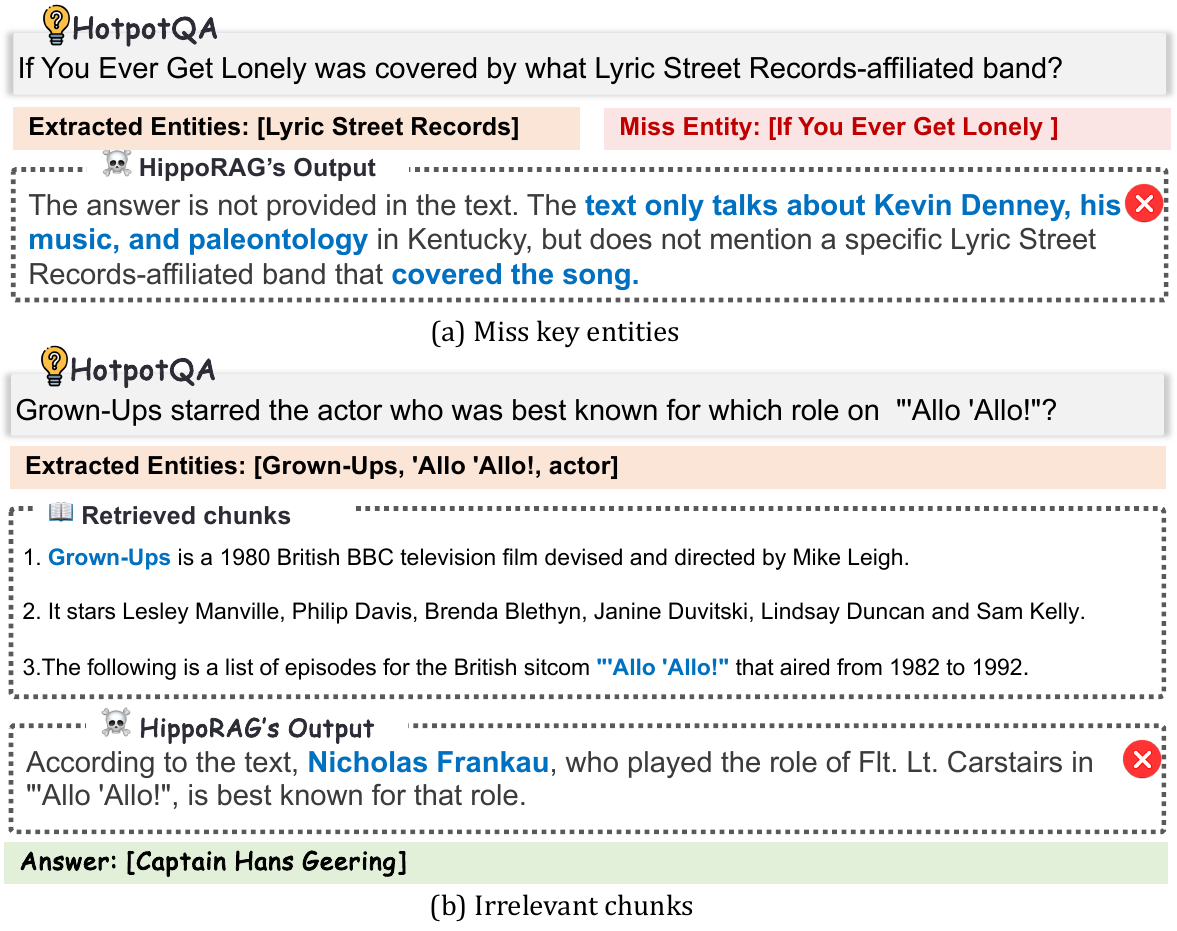}
    \caption{\change{The failure cases of {\tt HippoRAG}.}}
    \label{fig:fail:hippo}
\end{figure}

{
\begin{tikzpicture}
\filldraw (0,0) -- (-0.15,0.08) -- (-0.15,-0.08) -- cycle ; 
\end{tikzpicture}
\textbf{Exp.8. Failure cases of graph-based RAG methods.}  We would like to clarify that the results presented in our paper represent the average performance across all questions within each dataset.
Thus, not all graph-based RAG methods consistently outperform the baseline {\tt VanillaRAG} on every question.
We conduct a detailed failure case analysis focusing on the top-performing methods. Specifically, we examine why {\tt RAPTOR}, {\tt HippoRAG}, {\tt RAPTOR} and {\tt LGraphRAG} sometimes fall short in specific QA tasks.
Please refer to the detailed analysis provided below.

\noindent \begin{tikzpicture}
\filldraw (0,0) -- (-0.15,0.08) -- (-0.15,-0.08) -- cycle ; 
\end{tikzpicture} {\bf The key failure reasons of {\tt HippoRAG}:}

\begin{enumerate}

    \item {\it Incorrect or incomplete entity extraction from the question.} Consider the example in \Cref{fig:fail:hippo}(a), where {\tt HippoRAG} extracts the entity ``Lyric Street Records'' from the question and then applies Personalized PageRank based on this entity to retrieve relevant chunks.
However, to answer this question correctly, two key entities are required: ``Lyric Street Records'' and ``If You Ever Get Lonely''. Since {\tt HippoRAG} fails to extract the latter, it retrieves chunks that are insufficient to provide a correct answer.

    \item {\it Retrieval of irrelevant chunks by Personalized PageRank.}
In another example shown in \Cref{fig:fail:hippo}(b), {\tt HippoRAG} correctly extracts the relevant entities [``Grown-ups'', ``Allo 'Allo!'', ``actor''] from the question.
However, its chunk retriever strategy—Personalized PageRank—tends to favor chunks where these entities appear frequently, regardless of whether the content is semantically relevant to the question.
As a result, the retrieved chunks may not align with the actual intent of the question, leading to an incorrect final answer.
\end{enumerate}

\noindent \begin{tikzpicture}
\filldraw (0,0) -- (-0.15,0.08) -- (-0.15,-0.08) -- cycle ; 
\end{tikzpicture}{\bf The key failure reason of {\tt Raptor}:}

\begin{itemize}
    \item {\it Low-quality cluster summaries.} As illustrated in Figure~\ref{fig:fail:raptor}, {\tt Raptor} retrieves cluster summaries generated from groups of similar chunks. However, chunks within the same cluster may mention various loosely related facts that are topically similar but not logically unified. 
    For example, the summary ``Top Scorers of 2023'' in Figure~\ref{fig:fail:raptor} contains some loosely related facts about scorers in 2023, which are too general to provide a precise answer.
    When summarizing such content, the LLM tends to produce generic or fragmented summaries that fail to capture the key information required to answer the question.

\end{itemize}

\noindent \begin{tikzpicture}
\filldraw (0,0) -- (-0.15,0.08) -- (-0.15,-0.08) -- cycle ; 
\end{tikzpicture}{\bf The key failure reasons of {\tt LGraphRAG}:}

\begin{itemize}
    \item  {\it Irrelevant community reports retrieved by {\tt Entity} operator.}
Consider the example in \Cref{fig:fail:graphrag1}, where {\tt LGraphRAG} first extracts entities such as [``The Verge'', ``Google'', ``Mastercard'', $\cdots$] from the question.  
It then applies the {\tt Entity} operator to retrieve communities whose reports contain these entities.  
Among them, communities with frequent mentions of ``Google'' are prioritized.
However, these retrieved communities turn out to be irrelevant to the actual question, as the method relies solely on surface-level entity frequency while ignoring semantic relevance.

    \item {\it Irrelevant chunks retrieved by {\tt Occurrence} operator.} Consider the example in \Cref{fig:fail:graphrag2}, where {\tt LGraphRAG} extracts relationships such as [(``OpenAI'', ``pricing schema''), (``Eater'', ``wine''), (`greenmonday'', ``newegg'')] from the question.
It then applies the {\tt Occurrence} operator to retrieve chunks that contain the relationship (``OpenAI'', ``pricing schema'') with high frequency.
Based on these chunks, {\tt LGraphRAG} generates the incorrect answer ``OpenAI''.
The key reason is that the retrieved chunks, despite their frequent mentions of certain relationships, are not semantically relevant to the question. The method relies on co-occurrence frequency rather than actual contextual relevance.
\end{itemize}
}

\begin{figure}[]
    \centering
    \includegraphics[width=1.0\linewidth]{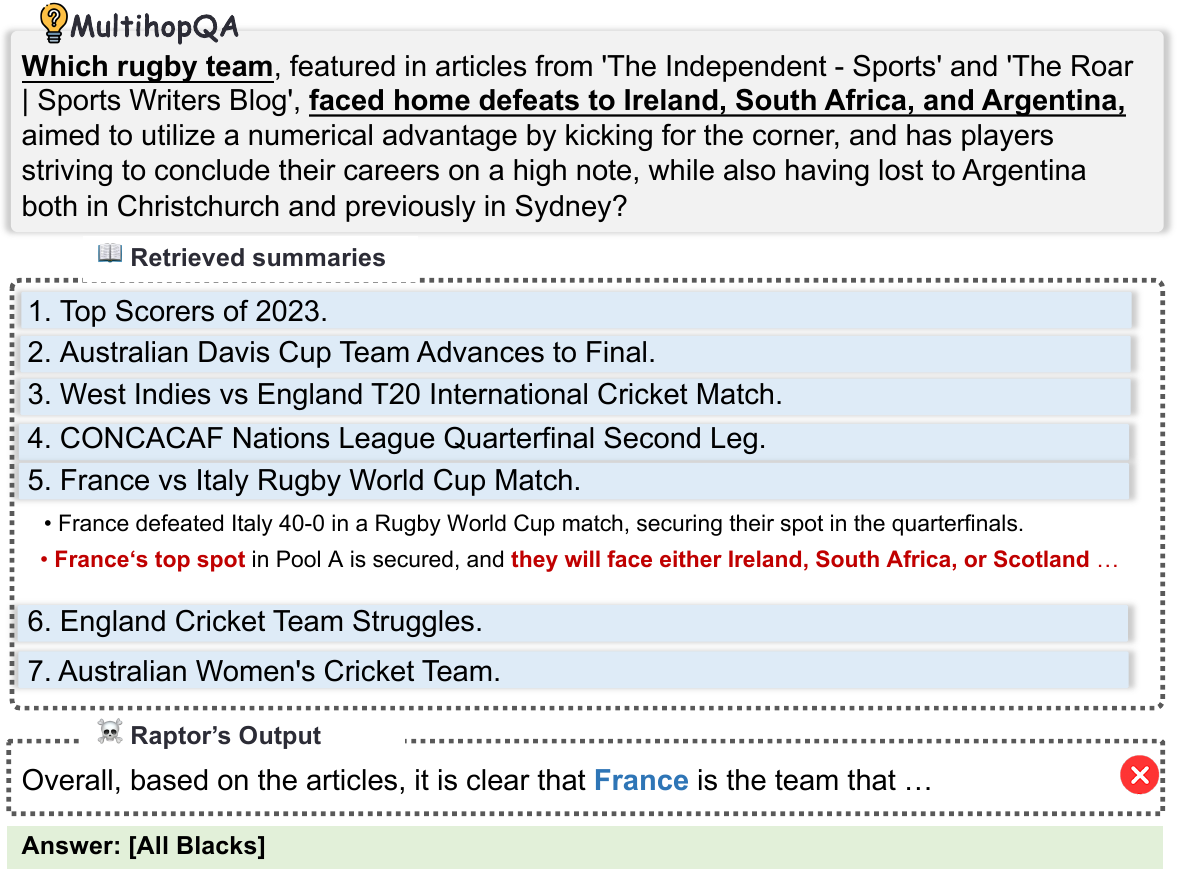}
    \caption{\change{The failure case of {\tt Raptor}.}}
    \label{fig:fail:raptor}
\end{figure}

\begin{figure}[]
    \centering
    \includegraphics[width=1.0\linewidth]{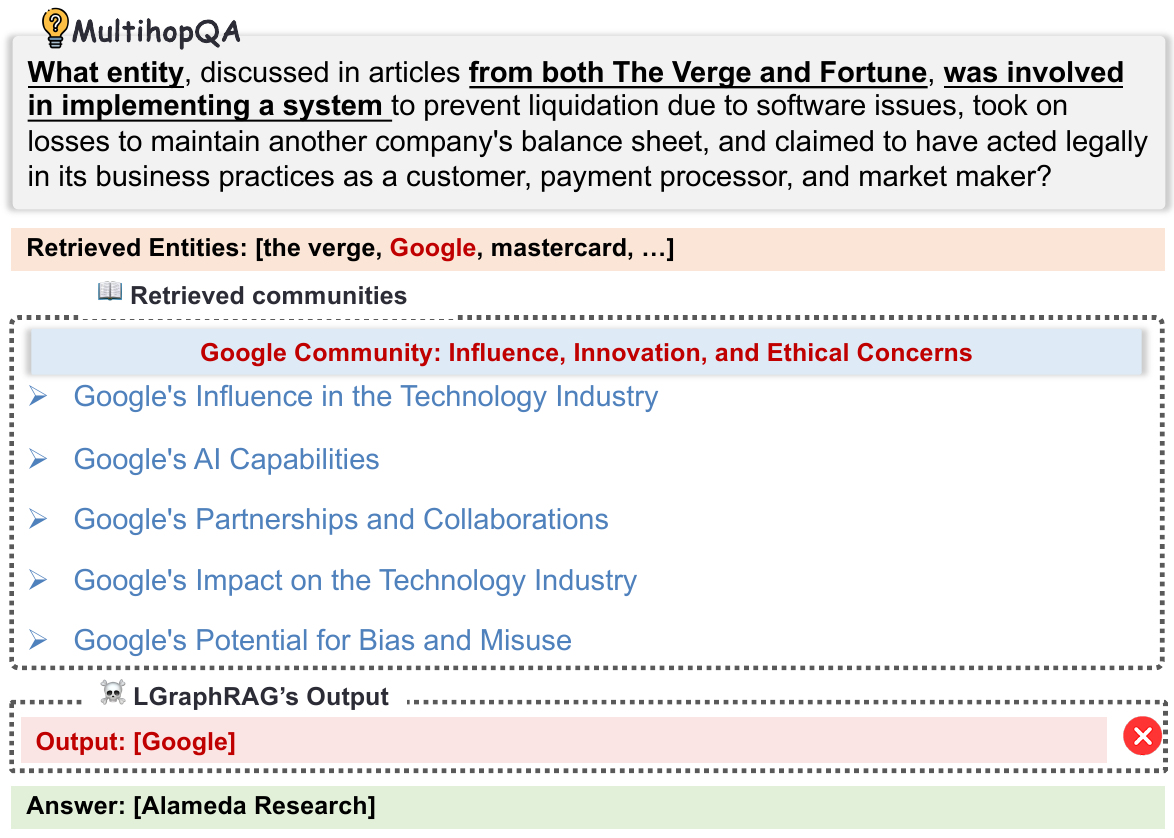}
    \caption{\change{The failure case of {\tt LGraphRAG} (I).}}
    \label{fig:fail:graphrag1}
\end{figure}

\begin{figure}[]
    \centering
    \includegraphics[width=1.0\linewidth]{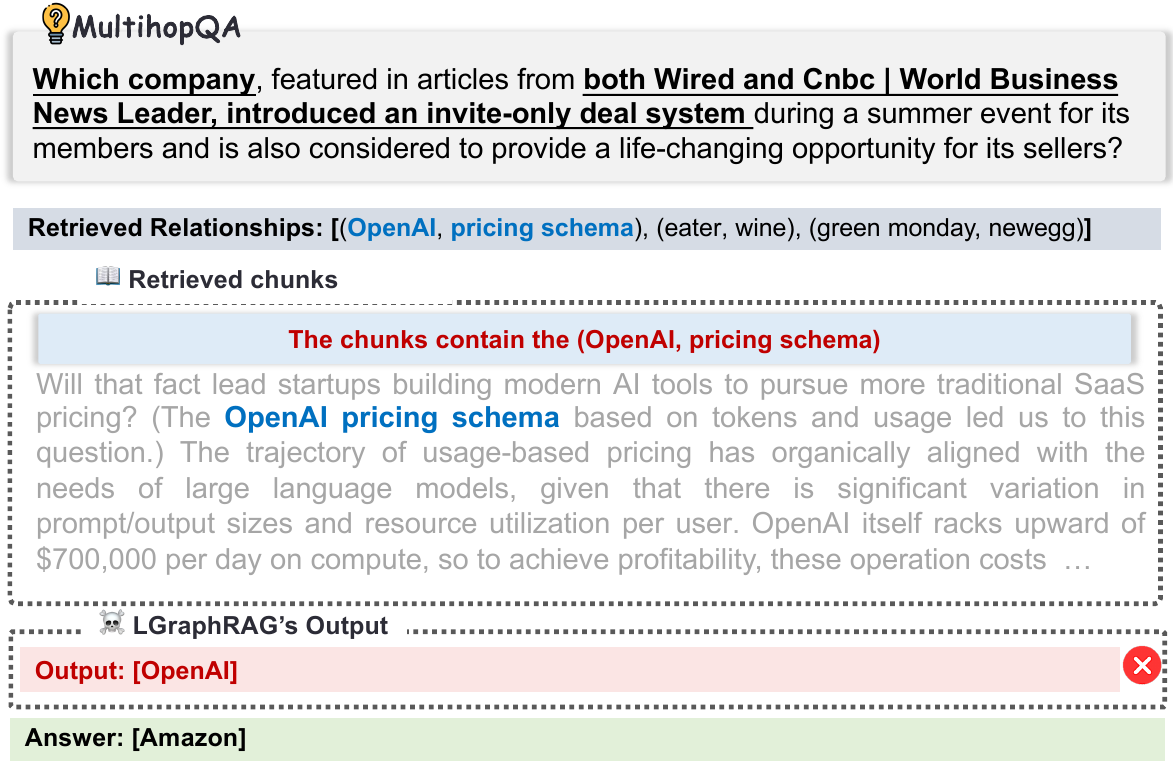}
    \caption{\change{The failure case of {\tt LGraphRAG} (II).}}
    \label{fig:fail:graphrag2}
\end{figure}

\begin{figure*}[]
\centering
\setlength{\abovecaptionskip}{-0.01cm}
\setlength{\belowcaptionskip}{-0.3cm}
\subfigure[Mix]{

    }

    \caption{Token cost of the graph building on abstract QA datasets.}
    \label{fig:graph_build_abs}
\end{figure*}

\begin{figure}[h]
    \centering
       	%
    	\caption{Token cost of index construction in abstract QA.}
    \label{fig:abs:index}
\end{figure}

\begin{figure*}[]
\centering
\setlength{\abovecaptionskip}{-0.1cm}
\setlength{\belowcaptionskip}{-0.4cm}
\renewcommand{\arraystretch}{1.1} 

\subfigure[Comprehensiveness]{
    \resizebox{0.23\textwidth}{!}{ 
    \begin{tabular}{c*{5}{c}}
& VR & RA & GS & LR & FG \\ 
{VR} & \colorcell{50} & \colorcell{27} & \colorcell{32} & \colorcell{78} & \colorcell{56} \\ 
{RA} & \colorcell{73} & \colorcell{50} & \colorcell{61} & \colorcell{88} & \colorcell{90} \\ 
{GS} & \colorcell{68} & \colorcell{39} & \colorcell{50} & \colorcell{81} & \colorcell{66} \\ 
{LR} & \colorcell{27} & \colorcell{12} & \colorcell{19} & \colorcell{50} & \colorcell{29} \\ 
{FG} & \colorcell{44} & \colorcell{10} & \colorcell{34} & \colorcell{71} & \colorcell{50} \\ 
    \end{tabular}
    }
}
\subfigure[Diversity]{
    \resizebox{0.23\textwidth}{!}{ 
    \begin{tabular}{c*{5}{c}}
& VR & RA & GS & LR & FG \\ 
{VR} & \colorcell{50} & \colorcell{42} & \colorcell{17} & \colorcell{82} & \colorcell{69} \\ 
{RA} & \colorcell{58} & \colorcell{50} & \colorcell{18} & \colorcell{87} & \colorcell{82} \\ 
{GS} & \colorcell{83} & \colorcell{82} & \colorcell{50} & \colorcell{97} & \colorcell{92} \\ 
{LR} & \colorcell{18} & \colorcell{13} & \colorcell{3} & \colorcell{50} & \colorcell{31} \\ 
{FG} & \colorcell{31} & \colorcell{18} & \colorcell{8} & \colorcell{69} & \colorcell{50} \\ 
    \end{tabular}
    }
}
\subfigure[Empowerment]{
    \resizebox{0.23\textwidth}{!}{ 
    \begin{tabular}{c*{5}{c}}
&VR & RA & GS & LR & FG \\ 
{VR} & \colorcell{50} & \colorcell{34} & \colorcell{24} & \colorcell{82} & \colorcell{74} \\ 
{RA} & \colorcell{66} & \colorcell{50} & \colorcell{38} & \colorcell{92} & \colorcell{91} \\ 
{GS} & \colorcell{76} & \colorcell{62} & \colorcell{50} & \colorcell{93} & \colorcell{89} \\ 
{LR} & \colorcell{18} & \colorcell{8} & \colorcell{7} & \colorcell{50} & \colorcell{27} \\ 
{FG} & \colorcell{26} & \colorcell{9} & \colorcell{11} & \colorcell{73} & \colorcell{50} \\ 
    \end{tabular}
    }
}
\subfigure[Overall]{
    \resizebox{0.23\textwidth}{!}{ 
    \begin{tabular}{c*{5}{c}}
&VR & RA & GS & LR & FG \\ 
{VR} & \colorcell{50} & \colorcell{27} & \colorcell{30} & \colorcell{78} & \colorcell{67} \\ 
{RA} & \colorcell{73} & \colorcell{50} & \colorcell{54} & \colorcell{90} & \colorcell{92} \\ 
{GS} & \colorcell{70} & \colorcell{46} & \colorcell{50} & \colorcell{86} & \colorcell{71} \\ 
{LR} & \colorcell{22} & \colorcell{10} & \colorcell{14} & \colorcell{50} & \colorcell{29} \\ 
{FG} & \colorcell{33} & \colorcell{8} & \colorcell{19} & \colorcell{71} & \colorcell{50} \\ 
    \end{tabular}
    }
}
\caption{\change{The abstract QA results on the MultihopSum dataset (chunk size = 600).}}
\label{fig:winmap_600}
\end{figure*}

\begin{figure*}[]
\centering
\setlength{\abovecaptionskip}{-0.1cm}
\setlength{\belowcaptionskip}{-0.4cm}
\renewcommand{\arraystretch}{1.1} 

\subfigure[Comprehensiveness]{
    \resizebox{0.23\textwidth}{!}{ 
    \begin{tabular}{c*{5}{c}}
& VR & RA & GS & LR & FG \\ 
VR & \colorcell{50} & \colorcell{50} & \colorcell{2} & \colorcell{46} & \colorcell{95} \\ 
RA & \colorcell{50} & \colorcell{50} & \colorcell{47} & \colorcell{48} & \colorcell{94} \\ 
GS & \colorcell{78} & \colorcell{53} & \colorcell{50} & \colorcell{79} & \colorcell{96} \\ 
LR & \colorcell{54} & \colorcell{52} & \colorcell{21} & \colorcell{50} & \colorcell{92} \\ 
FG & \colorcell{5} & \colorcell{6} & \colorcell{4} & \colorcell{8} & \colorcell{50} \\ 
    \end{tabular}
    }
}
\subfigure[Diversity]{
    \resizebox{0.23\textwidth}{!}{ 
    \begin{tabular}{c*{5}{c}}
& VR & RA & GS & LR & FG \\ 
VR & \colorcell{50} & \colorcell{64} & \colorcell{58} & \colorcell{64} & \colorcell{93} \\ 
RA & \colorcell{36} & \colorcell{50} & \colorcell{42} & \colorcell{49} & \colorcell{85} \\ 
GS & \colorcell{42} & \colorcell{55} & \colorcell{50} & \colorcell{52} & \colorcell{92} \\ 
LR & \colorcell{36} & \colorcell{51} & \colorcell{48} & \colorcell{50} & \colorcell{88} \\ 
FG & \colorcell{7} & \colorcell{15} & \colorcell{8} & \colorcell{12} & \colorcell{50} \\ 
    \end{tabular}
    }
}
\subfigure[Empowerment]{
    \resizebox{0.23\textwidth}{!}{ 
    \begin{tabular}{c*{5}{c}}
& VR & RA & GS & LR & FG \\ 
VR & \colorcell{50} & \colorcell{52} & \colorcell{36} & \colorcell{39} & \colorcell{95} \\ 
RA & \colorcell{48} & \colorcell{50} & \colorcell{45} & \colorcell{45} & \colorcell{93} \\ 
GS & \colorcell{64} & \colorcell{54} & \colorcell{50} & \colorcell{41} & \colorcell{97} \\ 
LR & \colorcell{61} & \colorcell{55} & \colorcell{59} & \colorcell{50} & \colorcell{95} \\ 
FG & \colorcell{5} & \colorcell{7} & \colorcell{3} & \colorcell{5} & \colorcell{50} \\ 
    \end{tabular}
    }
}
\subfigure[Overall]{
    \resizebox{0.23\textwidth}{!}{ 
    \begin{tabular}{c*{5}{c}}
& VR & RA & GS & LR & FG \\ 
VR & \colorcell{50} & \colorcell{52} & \colorcell{44} & \colorcell{46} & \colorcell{95} \\ 
RA & \colorcell{48} & \colorcell{50} & \colorcell{45} & \colorcell{47} & \colorcell{94} \\ 
GS & \colorcell{56} & \colorcell{55} & \colorcell{50} & \colorcell{52} & \colorcell{97} \\ 
LR & \colorcell{54} & \colorcell{53} & \colorcell{48} & \colorcell{50} & \colorcell{93} \\ 
FG & \colorcell{5} & \colorcell{6} & \colorcell{3} & \colorcell{7} & \colorcell{50} \\ 

    \end{tabular}
    }
}
\caption{\change{The abstract QA results on the MultihopSum dataset (chunk size = 1200).}}

\label{fig:winmap_multihop}
\end{figure*}

\begin{figure*}[]
\centering
\renewcommand{\arraystretch}{1.1} 

\subfigure[Comprehensiveness]{
    \resizebox{0.23\textwidth}{!}{ 
    \begin{tabular}{c*{5}{c}}
&VR & RA & GS & LR & FG \\ 
{VR} & \colorcell{50} & \colorcell{22} & \colorcell{44} & \colorcell{60} & \colorcell{67} \\ 
{RA} & \colorcell{78} & \colorcell{50} & \colorcell{73} & \colorcell{83} & \colorcell{88} \\ 
{GS} & \colorcell{56} & \colorcell{27} & \colorcell{50} & \colorcell{61} & \colorcell{63} \\ 
{LR} & \colorcell{40} & \colorcell{17} & \colorcell{39} & \colorcell{50} & \colorcell{53} \\ 
{FG} & \colorcell{33} & \colorcell{12} & \colorcell{37} & \colorcell{47} & \colorcell{50} \\ 
    \end{tabular}
    }
}
\subfigure[Diversity]{
    \resizebox{0.23\textwidth}{!}{ 
    \begin{tabular}{c*{5}{c}}
&VR & RA & GS & LR & FG \\ 
{VR} & \colorcell{50} & \colorcell{32} & \colorcell{17} & \colorcell{67} & \colorcell{69} \\ 
{RA} & \colorcell{68} & \colorcell{50} & \colorcell{28} & \colorcell{82} & \colorcell{82} \\ 
{GS} & \colorcell{83} & \colorcell{72} & \colorcell{50} & \colorcell{90} & \colorcell{90} \\ 
{LR} & \colorcell{33} & \colorcell{18} & \colorcell{10} & \colorcell{50} & \colorcell{53} \\ 
{FG} & \colorcell{31} & \colorcell{18} & \colorcell{10} & \colorcell{47} & \colorcell{50} \\ 
    \end{tabular}
    }
}
\subfigure[Empowerment]{
    \resizebox{0.23\textwidth}{!}{ 
    \begin{tabular}{c*{5}{c}}
&VR & RA & GS & LR & FG \\ 
{VR} & \colorcell{50} & \colorcell{26} & \colorcell{28} & \colorcell{66} & \colorcell{75} \\ 
{RA} & \colorcell{74} & \colorcell{50} & \colorcell{45} & \colorcell{83} & \colorcell{91} \\ 
{GS} & \colorcell{72} & \colorcell{55} & \colorcell{50} & \colorcell{86} & \colorcell{89} \\ 
{LR} & \colorcell{34} & \colorcell{17} & \colorcell{14} & \colorcell{50} & \colorcell{57} \\ 
{FG} & \colorcell{25} & \colorcell{9} & \colorcell{11} & \colorcell{43} & \colorcell{50} \\ 
    \end{tabular}
    }
}
\subfigure[Overall]{
    \resizebox{0.23\textwidth}{!}{ 
    \begin{tabular}{c*{5}{c}}
&VR & RA & GS & LR & FG \\ 
{VR} & \colorcell{50} & \colorcell{23} & \colorcell{37} & \colorcell{62} & \colorcell{68} \\ 
{RA} & \colorcell{77} & \colorcell{50} & \colorcell{66} & \colorcell{83} & \colorcell{90} \\ 
{GS} & \colorcell{63} & \colorcell{34} & \colorcell{50} & \colorcell{70} & \colorcell{72} \\ 
{LR} & \colorcell{38} & \colorcell{17} & \colorcell{30} & \colorcell{50} & \colorcell{54} \\ 
{FG} & \colorcell{32} & \colorcell{10} & \colorcell{28} & \colorcell{46} & \colorcell{50} \\ 
    \end{tabular}
    }
}
\caption{\change{The abstract QA results on the MultihopSum dataset (chunk size = 2400).}}
\label{fig:winmap_2400}
\end{figure*}
\begin{figure*}[]
    \centering
    \setlength{\abovecaptionskip}{-0.1cm}
    \setlength{\belowcaptionskip}{-0.4cm}
    \quad \ref{quality_ch}\\
    \subfigure[Comprehensiveness]{
    %
    }
    
    \caption{\change{Comparison of {\tt CheapRAG} and other methods under different chunk sizes on the MultihopSum dataset.}}
    \label{fig:winwinchunk}
\end{figure*}

\begin{figure*}[]
\centering
\setlength{\abovecaptionskip}{-0.1cm}
\setlength{\belowcaptionskip}{-0.4cm}
\renewcommand{\arraystretch}{1.1} 

\subfigure[Comprehensiveness]{
    \resizebox{0.23\textwidth}{!}{ 
    \begin{tabular}{c*{5}{c}}
&VR & RA & GS & LR & FG \\ 
{VR} & \colorcell{50} & \colorcell{17} & \colorcell{9} & \colorcell{14} & \colorcell{62} \\ 
{RA} & \colorcell{83} & \colorcell{50} & \colorcell{24} & \colorcell{55} & \colorcell{86} \\ 
{GS} & \colorcell{91} & \colorcell{76} & \colorcell{50} & \colorcell{77} & \colorcell{93} \\ 
{LR} & \colorcell{86} & \colorcell{45} & \colorcell{23} & \colorcell{50} & \colorcell{91} \\ 
{FG} & \colorcell{38} & \colorcell{14} & \colorcell{7} & \colorcell{9} & \colorcell{50} \\ 
    \end{tabular}
    }
}
\subfigure[Diversity]{
    \resizebox{0.23\textwidth}{!}{ 
    \begin{tabular}{c*{5}{c}}
&VR & RA & GS & LR & FG \\ 
{VR} & \colorcell{50} & \colorcell{22} & \colorcell{9} & \colorcell{22} & \colorcell{73} \\ 
{RA} & \colorcell{78} & \colorcell{50} & \colorcell{26} & \colorcell{60} & \colorcell{86} \\ 
{GS} & \colorcell{91} & \colorcell{74} & \colorcell{50} & \colorcell{83} & \colorcell{95} \\ 
{LR} & \colorcell{78} & \colorcell{40} & \colorcell{17} & \colorcell{50} & \colorcell{90} \\ 
{FG} & \colorcell{27} & \colorcell{14} & \colorcell{5} & \colorcell{10} & \colorcell{50} \\ 
    \end{tabular}
    }
}
\subfigure[Empowerment]{
    \resizebox{0.23\textwidth}{!}{ 
    \begin{tabular}{c*{5}{c}}
&VR & RA & GS & LR & FG \\ 
{VR} & \colorcell{50} & \colorcell{26} & \colorcell{12} & \colorcell{26} & \colorcell{74} \\ 
{RA} & \colorcell{74} & \colorcell{50} & \colorcell{24} & \colorcell{63} & \colorcell{90} \\ 
{GS} & \colorcell{88} & \colorcell{76} & \colorcell{50} & \colorcell{84} & \colorcell{94} \\ 
{LR} & \colorcell{74} & \colorcell{37} & \colorcell{16} & \colorcell{50} & \colorcell{91} \\ 
{FG} & \colorcell{26} & \colorcell{10} & \colorcell{6} & \colorcell{9} & \colorcell{50} \\ 
    \end{tabular}
    }
}
\subfigure[Overall]{
    \resizebox{0.23\textwidth}{!}{ 
    \begin{tabular}{c*{5}{c}}
&VR & RA & GS & LR & FG \\ 
{VR} & \colorcell{50} & \colorcell{19} & \colorcell{10} & \colorcell{18} & \colorcell{68} \\ 
{RA} & \colorcell{81} & \colorcell{50} & \colorcell{24} & \colorcell{61} & \colorcell{88} \\ 
{GS} & \colorcell{90} & \colorcell{76} & \colorcell{50} & \colorcell{80} & \colorcell{93} \\ 
{LR} & \colorcell{82} & \colorcell{39} & \colorcell{20} & \colorcell{50} & \colorcell{91} \\ 
{FG} & \colorcell{32} & \colorcell{12} & \colorcell{7} & \colorcell{9} & \colorcell{50} \\ 
    \end{tabular}
    }
}
\caption{\change{The abstract QA results on the MultihopSum dataset (LLM backbone = Qwen-2.5-32B).}}
\label{fig:winmap_32b}
\end{figure*}

\begin{figure*}[]
\centering
\setlength{\abovecaptionskip}{-0.1cm}
\setlength{\belowcaptionskip}{-0.4cm}
\renewcommand{\arraystretch}{1.1} 

\subfigure[Comprehensiveness]{
    \resizebox{0.23\textwidth}{!}{ 
    \begin{tabular}{c*{5}{c}}
&VR & RA & GS & LR & FG \\ 
{VR} & \colorcell{50} & \colorcell{10} & \colorcell{18} & \colorcell{16} & \colorcell{70} \\ 
{RA} & \colorcell{90} & \colorcell{50} & \colorcell{49} & \colorcell{68} & \colorcell{96} \\ 
{GS} & \colorcell{82} & \colorcell{51} & \colorcell{50} & \colorcell{63} & \colorcell{90} \\ 
{LR} & \colorcell{84} & \colorcell{32} & \colorcell{37} & \colorcell{50} & \colorcell{94} \\ 
{FG} & \colorcell{30} & \colorcell{4} & \colorcell{10} & \colorcell{6} & \colorcell{50} \\ 
    \end{tabular}
    }
}
\subfigure[Diversity]{
    \resizebox{0.23\textwidth}{!}{ 
    \begin{tabular}{c*{5}{c}}
&VR & RA & GS & LR & FG \\ 
{VR} & \colorcell{50} & \colorcell{15} & \colorcell{1} & \colorcell{19} & \colorcell{67} \\ 
{RA} & \colorcell{85} & \colorcell{50} & \colorcell{5} & \colorcell{68} & \colorcell{91} \\ 
{GS} & \colorcell{99} & \colorcell{95} & \colorcell{50} & \colorcell{99} & \colorcell{99} \\ 
{LR} & \colorcell{81} & \colorcell{32} & \colorcell{1} & \colorcell{50} & \colorcell{88} \\ 
{FG} & \colorcell{33} & \colorcell{9} & \colorcell{1} & \colorcell{12} & \colorcell{50} \\ 
    \end{tabular}
    }
 }
\subfigure[Empowerment]{
    \resizebox{0.23\textwidth}{!}{ 
    \begin{tabular}{c*{5}{c}}
&VR & RA & GS & LR & FG \\ 
{VR} & \colorcell{50} & \colorcell{13} & \colorcell{4} & \colorcell{29} & \colorcell{81} \\ 
{RA} & \colorcell{87} & \colorcell{50} & \colorcell{20} & \colorcell{74} & \colorcell{97} \\ 
{GS} & \colorcell{96} & \colorcell{80} & \colorcell{50} & \colorcell{91} & \colorcell{99} \\ 
{LR} & \colorcell{71} & \colorcell{26} & \colorcell{9} & \colorcell{50} & \colorcell{93} \\ 
{FG} & \colorcell{19} & \colorcell{3} & \colorcell{1} & \colorcell{7} & \colorcell{50} \\ 
    \end{tabular}
    }
}
\subfigure[Overall]{
    \resizebox{0.23\textwidth}{!}{ 
    \begin{tabular}{c*{5}{c}}
&VR & RA & GS & LR & FG \\ 
{VR} & \colorcell{50} & \colorcell{12} & \colorcell{6} & \colorcell{18} & \colorcell{78} \\ 
{RA} & \colorcell{88} & \colorcell{50} & \colorcell{30} & \colorcell{76} & \colorcell{97} \\ 
{GS} & \colorcell{94} & \colorcell{70} & \colorcell{50} & \colorcell{83} & \colorcell{99} \\ 
{LR} & \colorcell{82} & \colorcell{24} & \colorcell{17} & \colorcell{50} & \colorcell{95} \\ 
{FG} & \colorcell{22} & \colorcell{3} & \colorcell{1} & \colorcell{5} & \colorcell{50} \\ 
    \end{tabular}
    }
}
\caption{\change{The abstract QA results on the MultihopSum dataset (LLM backbone = Llama-3-70B).}}
\label{fig:winmap_70b}
\end{figure*}

\begin{figure*}[]
\centering
\setlength{\abovecaptionskip}{-0.1cm}
\setlength{\belowcaptionskip}{-0.4cm}
\renewcommand{\arraystretch}{1.1} 

\subfigure[Comprehensiveness]{
    \resizebox{0.23\textwidth}{!}{ 
    \begin{tabular}{c*{5}{c}}
&VR & RA & GS & LR & FG \\ 
{VR} & \colorcell{50} & \colorcell{66} & \colorcell{4} & \colorcell{12} & \colorcell{44} \\ 
{RA} & \colorcell{34} & \colorcell{50} & \colorcell{5} & \colorcell{9} & \colorcell{31} \\ 
{GS} & \colorcell{96} & \colorcell{95} & \colorcell{50} & \colorcell{81} & \colorcell{94} \\ 
{LR} & \colorcell{88} & \colorcell{91} & \colorcell{19} & \colorcell{50} & \colorcell{82} \\ 
{FG} & \colorcell{56} & \colorcell{69} & \colorcell{6} & \colorcell{18} & \colorcell{50} \\ 
    \end{tabular}
    }
}
\subfigure[Diversity]{
    \resizebox{0.23\textwidth}{!}{ 
    \begin{tabular}{c*{5}{c}}
&VR & RA & GS & LR & FG \\ 
{VR} & \colorcell{50} & \colorcell{61} & \colorcell{6} & \colorcell{15} & \colorcell{54} \\ 
{RA} & \colorcell{39} & \colorcell{50} & \colorcell{6} & \colorcell{14} & \colorcell{39} \\ 
{GS} & \colorcell{94} & \colorcell{94} & \colorcell{50} & \colorcell{85} & \colorcell{95} \\ 
{LR} & \colorcell{85} & \colorcell{86} & \colorcell{15} & \colorcell{50} & \colorcell{87} \\ 
{FG} & \colorcell{46} & \colorcell{61} & \colorcell{5} & \colorcell{13} & \colorcell{50} \\ 
    \end{tabular}
    }
}
\subfigure[Empowerment]{
    \resizebox{0.23\textwidth}{!}{ 
    \begin{tabular}{c*{5}{c}}
&VR & RA & GS & LR & FG \\ 
{VR} & \colorcell{50} & \colorcell{79} & \colorcell{8} & \colorcell{31} & \colorcell{74} \\ 
{RA} & \colorcell{21} & \colorcell{50} & \colorcell{4} & \colorcell{14} & \colorcell{40} \\ 
{GS} & \colorcell{92} & \colorcell{96} & \colorcell{50} & \colorcell{92} & \colorcell{95} \\ 
{LR} & \colorcell{69} & \colorcell{86} & \colorcell{8} & \colorcell{50} & \colorcell{86} \\ 
{FG} & \colorcell{26} & \colorcell{60} & \colorcell{5} & \colorcell{14} & \colorcell{50} \\ 
    \end{tabular}
    }
}
\subfigure[Overall]{
    \resizebox{0.23\textwidth}{!}{ 
    \begin{tabular}{c*{5}{c}}
&VR & RA & GS & LR & FG \\ 
{VR} & \colorcell{50} & \colorcell{72} & \colorcell{4} & \colorcell{13} & \colorcell{60} \\ 
{RA} & \colorcell{28} & \colorcell{50} & \colorcell{4} & \colorcell{10} & \colorcell{37} \\ 
{GS} & \colorcell{96} & \colorcell{96} & \colorcell{50} & \colorcell{90} & \colorcell{95} \\ 
{LR} & \colorcell{87} & \colorcell{90} & \colorcell{10} & \colorcell{50} & \colorcell{87} \\ 
{FG} & \colorcell{40} & \colorcell{63} & \colorcell{5} & \colorcell{13} & \colorcell{50} \\ 
    \end{tabular}
    }
}
\caption{\change{The abstract QA results on the MultihopSum dataset (LLM backbone = GPT-4o-mini).}}
\label{fig:winmap_4o-mini}
\end{figure*}

\begin{figure*}[]
\centering
\setlength{\abovecaptionskip}{-0.1cm}
\setlength{\belowcaptionskip}{-0.3cm}
\renewcommand{\arraystretch}{1.1} 

\subfigure[LLM backbone =  Llama-3-8B]{
    \resizebox{0.23\textwidth}{!}{ 
    \begin{tabular}{c*{5}{c}}
& Com. & Div. & Emp. & Overall  \\ 
VR & \colorcell{74} & \colorcell{47} & \colorcell{64} & \colorcell{68}  \\ 
RA & \colorcell{68} & \colorcell{48} & \colorcell{64} & \colorcell{67} \\ 
GS & \colorcell{72} & \colorcell{43} & \colorcell{65} & \colorcell{68} \\ 
LR & \colorcell{66}  & \colorcell{45} & \colorcell{56} & \colorcell{64} \\ 
FG & \colorcell{100} & \colorcell{86} & \colorcell{98} & \colorcell{99} \\ 
    \end{tabular}
    }
    }
\subfigure[LLM backbone = Qwen-2.5-32B]{
    \resizebox{0.23\textwidth}{!}{ 
    \begin{tabular}{c*{5}{c}}
& Com. & Div. & Emp. & Overall  \\ 
{VR} & \colorcell{99} & \colorcell{99} & \colorcell{99} & \colorcell{99} \\ 
{RA} & \colorcell{87} & \colorcell{96} & \colorcell{91} & \colorcell{93} \\ 
{GS} & \colorcell{90} & \colorcell{81} & \colorcell{60} & \colorcell{82} \\ 
{LR} & \colorcell{99} & \colorcell{99} & \colorcell{97} & \colorcell{99} \\ 
{FG} & \colorcell{99} & \colorcell{99} & \colorcell{99} & \colorcell{99} \\     \end{tabular}
    }
}
\subfigure[LLM backbone = Llama-3-70B]{
    \resizebox{0.23\textwidth}{!}{ 
    \begin{tabular}{c*{5}{c}}
& Com. & Div. & Emp. & Overall  \\ 
{VR} & \colorcell{98} & \colorcell{97} & \colorcell{91} & \colorcell{96} \\ 
{RA} & \colorcell{93} & \colorcell{77} & \colorcell{61} & \colorcell{73} \\ 
{GS} & \colorcell{95} & \colorcell{40} & \colorcell{76} & \colorcell{88} \\ 
{LR} & \colorcell{98} & \colorcell{94} & \colorcell{76} & \colorcell{88} \\ 
{FG} & \colorcell{99} & \colorcell{99} & \colorcell{98} & \colorcell{99} \\     \end{tabular}
    }
}
\subfigure[LLM backbone = GPT-4o-mini]{
    \resizebox{0.23\textwidth}{!}{ 
    \begin{tabular}{c*{5}{c}}
& Com. & Div. & Emp. & Overall  \\ 
{VR} & \colorcell{99} & \colorcell{99} & \colorcell{99} & \colorcell{99} \\ 
{RA} & \colorcell{99} & \colorcell{99} & \colorcell{99} & \colorcell{99} \\ 
{GS} & \colorcell{64} & \colorcell{90} & \colorcell{72} & \colorcell{80} \\ 
{LR} & \colorcell{93} & \colorcell{99} & \colorcell{96} & \colorcell{98} \\ 
{FG} & \colorcell{99} & \colorcell{99} & \colorcell{99} & \colorcell{99} \\     \end{tabular}
    }
}
\caption{\change{Comparison of our newly designed method on MultihopSum dataset.}}

\label{fig:app:new_sota:abs:llm}
\end{figure*}

\subsection{Results on abstract QA tasks}
In this subsection, we present the  addition results on abstract QA tasks.

%
  \textbf{Exp.1. Token costs of graph and index building.} 
The token costs of the graph and index building across all abstract QA datasets are shown in Figures~\ref{fig:graph_build_abs} and \ref{fig:abs:index}, respectively.
The conclusions are highly similar to those Exp.2 in Section \ref{sec:exp:speci}.

{
%
  \textbf{Exp.2. Effect of chunk size.} We report the performance of various RAG methods on abstract QA tasks under different chunk sizes in Figures~\ref{fig:winmap_600} to~\ref{fig:winmap_2400}. Our key observations are as follows:
(1) The performance of {\tt GGraphRAG} remains stable across different chunk sizes, likely due to its use of the {\tt Map-Reduce} strategy for final answer synthesis, which mitigates the influence of chunk granularity.
(2) In contrast, methods like {\tt FastGraphRAG} and {\tt VanillaRAG} show greater variance across chunk sizes, as their performance relies heavily on the granularity of individual chunks—smaller chunks tend to provide more precise information, directly impacting retrieval and generation quality.
(3) Regardless of chunk size, {\tt RAPTOR} and {\tt GGraphRAG} consistently achieve the best performance, reaffirming our earlier conclusion that high-level structural information is essential for abstract QA tasks.

Additionally, we evaluate our newly proposed method {\tt CheapRAG} across different chunk sizes. As shown in Figure~\ref{fig:winwinchunk}, {\tt CheapRAG} generally outperforms the five baselines. Notably, under the 600-token setting, {\tt CheapRAG} surpasses {\tt GGraphRAG} in more cases. We attribute this to the higher precision of smaller chunks, which enhances the effectiveness of semantic similarity-based retrieval in {\tt CheapRAG}, compared to the entity frequency-based retrieval strategy used in {\tt GGraphRAG}.

\begin{tikzpicture}
\filldraw (0,0) -- (-0.15,0.08) -- (-0.15,-0.08) -- cycle ; 
\end{tikzpicture}  \textbf{Exp.3. Effect of LLM backbone.}  We evaluate all methods that support abstract QA on the MultihopSum dataset, using different LLM backbones. Results for Llama-3-8B are shown in Figure~\ref{fig:winmap_multihop}, while Figures~\ref{fig:winmap_32b} to \ref{fig:winmap_4o-mini} present results for the other models.
We observe that, across different backbones, the performance of each method on abstract QA tasks remains relatively stable—especially when compared to the fluctuations seen in specific QA tasks.
We note that all methods still lag behind {\tt GGraphRAG}, further highlighting that community-level information is particularly beneficial for abstract QA tasks.
Moreover, we compare our newly proposed method, {\tt CheapRAG}, with five strong baselines under different LLM backbones. As shown in Figure~\ref{fig:app:new_sota:abs:llm}, {\tt CheapRAG} exhibits remarkable performance improvements as the model capacity increases. Notably, under the GPT-4o-mini backbone, {\tt CheapRAG} achieves near-universal wins across all evaluated cases, clearly demonstrating its strong generalization ability and effectiveness.

\begin{tikzpicture}
\filldraw (0,0) -- (-0.15,0.08) -- (-0.15,-0.08) -- cycle ; 
\end{tikzpicture}  \textbf{Exp.4. More analysis.} We have further analyzed the per-metric results across different methods, and summarize the key insights as follows:

\begin{itemize}
    \item {\it Comprehensiveness:}  {\tt GGraphRAG} consistently achieves the highest scores, highlighting the strength of community-level retrieval in capturing global context. By grouping semantically related content, communities help reduce fragmented evidence and support more holistic answers.

    \item {\it Diversity:} Both {\tt RAPTOR} and {\tt GGraphRAG} perform strongly by aggregating information across multiple clusters or communities. This enables the generation of responses that span diverse subtopics while maintaining relevance.

    \item {\it Empowerment:}  {\tt GGraphRAG} and {\tt LightRAG} jointly lead on this metric. Their retrieval strategies incorporate structured elements—entities, relations, and keywords—that provide concrete grounding for the model to generate actionable, role-relevant responses. This better supports practical decision-making in activity-centered QA.

\item \textit{Overall:} {\tt GGraphRAG} consistently ranks first, with {\tt RAPTOR} typically second. This highlights the advantage of leveraging high-level summarized information—such as community reports and cluster-level chunks—with the former generally proving more effective. Additionally, the results support the effectiveness of the {\tt Map-Reduce} mechanism in filtering out irrelevant information during retrieval.    
\end{itemize}
}

\subsection{Evaluation metrics} \label{sec:metrics}

This section outlines the metrics used for evaluation.

$\bullet$ \textbf{Metrics for specific QA Tasks.} We use accuracy as the evaluation metric, based on whether the gold answers appear in the model's generated outputs, rather than requiring an exact match, following the approach in \cite{schick2024toolformer,mallen2022not,asai2023self}. This choice is motivated by the uncontrollable nature of LLM outputs, which often makes it difficult to achieve exact matches with standard answers. Similarly, we prefer recall over precision as it better reflects the accuracy of the generated responses. 

$\bullet$ \textbf{Metrics for abstract QA Tasks.} Building on existing work, we use an LLM to generate abstract questions, as shown in Figure \ref{fig:prompt_summary}.
\change{We adopt 125 questions following the prior works, such as {\tt GraphRAG}~\cite{edge2024local}, {\tt LightRAG}~\cite{guo2024lightrag}, and {\tt FastGraphRAG}~\cite{fastgraphrag}.
In their setup, the number 125 comes from generating $N{=}5$ user roles, each with $N$ associated tasks, and for each (user, task) pair, $N$ abstract questions—yielding $5 \times 5 \times 5 = 125$ questions per dataset. We follow this standard for consistency and comparability across studies.
The reasons of selecting GPT-4o are twofold: (1) at the time when conducted our experiments, GPT-4o was one of the most advanced LLM, demonstrating strong zero-shot capabilities and superior performance in long-context understanding compared to other models, and (2) it provides the highest fluency, coherence, and factual consistency, ensuring that the generated questions are both challenging and realistic. This effectively supports our exploration of real-world data sensemaking scenarios.
}

Defining ground truth for abstract questions, especially those involving complex high-level semantics, presents significant challenges. To address this, we adopt an LLM-based multi-dimensional comparison method, inspired by \cite{edge2024local,guo2024lightrag}, which evaluates comprehensiveness, diversity, empowerment, and overall quality. We use a robust LLM, specifically GPT-4o, to rank each baseline in comparison to our method. The evaluation prompt used is shown in Figure \ref{fig:eval_summary}.
\change{These four metrics are defined \cite{fastgraphrag,edge2024local,guo2024lightrag} as follows:
\begin{itemize}
    \item \textit{Comprehensiveness}. How much detail does the answer provide to cover all aspects and details of the question?  
    \item \textit{Diversity}. How varied and rich is the answer in providing different perspectives and insights on the question?  
    \item \textit{Empowerment}. How well does the answer help the reader understand and make informed judgments about the topic?
    \item \textit{Overall}. Select an overall winner based on these categories.
\end{itemize}
To better illustrate these dimensions, Figure~\ref{fig:eval_dimensions_example} presents examples of both good and bad answers with respect to comprehensiveness, diversity, and empowerment.}

{
\textbf{Head-to-head comparison.} To evaluate abstract QA tasks by head-to-head comparison using an LLM evaluator, selecting four target metrics capturing qualities that are desirable for abstract questions.
The answer to an abstract question is not a collection of details from specific texts, but rather a high-level understanding of the dataset's contents relevant to the query.
A good response to an abstract question should perform well across the following four metrics, including ``Comprehensiveness'', ``Diversity'', ``Empowerment'', and ``Overall''.

Head-to-head win results illustrate the relative performance among different methods. 
Each value indicates the percentage of test cases where the row method outperforms the column method - higher values indicate better performance.
For example, Figure~\ref{fig:enter-label} demonstrates an example head-to-head result under the ``comprehensive'' dimension on the Mix dataset.
The value $30$ in the first row and third column indicates that \texttt{VanillaRAG} outperforms \texttt{GGraphRAG} in 30\% of cases, suggesting that \texttt{VanillaRAG} performs worse than \texttt{GGraphRAG} on this metric.
Based on the overall comparison in the figure, the relative ranking of the five methods is: 1. \texttt{HLightRAG} (LR), 2. \texttt{GGraphRAG} (GS), 3. \texttt{VanillaRAG} (VR), 4. \texttt{RAPTOR} (RA), 5. \texttt{FastGraphRAG} (FG).
}

\begin{figure}[]
    \centering
     \setlength{\abovecaptionskip}{-0.01em}
        \setlength{\belowcaptionskip}{-0.1em}
    \includegraphics[scale=0.43]{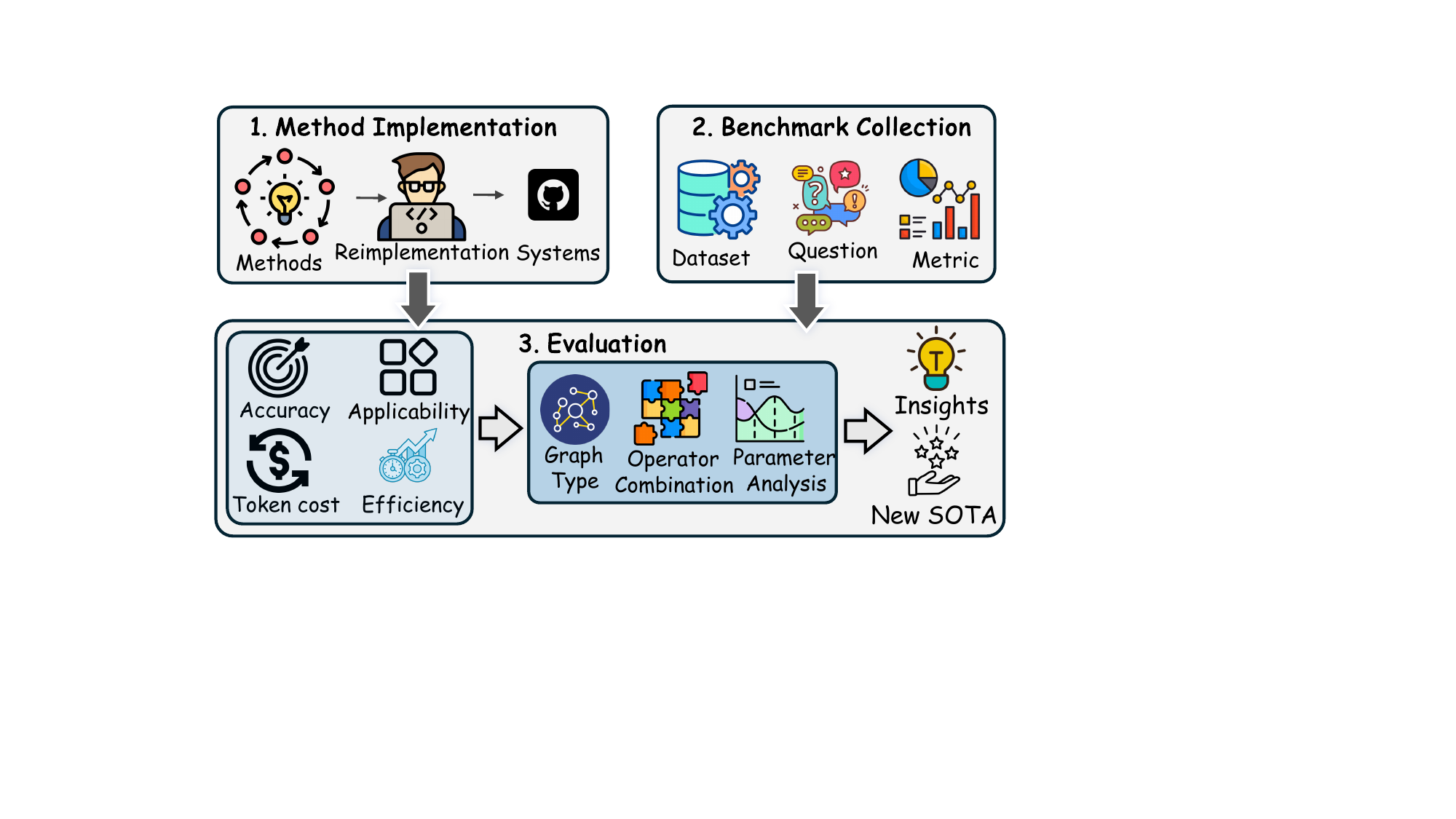}
    \caption{Workflow of our empirical study.}
    \label{fig:workflow}
\end{figure}

\subsection{Implementation details}
In this subsection, we present more details about our system implementation. 
Specifically, we use HNSW~\cite{malkov2018efficient} from Llama-index~\cite{llamaindex} (a well-known open-source project) as the default vector database for efficient vector search.
In addition, for each method, we optimize efficiency by batching or parallelizing operations such as encoding nodes or chunks, and computing personalized page rank, among others, during the retrieval stage.

{
\textbf{Why Llama-3-8B?} Beyond the widespread use of Llama-3-8B in existing works, our choice is further motivated by the following considerations:

\begin{enumerate}

    
    \item \textit{Strong Capability}: Llama-3-8B  exhibits solid language understanding and reasoning abilities \cite{dubey2024llama}, which are crucial for RAG workflows. Its effectiveness in integrating retrieved content with internal knowledge enables accurate and contextually relevant response generation.
    \item \textit{Practical Efficiency}: The 8B variant offers a favorable trade-off between performance and resource efficiency. It supports FP16 inference within 20 GB of GPU memory, allowing deployment on widely available hardware (e.g., a single 24 GB GPU).

         \item \textit{Open source LLM}: As an open-source LLM, it can be deployed locally, saving the costs of paying for API calls required by proprietary models such as GPT-4o.

\end{enumerate}

\textbf{Workflow of our evaluation.} We present the first open-source
testbed for graph-based RAG methods, which (1) collects and reimplements 12 representative methods within a unified framework
(as depicted in Section 3). (2) supports a fine-grained comparison
over the building blocks of the retrieval stage with up to 100+ variants, and (3) provides a comprehensive evaluation over 11 datasets
with various metrics in different scenarios, we summarize the workflow of our empirical study in Figure \ref{fig:workflow}.

\subsection{Criteria of selection dataset}

Our work aims to systematically analyze various graph-based RAG methods and provide in-depth insights.
To this end, we select datasets based on the following criteria:

\begin{enumerate}
  
   \item \textit{Widely used in the existing works:} All selected datasets are extensively used in the RAG and LLM research communities. For example, the HotpotQA dataset has been cited more than 2,800 times.

    \item \textit{Diverse corpus domains:} Our selected datasets cover a broad range of corpus domains. For instance, MultihopRAG consists of long English news articles; HotpotQA includes short passages from Wikipedia; and Quality comprises a diverse mixture of sources such as fiction from Project Gutenberg and articles from Slate magazine, with each document containing at least 2,000 tokens. For abstract questions, we include datasets spanning various domains such as agriculture, computer science, and legal texts.

        \item \textit{Diverse in task types:} Question answering (QA) tasks are typically categorized into abstract and specific questions \cite{guo2024lightrag}, in which the specific questions can be further divided by complexity into simple and complex types \cite{sarthi2024raptor,asai2023self}. Specifically, simple questions typically require only one or two text chunks for an answer, without the need for multi-hop reasoning. In contrast, complex questions necessitate reasoning across multiple chunks, understanding implicit relations, and synthesizing information.

\end{enumerate}
}

\section{More Discussions.}

\subsection{New operators}

Here, we introduce the operators that are not used in existing graph-based RAG methods but are employed in our newly designed state-of-the-art methods.

\textbf{Chunk type.} We include a new operator \texttt{VDB} of chunk type, which is used in our {\tt VGraphRAG} method.
This operator is the same as the chunk retrieval strategy of \texttt{VanillaRAG}.

\textbf{Community type.}  We also include a new operator \texttt{VDB} of community type, retrieving the top-$k$ communities by vector searching, where the embedding of each community is generated by encoding its community report.
    

\subsection{ More Lessons and Opportunities}

In this subsection, we show the more lessons and opportunities learned from our study. 

\noindent \textbf{Lessons}

\noindent \begin{tikzpicture}
\filldraw (0,0) -- (-0.15,0.08) -- (-0.15,-0.08) -- cycle ; 
\end{tikzpicture} \textbf{\underline{L6.}} For large datasets, both versions of the \texttt{GraphRAG} methods incur unacceptable token costs, as they contain a large number of communities, leading to high costs for generating community reports.

\noindent \begin{tikzpicture}
\filldraw (0,0) -- (-0.15,0.08) -- (-0.15,-0.08) -- cycle ; 
\end{tikzpicture} \textbf{\underline{L7.}} Regardless of whether the questions are specific or abstract, they all rely on an external corpus (i.e., documents). For such questions, merely using graph-structure information (nodes, edges, or subgraphs) is insufficient to achieve good performance.

\noindent \begin{tikzpicture}
\filldraw (0,0) -- (-0.15,0.08) -- (-0.15,-0.08) -- cycle ; 
\end{tikzpicture} \textbf{\underline{L8.}} Methods designed for knowledge reasoning tasks, such as \texttt{DALK}, \texttt{ToG}, and \texttt{G-retriever}, do not perform well on document-based QA tasks.
This is because these methods are better suited for extracting reasoning rules or paths from well-constructed KGs. However, when KGs are built from raw text corpora, they may not accurately capture the correct reasoning rules, leading to suboptimal performance in document-based QA tasks.

\noindent \begin{tikzpicture}
\filldraw (0,0) -- (-0.15,0.08) -- (-0.15,-0.08) -- cycle ; 
\end{tikzpicture} \textbf{\underline{L9.}} The effectiveness of RAG methods is highly impacted by the relevance of the retrieved elements to the given question. 
That is, if the retrieved information is irrelevant or noisy, it may degrade the LLM's performance.
When designing new graph-based RAG methods, it is crucial to evaluate whether the retrieval strategy effectively retrieves relevant information for the given question.

\noindent  \textbf{Opportunities}

\noindent \begin{tikzpicture}
\filldraw (0,0) -- (-0.15,0.08) -- (-0.15,-0.08) -- cycle ; 
\end{tikzpicture}   \underline{\textbf{O5.}} 
In real applications, external knowledge sources are not limited to text corpora; they may also include PDFs, HTML pages, tables, and other structured or semi-structured data.  
A promising future research direction is to explore graph-based RAG methods for heterogeneous knowledge sources.

\noindent \begin{tikzpicture}
\filldraw (0,0) -- (-0.15,0.08) -- (-0.15,-0.08) -- cycle ; 
\end{tikzpicture} \textbf{\underline{O6.}} An interesting future research direction is to explore more graph-based RAG applications. For example, applying graph-based RAG to scientific literature retrieval can help researchers efficiently extract relevant studies and discover hidden relationships between concepts. Another potential application is legal document analysis, where graph structures can capture case precedents and legal interpretations to assist in legal reasoning.

\noindent \begin{tikzpicture}
\filldraw (0,0) -- (-0.15,0.08) -- (-0.15,-0.08) -- cycle ; 
\end{tikzpicture} \textbf{\underline{O7.}} The users may request multiple questions simultaneously, but existing graph-based RAG methods process them sequentially.  
Hence, a promising future direction is to explore efficient scheduling strategies that optimize multi-query handling. This could involve batching similar questions or parallelizing retrieval.

\noindent \begin{tikzpicture}
\filldraw (0,0) -- (-0.15,0.08) -- (-0.15,-0.08) -- cycle ; 
\end{tikzpicture} \textbf{\underline{O8.}} Different types of questions require different levels of information, yet all existing graph-based RAG methods rely on fixed, predefined rules. How to design an adaptive mechanism that can address these varying needs remains an open question.

\noindent \begin{tikzpicture}
\filldraw (0,0) -- (-0.15,0.08) -- (-0.15,-0.08) -- cycle ; 
\end{tikzpicture} \textbf{\underline{O9.}}  Existing methods do not fully leverage the graph structure; they typically rely on simple graph patterns (e.g., nodes, edges, or $k$-hop paths). Although {\tt GraphRAG} adopts a hierarchical community structure (detecting by the Leiden algorithm), this approach does not consider node attributes, potentially compromising the quality of the communities. 
That is, determining which graph structures are superior remains an open question.

\noindent \begin{tikzpicture}
\filldraw (0,0) -- (-0.15,0.08) -- (-0.15,-0.08) -- cycle ; 
\end{tikzpicture} \textbf{\underline{O10.}}  The well-known graph database systems, such as Neo4j \cite{neo4j} and Nebula \cite{nebula}, support transferring the corpus into a knowledge
graph via LLM. However, enabling these popular systems to support the diverse operators required by various graph-based RAG
methods presents an exciting opportunity. 

\subsection{Benefit of our framework}

Our framework offers exceptional flexibility by enabling the combination of different methods at various stages. This modular design allows different algorithms to be seamlessly integrated, ensuring that each stage—such as {\it graph building}, and {\it retrieval\&generation}—can be independently optimized and recombined. For example, methods like {\tt HippoRAG}, which typically rely on KG, can easily be adapted to use {\tt RKG} instead, based on specific domain needs.

In addition, our operator design allows for simple modifications—often just a few lines of code—to create entirely new graph-based RAG methods. By adjusting the retrieval stage or swapping components, researchers can quickly test and implement new strategies, significantly accelerating the development cycle of graph-based RAG methods.

The modular nature of our framework is further reinforced by the use of retrieval elements (such as node, relationship, or subgraph) coupled with retrieval operators. This combination enables us to easily design new operators tailored to specific tasks. For example, by modifying the strategy for retrieving given elements, we can create customized operators that suit different application scenarios.

By systematically evaluating the effectiveness of various retrieval components under our unified framework, we can identify the most efficient combinations of graph construction, indexing, and retrieval strategies. This approach enables us to optimize retrieval performance across a range of use cases, allowing for both the enhancement of existing methods and the creation of novel, state-of-the-art techniques.

Finally, our framework contributes to the broader research community by providing a standardized methodology to assess graph-based RAG approaches. The introduction of a unified evaluation testbed ensures reproducibility, promotes fair a benchmark, and facilitates future innovations in RAG-based LLM applications.

\subsection{Limitations}

In our empirical study, we put considerable effort into evaluating the performance of existing graph-based RAG methods from various angles. 
However, our study still has some limitations, primarily due to resource constraints.
{\it (1) Token Length Limitation:} The primary experiments are conducted using Llama-3-8B with a token window size of 8k. This limitation on token length restricted the model’s ability to process longer input sequences, which could potentially impact the overall performance of the methods, particularly in tasks that require extensive context. 
Larger models with larger token windows could better capture long-range dependencies and deliver more robust results. This constraint is a significant factor that may affect the generalizability of our findings.
{\it (2) Limited Knowledge Datasets:} Our study did not include domain-specific knowledge datasets, which are crucial for certain applications. Incorporating such datasets could provide more nuanced insights and allow for a better evaluation of how these methods perform in specialized settings.
{\it (3) Resource Constraints:} Due to resource limitations, the largest model we utilized is Llama-3-70B, and the entire paper consumes nearly 10 billion tokens. Running larger models, such as GPT-4o (175B parameters or beyond), would incur significantly higher costs, potentially reaching several hundred thousand dollars depending on usage.
While we admit that introducing more powerful models could further enhance performance, the 70B model is already a strong choice, balancing performance and resource feasibility. 
That is to say, exploring the potential of even larger models in future work could offer valuable insights and further refine the findings.
{\it (4) Prompt Sensitivity:} The performance of each method is highly affected by its prompt design. Due to resource limitations, we did not conduct prompt ablation studies and instead used the available prompts from the respective papers. 
Actually, a fairer comparison would mitigate this impact by using prompt tuning tools, such as DSPy~\cite{khattab2023dspy}, to customize the prompts and optimize the performance of each method.

These limitations highlight areas for future exploration, and overcoming these constraints would enable a more thorough and reliable evaluation of graph-based RAG methods, strengthening the findings and advancing the research.
\begin{figure*}[] 
\begin{AIbox}{Prompt for generating abstract questions}
{\bf Prompt:} \\
{
Given the following description of a dataset:

\{description\}

Please identify 5 potential users who would engage with this dataset. For each user, list 5 tasks they would perform with this dataset. Then, for each (user, task) combination, generate 5 questions that require a high-level understanding of the entire dataset.

Output the results in the following structure:
\begin{description}
    \item[- User 1: \text{[user description]}]
    \begin{description}
        \item \item[- Task 1: \text{[task description]}]
        \begin{description}  
            \item \item  - Question 1:
            \item  - Question 2:
            \item  - Question 3:
            \item  - Question 4:
            \item  - Question 5:
        \end{description}  
       \item[- Task 2: \text{[task description]}]
            \item ...
       \item[- Task 5: \text{[task description]}]
    \end{description}   
\end{description}    

\begin{description}
    \item[- User 2: \text{[user description]}]
        \item...
\end{description}

\begin{description}
    \item[- User 5: \text{[user description]}]
        \item...
\end{description} 
    
    Note that there are 5 users and 5 tasks for each user, resulting in 25 tasks in total. Each task should have 5 questions, resulting in 125 questions in total.
    The Output should present the whole tasks and questions for each user.
    
    Output:

}


\end{AIbox} 
\caption{The prompt for generating abstract questions.}
\label{fig:prompt_summary}
\end{figure*}

\begin{figure*}[] 
\begin{AIbox}{Prompt for LLM-based multi-dimensional comparison}
{\bf Prompt:} \\
{
You will evaluate two answers to the same question based on three criteria: {\bf Comprehensiveness}, {\bf Diversity}, {\bf Empowerment}, and {\bf Directness}.
\begin{itemize}
    \item Comprehensiveness: How much detail does the answer provide to cover all aspects and details of the question?
    \item Diversity: How varied and rich is the answer in providing different perspectives and insights on the question?
    \item Empowerment: How well does the answer help the reader understand and make informed judgments about the topic?
    \item Directness: How specifically and clearly does the answer address the question?
\end{itemize}

For each criterion, choose the better answer (either Answer 1 or Answer 2) and explain why. Then, select an overall winner based on these four categories.

Here is the {\bf question}: 
\begin{verbatim}
    Question: {query}
\end{verbatim}

Here are the two answers:
\begin{verbatim}
    Answer 1: {answer1}
    Answer 2: {answer2}
\end{verbatim}

Evaluate both answers using the four criteria listed above and provide detailed explanations for each criterion. Output your evaluation in the following JSON format:
\begin{verbatim}
{
    "Comprehensiveness": {
        "Winner": "[Answer 1 or Answer 2]",
        "Explanation": "[Provide one sentence explanation here]"
    },
    "Diversity": {
        "Winner": "[Answer 1 or Answer 2]",
        "Explanation": "[Provide one sentence explanation here]"
    },
    "Empowerment": {
        "Winner": "[Answer 1 or Answer 2]",
        "Explanation": "[Provide one sentence explanation here]"
    },
    "Overall Winner": {
        "Winner": "[Answer 1 or Answer 2]",
        "Explanation": "[Briefly summarize why this answer is the overall winner]"
    }
}
\end{verbatim}

Output:

}

\end{AIbox} 
\caption{The prompt for the evaluation of abstract QA.}
\label{fig:eval_summary}
\end{figure*}

\newenvironment{shadequote}%
{\begin{snugshade}\begin{quote}}
{\hfill\end{quote}\end{snugshade}}
\definecolor{shadecolor}{rgb}{0.9,0.9,0.9}

\begin{figure*}[]
    \centering
    \begin{tcolorbox}[colback=gray!5, colframe=gray!50, title=An example abstract question]
\textit{Question:} How do the syntactic and semantic differences between functional and object-oriented paradigms affect program design?
    \end{tcolorbox}
\vspace{-0.8em}

\begin{tcolorbox}[title=Responses of different methods]

\textbf{Comprehensiveness:} \\
\noindent \begin{tikzpicture}
\filldraw (0,0) -- (-0.15,0.08) -- (-0.15,-0.08) -- cycle ; 
\end{tikzpicture} \textbf{{Why it matters:}} \textit{A comprehensive answer should mention key differences and explain their design implications.}

\noindent \begin{tikzpicture}
\filldraw (0,0) -- (-0.15,0.08) -- (-0.15,-0.08) -- cycle ; 
\end{tikzpicture} {\bf \textcolor{cyan}{Comprehensive Answer:}} 
Functional programming emphasizes \textit{\textcolor{cyan}{pure functions, immutability, and recursion}}, reducing side effects and improving testability. Object-oriented programming (OOP) uses \textit{\textcolor{cyan}{classes, encapsulation, and mutable state}}, supporting modularity and reuse. These paradigms shape program design in \textit{\textcolor{cyan}{syntax}} (e.g., \textit{\textcolor{cyan}{method chaining vs. composition}}) and \textit{\textcolor{cyan}{semantics}} (e.g., \textit{\textcolor{cyan}{stateful vs. stateless logic}}). For example, FP promotes \textit{\textcolor{cyan}{declarative flows}} and \textit{\textcolor{cyan}{better concurrency}}, while OOP mirrors \textit{\textcolor{cyan}{real-world models}}. \textit{\textcolor{cyan}{Understanding both helps in choosing suitable architecture patterns.}}

\noindent \begin{tikzpicture}
\filldraw (0,0) -- (-0.15,0.08) -- (-0.15,-0.08) -- cycle ; 
\end{tikzpicture}  {\bf \textcolor{gray!60!black}{Incomprehensive Answer:}} 
Functional programming focuses on using \textit{\textcolor{gray}{functions}}, and object-oriented programming uses \textit{\textcolor{gray}{objects and classes}}. These differences change how you write code, but \textit{\textcolor{gray}{both are just different ways to program}}. Most people use OOP because it’s more common. Functional programming is used sometimes, but not always necessary. \textit{\textcolor{gray}{It depends on what you’re doing}}, but overall, they just represent two styles.

\begin{tcolorbox}[colback=gray!10, colframe=gray!20, boxrule=0pt, sharp corners]
\noindent \begin{tikzpicture}
\filldraw (0,0) -- (-0.15,0.08) -- (-0.15,-0.08) -- cycle ; 
\end{tikzpicture}  \textbf{Explanation:} {\it The second response omits key distinctions like mutability and side effects, and lacks detail on how such differences influence program structure or architecture.}
\end{tcolorbox}

\vspace{0.5em}\hrule\vspace{0.5em}

\textbf{Diversity:} \\
\noindent \begin{tikzpicture}
\filldraw (0,0) -- (-0.15,0.08) -- (-0.15,-0.08) -- cycle ; 
\end{tikzpicture} \textbf{{Why it matters:}} \textit{A diverse answer should present multiple angles, use cases, and comparisons.}

\noindent \begin{tikzpicture}
\filldraw (0,0) -- (-0.15,0.08) -- (-0.15,-0.08) -- cycle ; 
\end{tikzpicture} {\bf \textcolor{cyan}{Diverse Answer:}} 
Functional and object-oriented paradigms differ across many dimensions. Syntactically, FP uses \textit{\textcolor{cyan}{higher-order functions}} and \textit{\textcolor{cyan}{expressions}}, while OOP relies on \textit{\textcolor{cyan}{method calls}} and \textit{\textcolor{cyan}{class hierarchies}}. Semantically, FP enforces \textit{\textcolor{cyan}{immutability}} and \textit{\textcolor{cyan}{stateless computation}}, whereas OOP supports \textit{\textcolor{cyan}{encapsulated state}} and side effects. These lead to varied design strategies: FP suits \textit{\textcolor{cyan}{parallel computation}}, \textit{\textcolor{cyan}{mathematical modeling}}, and \textit{\textcolor{cyan}{data transformation}}; OOP works well for \textit{\textcolor{cyan}{user interfaces}}, \textit{\textcolor{cyan}{simulations}}, and \textit{\textcolor{cyan}{domain modeling}}. \textit{\textcolor{cyan}{Hybrid designs}} are also common in modern software.

\noindent \begin{tikzpicture}
\filldraw (0,0) -- (-0.15,0.08) -- (-0.15,-0.08) -- cycle ; 
\end{tikzpicture} {\bf \textcolor{gray!60!black}{Low Diversity Answer:}} 
Functional programming is good for \textit{\textcolor{gray}{parallelism}} because it doesn’t use shared state. Object-oriented programming helps organize code using \textit{\textcolor{gray}{classes}}. In general, most developers stick to OOP because it’s easier and more practical. \textit{\textcolor{gray}{Functional programming is more theoretical}}. While they work differently, the main thing to know is that \textit{\textcolor{gray}{functional is more mathematical}}, and object-oriented is more practical for apps.

\begin{tcolorbox}[colback=gray!10, colframe=gray!20, boxrule=0pt, sharp corners]
\noindent \begin{tikzpicture}
\filldraw (0,0) -- (-0.15,0.08) -- (-0.15,-0.08) -- cycle ; 
\end{tikzpicture} \textbf{Explanation:} {\it The second answer focuses on only one or two differences and lacks variety in perspectives, missing concrete use cases and deeper design implications.}
\end{tcolorbox}

\vspace{0.5em}\hrule\vspace{0.5em}

\textbf{Empowerment}\\
\noindent \begin{tikzpicture}
\filldraw (0,0) -- (-0.15,0.08) -- (-0.15,-0.08) -- cycle ; 
\end{tikzpicture} \textbf{Why it matters:} {\it An empowering answer helps readers apply the concepts to make better design decisions.}

\noindent \begin{tikzpicture}
\filldraw (0,0) -- (-0.15,0.08) -- (-0.15,-0.08) -- cycle ; 
\end{tikzpicture} {\bf \textcolor{cyan}{Empowerment Answer:}} 
Understanding the core trade-offs between functional and object-oriented paradigms equips developers to design more robust systems. For instance, choosing \textit{\textcolor{cyan}{functional principles like immutability}} can minimize bugs in \textit{\textcolor{cyan}{concurrent applications}}. OOP’s modeling of \textit{\textcolor{cyan}{real-world entities}} simplifies maintenance in large systems. Awareness of these options enables \textit{\textcolor{cyan}{informed architectural decisions}}, such as using a \textit{\textcolor{cyan}{functional core within an object-oriented shell}}. This empowers developers to \textit{\textcolor{cyan}{select the right paradigm or blend}} based on project needs.

\noindent \begin{tikzpicture}
\filldraw (0,0) -- (-0.15,0.08) -- (-0.15,-0.08) -- cycle ; 
\end{tikzpicture} {\bf \textcolor{gray!60!black}{Low Empowerment Answer:}} 
Object-oriented programming is the \textit{\textcolor{gray}{standard approach}}, so you should usually stick to that. \textit{\textcolor{gray}{Functional programming is more academic}} and harder to understand. Unless you’re working on something very technical, like machine learning or math-heavy software, \textit{\textcolor{gray}{it’s not really useful}}. Most beginners don’t need to worry about it. \textit{\textcolor{gray}{Just learn OOP and you’ll be fine}} for most jobs or projects.

\begin{tcolorbox}[colback=gray!10, colframe=gray!20, boxrule=0pt, sharp corners]
\noindent \begin{tikzpicture}
\filldraw (0,0) -- (-0.15,0.08) -- (-0.15,-0.08) -- cycle ; 
\end{tikzpicture} \textbf{Explanation:} {\it The second answer discourages understanding by dismissing functional programming as impractical, instead of helping the reader see when and why each paradigm is appropriate.}
\end{tcolorbox}

\end{tcolorbox}
\vspace{-2em}
\caption{\change{Representative good and bad answers across three evaluation dimensions. \textcolor{cyan}{Cyan italics} highlight informative content in good answers; \textcolor{gray}{gray italics} indicate vague or unhelpful points in bad answers.}}

\label{fig:eval_dimensions_example}
\end{figure*}

\begin{figure}[]
    \centering
    \includegraphics[width=0.8\linewidth]{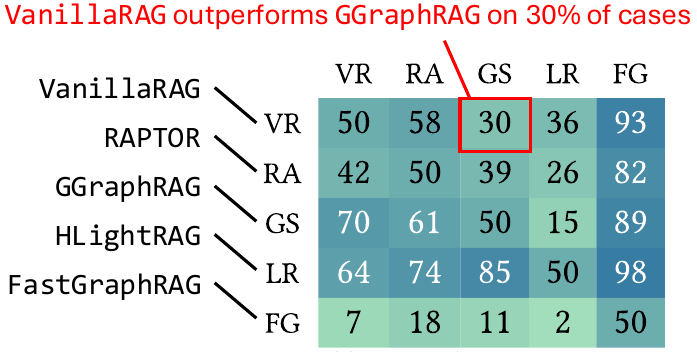}
    \caption{\change{An example head-to-head comparison result.}}
    \label{fig:enter-label}
\end{figure}


{
\subsection{Necessity of chunk splitting in RAG systems}

Splitting the input corpus into smaller chunks is a necessary step for all RAG methods, including both non-graph and graph-based variants.
The reasons can be summarized in three aspects below: 
\begin{enumerate}
    \item {\it Input length limit of LLMs.} Every LLM has its own input length limitation. For example, ChatGPT-3.5 supports up to 4,096 tokens, while Llama-3-8B allows up to 8,192 tokens. A token is a basic unit of text (roughly a word or subword). However, real-world corpora often contain tens of thousands to millions of tokens—e.g., the ALCE dataset exceeds 13 million tokens—far beyond the processing capability of any existing LLM. 
    Therefore, it is essential to split the corpus into smaller chunks.

     \item {\it Relevance filtering and noise reduction.}  
Even if the full corpus could fit into the LLM, chunking is still necessary because most queries only relate to a small part of the content. 
That is, given a query $Q$, inputting the entire corpus into the LLM introduces unnecessary noise. To alleviate this, RAG systems retrieve only the top-$k$ relevant chunks, which helps filter out irrelevant content and improves both accuracy and efficiency.

\item {\it Graph construction in graph-based RAG.} 
Chunking is equally essential and plays a critical role in all graph-based RAG methods. Below, we outline how different types of graphs rely on chunking:

\begin{itemize}
    \item {\it Knowledge graph (KG), Textual Knowledge graph (TKG), and Rich Knowledge graph (RKG):} These graphs use LLMs to extract entities and relationships from the input corpus. Due to the token limitation of the LLM, the corpus must first be split into chunks. The LLM is then applied independently to each chunk, and the resulting outputs are aggregated to construct the complete graph. This follows a divide-and-conquer paradigm.


\item  {\it Passage graph (PG):} In the passage graph, each chunk is treated as a node, and an edge is added between two nodes if their corresponding chunks share at least a certain number of common entities. The graph is constructed by comparing entity overlap across all pairs of chunks.
\end{itemize}
\end{enumerate} 

In summary, chunking is not only necessary due to token limits but also integral to the design of RAG systems.

}
~


Indeed, it is difficult to define a single ``best'' graph, as each type of graph has its own advantages and disadvantages.
To help clarify this, we summarize the key characteristics of each graph type in Table~\ref{tab:compare_graph}. 
For example, although tree can be constructed with minimal cost, it carries limited information. Methods such as {\tt RAPOTR}, which are based on tree, demonstrate competitive performance under lightweight settings. However, they may fall short in complex QA and abstract QA tasks due to limited semantic coverage.
In contrast, methods such as {\tt VGraphRAG} and {\tt GGraphRAG}, which are built upon TKG, benefit from richer entity and relationship representations. This enhanced expressiveness enables more accurate reasoning in information-intensive tasks. However, these advantages come at the cost of increased token consumption and longer construction time.
Additionally, while PG avoids token usage during construction, it incurs significant time overhead due to exhaustive pairwise chunk comparisons.

\begin{table}[]
\centering
\footnotesize
\caption{\textbf{\change{Comparison of different types of index.}}}
\begin{tabular}{lccc}
\toprule
\textbf{Index} 
& \makecell[c]{\textbf{Index}\\\textbf{Size}} 
& \makecell[c]{\textbf{Token}\\\textbf{Consuming}} 
& \makecell[c]{\textbf{Construction}\\\textbf{Time}} \\
\midrule
{\bf Node Index} & \ding{72}
\ding{72}
 & N/A & \ding{72}
 \\
{\bf Relationship Index}   & \ding{72}
\ding{72}
\ding{72}
 & N/A & \ding{72}
\ding{72}
 \\
{\bf Community Index}   & \ding{72}
 & \ding{72}
\ding{72}
\ding{72}
\ding{72}
 & \ding{72}
\ding{72}
\ding{72}
\ding{72}
 \\
\bottomrule
\end{tabular}
\label{tab:compare_index}
\end{table}

Similar to the comparison of graphs, it is also challenging to identify a universally ``best'' index.
Instead, we summarize the characteristics of different indices in Table~\ref{tab:compare_index}.
We can see that the {\it relationship index} tends to have a larger size, whereas the {\it community index} is more compact but incurs the highest construction cost in terms of tokens and time.
Indeed, in our experiments, we report the token costs associated with constructing the {\it community index} for each dataset.
Notably, for some datasets, such as HotpotQA and ALCE, this cost is comparable to that of building the graph itself, exceeding $10^8$ tokens.
Despite their differences, all indices share a common goal of facilitating fast and effective retrieval within graph-based RAG systems.
The selection of a specific index should align with the retrieval strategy and the demands of the downstream QA task.
For instance, to answer abstract questions that rely on some high-level summaries, it is better to use the community-based indexes (e.g., the index of {\tt GraphRAG}) to retrieve relevant information since they often provide the summary reports for the communities.


\subsection{Relevance to the data management community}
We agree that graph-based RAG is very relevant to the information retrieval and NLP communities, but we also believe that it is highly relevant to the graph data management community, since it has received growing interest from both industry and academia.
In the industrial area, Ant-Group recently proposed Chat2Graph \cite{chat2graph}, a graph-native agent system that incorporates graph-based RAG as a core component.
Similarly, modern graph database systems such as Neo4j \cite{neo4jrag}, NebulaGraph \cite{nebula}, and  PostgreSQL \cite{postgresrag} are beginning to support graph-based RAG methods, further reinforcing the growing intersection between AI and graph data management.
In the academia area, interest in RAG is also rising within the data management field, as demonstrated by recent work such as {\tt Chameleon} \cite{jiang2024chameleon} (VLDB 2025) and {\tt Cache-Craft} \cite{agarwal2025cache} (SIGMOD 2025).

The core retrieval stage in graph-based RAG systems can be naturally framed as a query optimization process over graph databases, where the goal is to identify the most relevant subgraphs, paths, nodes, or relationships from a large graph $G$ given a natural language query $Q$.
This introduces new challenges for graph query processing and indexing, as many modern graph databases lack native support for the complex retrieval operators required by graph-based RAG workflows.
For instance, while retrieving $k$-hop neighbors is commonly supported, operators such as efficiently identifying communities containing specific entities, computing Steiner trees, or finding $k$-hop semantically relevant paths to the query $Q$  remain largely unsupported.
Moreover, retrieval operators that rely on LLM-based reasoning or semantic similarity are fundamentally beyond the capabilities of existing graph data systems.
Our work highlights these limitations and identifies opportunities for extending graph databases to such operators over large graph data—an emerging direction highly relevant to the graph data management community.

Lastly, our work has garnered significant attention from both academic and industrial areas.
Our open-source project has attracted significant attention (e.g., 1.4k+ GitHub stars \footnote{\url{https://github.com/JayLZhou/GraphRAG}}) and has been successfully adopted by {\it Huawei Cloud for domain-specific QA}, thanks to its modular architecture and ability to handle complex retrieval workflows efficiently.
In summary, as noted in the meta-review, we believe that with the increasing importance of graphs and the growing adoption of AI techniques such as RAG, it is both timely and relevant to the graph data management community.

\begin{figure*}[h]
    \centering
    \includegraphics[width=1.0\linewidth]{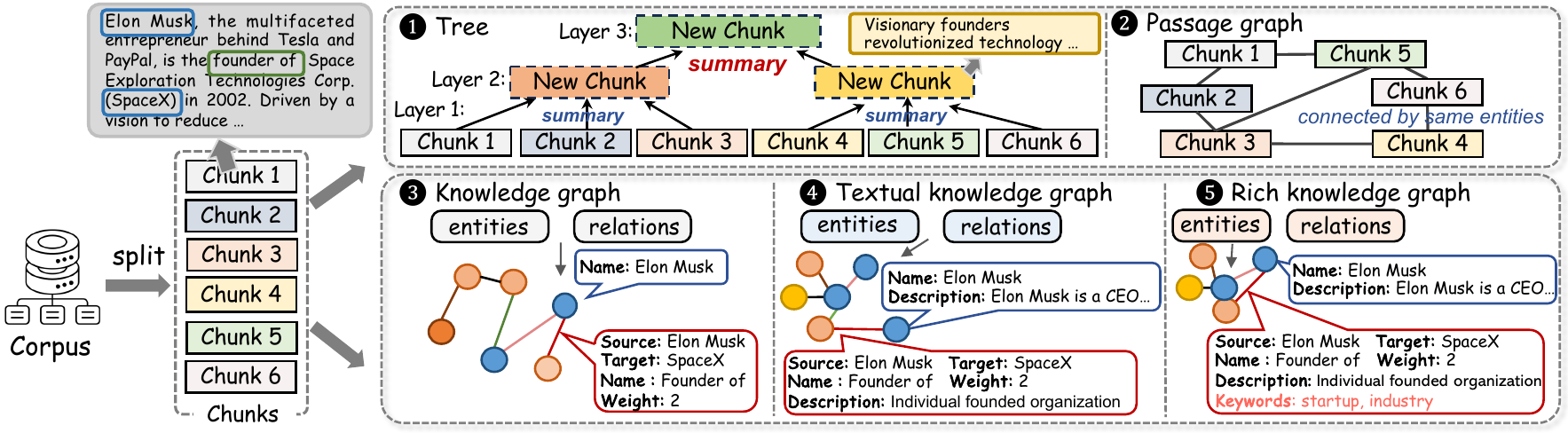}
    \caption{\change{Examples of five types of graphs.}}
    \label{fig:app:graph_example}
\end{figure*}

\begin{figure*}
    \centering
    \includegraphics[width=\linewidth]{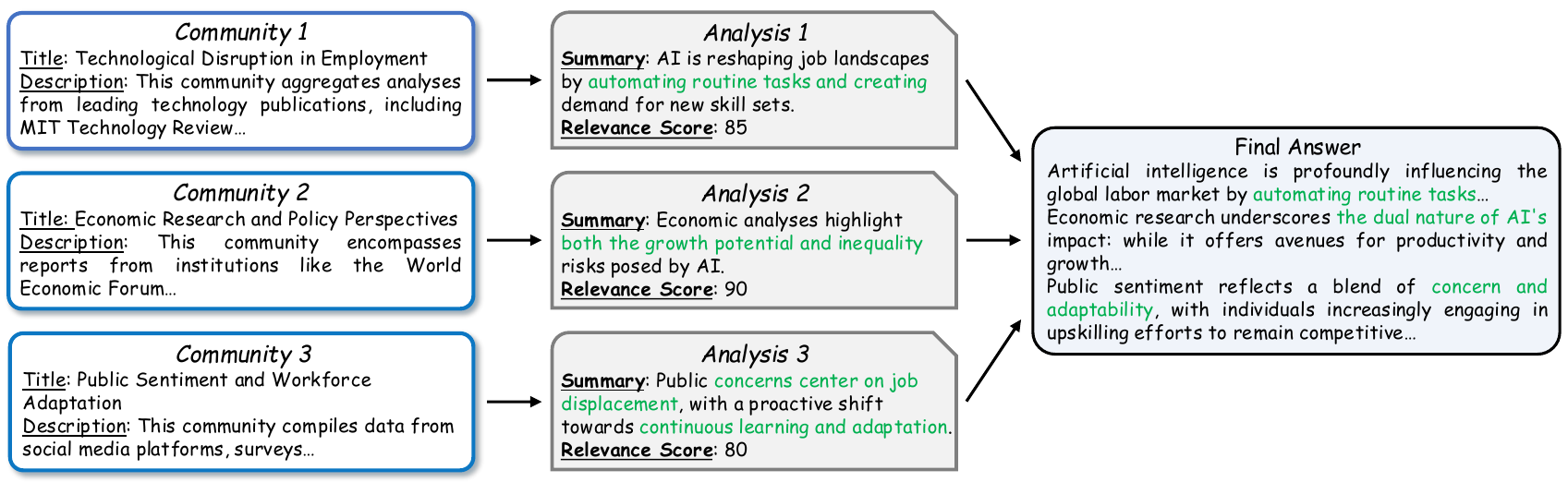}
    \caption{\change{An illustrative example of {\tt GGraphRAG}'s {\tt Map-Reduce} generation process. The analysis presents partial answers derived from each community, with the portions incorporated into the final answer highlighted in green.}}
    \label{fig:exp-mr}
\end{figure*}

{

\subsection{More examples}

$\bullet$ \textbf{An example of building graph.} As shown in Figure~\ref{fig:app:graph_example}, we consider a corpus from the IT domain, which is segmented into six chunks. We then illustrate how to construct different types of graphs based on this corpus:
\begin{itemize}
    \item {\it Tree:} We first apply the Gaussian Mixture clustering algorithm to group the six chunks into two clusters: (chunk 1, chunk 2, chunk 3) and (chunk 4, chunk 5, chunk 6). Each cluster forms a parent node in the second layer, where an LLM is used to summarize the content of all associated chunks. These two parent nodes are then clustered into a single root node (third layer), whose content is also summarized via LLM. The resulting structure is a three-layer tree with 6 leaf nodes (original chunks), 2 intermediate cluster nodes, and 1 root node.

    \item {\it Passage Graph:} We construct a passage graph where each chunk is treated as a node, and an edge is added between two nodes if their corresponding chunks share at least $\tau$ common entities.
For example, an edge is added between chunk 4 and chunk 6 since they both contain the same entities, e.g., ``ChatGPT'' and ``OpenAI''.

    \item {\it Knowledge Graph:} We use an LLM to extract entity–relation triples from each chunk. Each entity includes a name, while each relationship is represented by a name and an associated weight indicating how frequently it appears within the given chunk.    %
    For example, from chunk 1, we extract the entity ``Elon Musk'' and its relationship ``Founder of''.

    \item {\it Textural \& Rich Knowledge Graph:} The two types of graphs are constructed in a similar way: for each chunk, we use an LLM to extract entities and their relationships.
Each entity is represented with three attributes—name, type, and description—while each relationship includes a name, description, and weight.
The key difference lies in that the Rich Knowledge Graph further annotates each relationship with a set of keywords, providing more semantic cues for retrieval.
\end{itemize}

$\bullet$ \textbf{Details of operators.}
Specifically, we abstract three operators under the {\bf Chunk type}:  {\Large \ding{182}} {\texttt{Aggregator}},  {\Large \ding{183}} {\texttt{FromRel}}, and {\Large \ding{184}}  {\texttt{Occurrence}}.
\begin{itemize}
    \item   {\texttt{Aggregator}}: This operator relies on two matrices: a score vector $\mathcal{R} \in \mathbb{R}^{1 \times m}$ and an interaction matrix $\mathcal{M} \in \mathbb{R}^{m \times c}$, where $m$ is the number of relationships and $c$ is the number of chunks. Specifically, the $i$-th entry in $\mathcal{R}$ represents the score of the $i$-th relationship, while $\mathcal{M}_{i,j} = 1$ if the $i$-th relationship is extracted from the $j$-th chunk, and $0$ otherwise.
Afterwards,  the aggregated score of each chunk is computed via matrix multiplication:
$$
\Psi = \mathcal{R} \times \mathcal{M},
$$
where each entry in $\Psi \in \mathbb{R}^{1 \times c}$ represents the aggregated relationship score of a chunk. Based on $\Psi$, the top-$k$ chunks with the highest scores are selected for retrieval.

    \item  \texttt{FromRel}: For each relationship in the graph, we maintain a mapping to the set of chunks from which it is extracted.
    This allows us to efficiently retrieve relevant chunks given a set of relationships by computing the union of all chunks associated with these relationships.

     \item  \texttt{Occurrence}: 
     For every relationship, we identify its two associated entities. For each entity, we maintain a mapping to the set of chunks from which it is extracted, since an entity can appear in multiple chunks.
If a chunk contains both entities of a given relationship, its score (initially set to 0) is incremented by 1. After processing all relationships, we obtain a score for each chunk, and select the top-$k$ chunks based on these scores.
\end{itemize}

Besides, there are two operators under the \textbf{Community type}: {\Large \ding{182}} {\texttt{Entity}},  and {\Large \ding{183}} {\texttt Layer}:

\begin{itemize}
    \item \texttt{Entity}: This operator retrieves communities that contain the specified entities. Each community maintains a list of associated entities. Retrieved communities are then ranked based on their relevance scores (generated by the LLM), and the top-$k$ communities are returned.

    \item \texttt{Layer}: In {\tt GGraphRAG}, communities are detected using the Leiden algorithm, resulting in a hierarchical structure where higher layers represent more abstract and coarse-grained information. The \texttt{Layer} operator retrieves all communities at or below a specified layer, allowing access to more fine-grained community information.
\end{itemize}

$\bullet$ \textbf{An example of  {\tt Map-Reduce} strategy.}
In the {\tt Map-Reduce} phase of {\tt GGraphRAG}, each retrieved community is individually used to answer the question.
Specifically, for each community, the LLM is prompted to generate a partial answer along with a confidence score (ranging from 0 to 100), reflecting how well the community summary addresses the query.
After processing all communities, we obtain a set of partial answers, each represented as a pair: (answer, score).
These partial answers are then ranked in descending order by score and sequentially appended to the prompt for final answer generation.
We also provide an example to illustrate this.

\begin{example}
    Figure~\ref{fig:exp-mr} illustrates how the partial answers are generated and used in {\tt GGraphRAG}.
    Consider an abstract question: {\it ``What are the socio-economic impacts of artificial intelligence on the global labor market?''}
     In this example, three communities are retrieved, and {\tt GGraphRAG} generates partial answers from each one.
    For instance, using the first community, it produces a partial answer discussing the impact of AI on the job landscape, which receives a relevance score of 85.
    These partial answers are then aggregated, allowing {\tt GGraphRAG} to generate a final, comprehensive response.
\end{example}



%% file: ref.bib
@article{asai2023self,
  title={Self-rag: Learning to retrieve, generate, and critique through self-reflection},
  author={Asai, Akari and Wu, Zeqiu and Wang, Yizhong and Sil, Avirup and Hajishirzi, Hannaneh},
  journal={arXiv preprint arXiv:2310.11511},
  year={2023}
}

@article{li2023gslb,
  title={GSLB: the graph structure learning benchmark},
  author={Li, Zhixun and Wang, Liang and Sun, Xin and Luo, Yifan and Zhu, Yanqiao and Chen, Dingshuo and Luo, Yingtao and Zhou, Xiangxin and Liu, Qiang and Wu, Shu and others},
  journal={Advances in Neural Information Processing Systems},
  volume={36},
  pages={30306--30318},
  year={2023}
}

@article{zhang2025erarag,
  title={EraRAG: Efficient and Incremental Retrieval Augmented Generation for Growing Corpora},
  author={Zhang, Fangyuan and Huang, Zhengjun and Zhou, Yingli and Guo, Qintian and Li, Zhixun and Luo, Wensheng and Jiang, Di and Fang, Yixiang and Zhou, Xiaofang},
  journal={arXiv preprint arXiv:2506.20963},
  year={2025}
}

@article{xiang2025use,
  title={When to use graphs in rag: A comprehensive analysis for graph retrieval-augmented generation},
  author={Xiang, Zhishang and Wu, Chuanjie and Zhang, Qinggang and Chen, Shengyuan and Hong, Zijin and Huang, Xiao and Su, Jinsong},
  journal={arXiv preprint arXiv:2506.05690},
  year={2025}
}

@article{zhuang2025linearrag,
  title={LinearRAG: Linear Graph Retrieval Augmented Generation on Large-scale Corpora},
  author={Zhuang, Luyao and Chen, Shengyuan and Xiao, Yilin and Zhou, Huachi and Zhang, Yujing and Chen, Hao and Zhang, Qinggang and Huang, Xiao},
  journal={arXiv preprint arXiv:2510.10114},
  year={2025}
}

@article{chen2025you,
  title={You Don't Need Pre-built Graphs for RAG: Retrieval Augmented Generation with Adaptive Reasoning Structures},
  author={Chen, Shengyuan and Zhou, Chuang and Yuan, Zheng and Zhang, Qinggang and Cui, Zeyang and Chen, Hao and Xiao, Yilin and Cao, Jiannong and Huang, Xiao},
  journal={arXiv preprint arXiv:2508.06105},
  year={2025}
}

@article{li2023survey,
  title={A survey of graph meets large language model: Progress and future directions},
  author={Li, Yuhan and Li, Zhixun and Wang, Peisong and Li, Jia and Sun, Xiangguo and Cheng, Hong and Yu, Jeffrey Xu},
  journal={arXiv preprint arXiv:2311.12399},
  year={2023}
}

@article{edge2024local,
  title={From local to global: A graph rag approach to query-focused summarization},
  author={Edge, Darren and Trinh, Ha and Cheng, Newman and Bradley, Joshua and Chao, Alex and Mody, Apurva and Truitt, Steven and Larson, Jonathan},
  journal={arXiv preprint arXiv:2404.16130},
  year={2024}
}

@article{zhang2025survey,
  title={A Survey of Graph Retrieval-Augmented Generation for Customized Large Language Models},
  author={Zhang, Qinggang and Chen, Shengyuan and Bei, Yuanchen and Yuan, Zheng and Zhou, Huachi and Hong, Zijin and Dong, Junnan and Chen, Hao and Chang, Yi and Huang, Xiao},
  journal={arXiv preprint arXiv:2501.13958},
  year={2025}
}

@inproceedings{kamra2024enhancing,
  title={Enhancing Document Retrieval Using AI and Graph-Based RAG Techniques},
  author={Kamra, Vikas and Gupta, Lakshya and Arora, Dhruv and Yadav, Ashwin Kumar},
  booktitle={2024 5th International Conference on Communication, Computing \& Industry 6.0 (C2I6)},
  pages={1--7},
  year={2024},
  organization={IEEE}
}

@article{han2025rag,
  title={Rag vs. graphrag: A systematic evaluation and key insights},
  author={Han, Haoyu and Shomer, Harry and Wang, Yu and Lei, Yongjia and Guo, Kai and Hua, Zhigang and Long, Bo and Liu, Hui and Tang, Jiliang},
  journal={arXiv preprint arXiv:2502.11371},
  year={2025}
}

@article{brown2020language,
  title={Language models are few-shot learners},
  author={Brown, Tom and Mann, Benjamin and Ryder, Nick and Subbiah, Melanie and Kaplan, Jared D and Dhariwal, Prafulla and Neelakantan, Arvind and Shyam, Pranav and Sastry, Girish and Askell, Amanda and others},
  journal={Advances in neural information processing systems},
  volume={33},
  pages={1877--1901},
  year={2020}
}

@article{lewis2020retrieval,
  title={Retrieval-augmented generation for knowledge-intensive nlp tasks},
  author={Lewis, Patrick and Perez, Ethan and Piktus, Aleksandra and Petroni, Fabio and Karpukhin, Vladimir and Goyal, Naman and K{\"u}ttler, Heinrich and Lewis, Mike and Yih, Wen-tau and Rockt{\"a}schel, Tim and others},
  journal={Advances in neural information processing systems},
  volume={33},
  pages={9459--9474},
  year={2020}
}

@article{yang2018hotpotqa,
  title={HotpotQA: A dataset for diverse, explainable multi-hop question answering},
  author={Yang, Zhilin and Qi, Peng and Zhang, Saizheng and Bengio, Yoshua and Cohen, William W and Salakhutdinov, Ruslan and Manning, Christopher D},
  journal={arXiv preprint arXiv:1809.09600},
  year={2018}
}

@article{peng2024graph,
  title={Graph retrieval-augmented generation: A survey},
  author={Peng, Boci and Zhu, Yun and Liu, Yongchao and Bo, Xiaohe and Shi, Haizhou and Hong, Chuntao and Zhang, Yan and Tang, Siliang},
  journal={arXiv preprint arXiv:2408.08921},
  year={2024}
}

@article{yang2025qwen3,
  title={Qwen3 technical report},
  author={Yang, An and Li, Anfeng and Yang, Baosong and Zhang, Beichen and Hui, Binyuan and Zheng, Bo and Yu, Bowen and Gao, Chang and Huang, Chengen and Lv, Chenxu and others},
  journal={arXiv preprint arXiv:2505.09388},
  year={2025}
}

@article{agarwal2025cache,
  title={Cache-Craft: Managing Chunk-Caches for Efficient Retrieval-Augmented Generation},
  author={Agarwal, Shubham and Sundaresan, Sai and Mitra, Subrata and Mahapatra, Debabrata and Gupta, Archit and Sharma, Rounak and Kapu, Nirmal Joshua and Yu, Tong and Saini, Shiv},
  journal={arXiv preprint arXiv:2502.15734},
  year={2025}
}

@article{jiang2024chameleon,
  title={Chameleon: A Heterogeneous and Disaggregated Accelerator System for Retrieval-Augmented Language Models},
  author={Jiang, Wenqi and Zeller, Marco and Waleffe, Roger and Hoefler, Torsten and Alonso, Gustavo},
  journal={Proceedings of the VLDB Endowment},
  volume={18},
  number={1},
  pages={42--52},
  year={2024},
  publisher={VLDB Endowment}
}

@misc{chat2graph,
  author = {Chat2Graph},
  title = {About Chat2Graph: Graph Native Agentic System.},
  howpublished = {\url{https://chat2graph.vercel.app/chat2graph/en-us/introduction}},
  year = 2025,
}

@misc{LangGraph,
  author = {LangGraph},
  title = {LangGraph},
  howpublished = {\url{https://www.langchain.com/langgraph}},
  year = 2024,
}

@misc{llamaindex,
  author = {llamaindex},
  title = {llamaindex},
  howpublished = {\url{https://www.llamaindex.ai/}},
  year = 2023,
}

@misc{neo4j,
  author = {neo4j},
  title = {neo4j},
  howpublished = {\url{https://neo4j.com/}},
  year = 2006,
}

@misc{neo4jrag,
  author = {neo4j-rag},
  title = {Neo4j GraphRAG},
  howpublished = {\url{https://neo4j.com/docs/neo4j-graphrag-python/current/}},
  year = 2024,
}

@misc{databricksrag,
  author = {Databricks},
  title = {Building, Improving, and Deploying Knowledge Graph RAG Systems on Databricks},
  howpublished = {\url{https://neo4j.com/docs/neo4j-graphrag-python/current/}},
  year = 2024,
}

@misc{postgresrag,
  author = {PostgreSQL},
  title = {Azure Database for PostgreSQL Blog
},
  howpublished = {\url{https://techcommunity.microsoft.com/blog/adforpostgresql/introducing-the-graphrag-solution-for-azure-database-for-postgresql/4299871}},
  year = 2024,
}

@misc{nebula,
  author = {nebula},
  title = {nebula},
  howpublished = {\url{https://www.databricks.com/blog/building-improving-and-deploying-knowledge-graph-rag-systems-databricks}},
  year = 2024,
}

@misc{huawei,
    author={huawei},
    title = {Ascend GPU},
    howpublished = {\url{https://e.huawei.com/ph/products/computing/ascend}},
    year = 2019,
}

@inproceedings{wang2024knowledge,
  title={Knowledge graph prompting for multi-document question answering},
  author={Wang, Yu and Lipka, Nedim and Rossi, Ryan A and Siu, Alexa and Zhang, Ruiyi and Derr, Tyler},
  booktitle={Proceedings of the AAAI Conference on Artificial Intelligence},
  volume={38},
  number={17},
  pages={19206--19214},
  year={2024}
}

@inproceedings{sunthink,
  title={Think-on-Graph: Deep and Responsible Reasoning of Large Language Model on Knowledge Graph},
  author={Sun, Jiashuo and Xu, Chengjin and Tang, Lumingyuan and Wang, Saizhuo and Lin, Chen and Gong, Yeyun and Ni, Lionel and Shum, Heung-Yeung and Guo, Jian},
  booktitle={The Twelfth International Conference on Learning Representations},
    year = {2024}
}

@article{jeong2024adaptive,
  title={Adaptive-rag: Learning to adapt retrieval-augmented large language models through question complexity},
  author={Jeong, Soyeong and Baek, Jinheon and Cho, Sukmin and Hwang, Sung Ju and Park, Jong C},
  journal={arXiv preprint arXiv:2403.14403},
  year={2024}
}

@article{siriwardhana2023improving,
  title={Improving the domain adaptation of retrieval augmented generation (RAG) models for open domain question answering},
  author={Siriwardhana, Shamane and Weerasekera, Rivindu and Wen, Elliott and Kaluarachchi, Tharindu and Rana, Rajib and Nanayakkara, Suranga},
  journal={Transactions of the Association for Computational Linguistics},
  volume={11},
  pages={1--17},
  year={2023},
  publisher={MIT Press One Broadway, 12th Floor, Cambridge, Massachusetts 02142, USA~…}
}

@article{chen2024auto,
  title={Auto-Formula: Recommend Formulas in Spreadsheets using Contrastive Learning for Table Representations},
  author={Chen, Sibei and He, Yeye and Cui, Weiwei and Fan, Ju and Ge, Song and Zhang, Haidong and Zhang, Dongmei and Chaudhuri, Surajit},
  journal={Proceedings of the ACM on Management of Data},
  volume={2},
  number={3},
  pages={1--27},
  year={2024},
  publisher={ACM New York, NY, USA}
}

@article{chen2023haipipe,
  title={Haipipe: Combining human-generated and machine-generated pipelines for data preparation},
  author={Chen, Sibei and Tang, Nan and Fan, Ju and Yan, Xuemi and Chai, Chengliang and Li, Guoliang and Du, Xiaoyong},
  journal={Proceedings of the ACM on Management of Data},
  volume={1},
  number={1},
  pages={1--26},
  year={2023},
  publisher={ACM New York, NY, USA}
}

@article{song2025quite,
  title={QUITE: A Query Rewrite System Beyond Rules with LLM Agents},
  author={Song, Yuyang and Yan, Hanxu and Lao, Jiale and Wang, Yibo and Li, Yufei and Zhou, Yuanchun and Wang, Jianguo and Tang, Mingjie},
  journal={arXiv preprint arXiv:2506.07675},
  year={2025}
}

@article{lillm,
  title={LLM-R2: A Large Language Model Enhanced Rule-based Rewrite System for Boosting Query  Efficiency},
  author={Li, Zhaodonghui and Yuan, Haitao and Wang, Huiming and Cong, Gao and Bing, Lidong},
  journal={Proceedings of the VLDB Endowment},
  volume={1},
  number={18},
  pages={53--65},
  year={2025},
  publisher={VLDB Endowment}
}

@article{sun2024r,
  title={R-Bot: An LLM-based Query Rewrite System},
  author={Sun, Zhaoyan and Zhou, Xuanhe and Li, Guoliang},
  journal={arXiv preprint arXiv:2412.01661},
  year={2024}
}

@article{naeem2024retclean,
  title={RetClean: Retrieval-Based Data Cleaning Using LLMs and Data Lakes},
  author={Naeem, Zan Ahmad and Ahmad, Mohammad Shahmeer and Eltabakh, Mohamed and Ouzzani, Mourad and Tang, Nan},
  journal={Proceedings of the VLDB Endowment},
  volume={17},
  number={12},
  pages={4421--4424},
  year={2024},
  publisher={VLDB Endowment}
}

@article{qian2024unidm,
  title={UniDM: A Unified Framework for Data Manipulation with Large Language Models},
  author={Qian, Yichen and He, Yongyi and Zhu, Rong and Huang, Jintao and Ma, Zhijian and Wang, Haibin and Wang, Yaohua and Sun, Xiuyu and Lian, Defu and Ding, Bolin and others},
  journal={Proceedings of Machine Learning and Systems},
  volume={6},
  pages={465--482},
  year={2024}
}

@article{zhou2023d,
  title={D-bot: Database diagnosis system using large language models},
  author={Zhou, Xuanhe and Li, Guoliang and Sun, Zhaoyan and Liu, Zhiyuan and Chen, Weize and Wu, Jianming and Liu, Jiesi and Feng, Ruohang and Zeng, Guoyang},
  journal={Proceedings of the VLDB Endowment},
  volume={17},
  number={10},
  pages={2514--2527},
  year={2024},
  publisher={VLDB Endowment}
}

@article{singh2024panda,
  title={Panda: Performance debugging for databases using LLM agents},
  author={Singh, Vikramank and Vaidya, Kapil Eknath and Kumar, Vinayshekhar Bannihatti and Khosla, Sopan and Narayanaswamy, Murali and Gangadharaiah, Rashmi and Kraska, Tim},
  year={2024}
}

@article{lao2023gptuner,
  title={Gptuner: A manual-reading database tuning system via gpt-guided bayesian optimization},
  author={Lao, Jiale and Wang, Yibo and Li, Yufei and Wang, Jianping and Zhang, Yunjia and Cheng, Zhiyuan and Chen, Wanghu and Tang, Mingjie and Wang, Jianguo},
  journal={Proceedings of the VLDB Endowment},
  volume={17},
  number={8},
  pages={1939--1952},
  year={2024},
  publisher={VLDB Endowment}
}

@article{fan2024combining,
  title={Combining small language models and large language models for zero-shot nl2sql},
  author={Fan, Ju and Gu, Zihui and Zhang, Songyue and Zhang, Yuxin and Chen, Zui and Cao, Lei and Li, Guoliang and Madden, Samuel and Du, Xiaoyong and Tang, Nan},
  journal={Proceedings of the VLDB Endowment},
  volume={17},
  number={11},
  pages={2750--2763},
  year={2024},
  publisher={VLDB Endowment}
}

@article{zhou2024db,
  title={Db-gpt: Large language model meets database},
  author={Zhou, Xuanhe and Sun, Zhaoyan and Li, Guoliang},
  journal={Data Science and Engineering},
  volume={9},
  number={1},
  pages={102--111},
  year={2024},
  publisher={Springer}
}

@article{dong2022survey,
  title={A survey on in-context learning},
  author={Dong, Qingxiu and Li, Lei and Dai, Damai and Zheng, Ce and Ma, Jingyuan and Li, Rui and Xia, Heming and Xu, Jingjing and Wu, Zhiyong and Liu, Tianyu and others},
  journal={arXiv preprint arXiv:2301.00234},
  year={2022}
}

@article{ouyang2022training,
  title={Training language models to follow instructions with human feedback},
  author={Ouyang, Long and Wu, Jeffrey and Jiang, Xu and Almeida, Diogo and Wainwright, Carroll and Mishkin, Pamela and Zhang, Chong and Agarwal, Sandhini and Slama, Katarina and Ray, Alex and others},
  journal={Advances in neural information processing systems},
  volume={35},
  pages={27730--27744},
  year={2022}
}

@article{achiam2023gpt,
  title={Gpt-4 technical report},
  author={Achiam, Josh and Adler, Steven and Agarwal, Sandhini and Ahmad, Lama and Akkaya, Ilge and Aleman, Florencia Leoni and Almeida, Diogo and Altenschmidt, Janko and Altman, Sam and Anadkat, Shyamal and others},
  journal={arXiv preprint arXiv:2303.08774},
  year={2023}
}

@article{he2024g,
  title={G-retriever: Retrieval-augmented generation for textual graph understanding and question answering},
  author={He, Xiaoxin and Tian, Yijun and Sun, Yifei and Chawla, Nitesh V and Laurent, Thomas and LeCun, Yann and Bresson, Xavier and Hooi, Bryan},
  journal={arXiv preprint arXiv:2402.07630},
  year={2024}
}

@inproceedings{haveliwala2002topic,
  title={Topic-sensitive pagerank},
  author={Haveliwala, Taher H},
  booktitle={Proceedings of the 11th international conference on World Wide Web},
  pages={517--526},
  year={2002}
}

@article{li2024dalk,
  title={DALK: Dynamic Co-Augmentation of LLMs and KG to answer Alzheimer's Disease Questions with Scientific Literature},
  author={Li, Dawei and Yang, Shu and Tan, Zhen and Baik, Jae Young and Yun, Sukwon and Lee, Joseph and Chacko, Aaron and Hou, Bojian and Duong-Tran, Duy and Ding, Ying and others},
  journal={arXiv preprint arXiv:2405.04819},
  year={2024}
}

@article{guo2024lightrag,
  title={LightRAG: Simple and Fast Retrieval-Augmented Generation},
  author={Guo, Zirui and Xia, Lianghao and Yu, Yanhua and Ao, Tu and Huang, Chao},
  journal={arXiv e-prints},
  pages={arXiv--2410},
  year={2024}
}

@article{sarthi2024raptor,
  title={Raptor: Recursive abstractive processing for tree-organized retrieval},
  author={Sarthi, Parth and Abdullah, Salman and Tuli, Aditi and Khanna, Shubh and Goldie, Anna and Manning, Christopher D},
  journal={arXiv preprint arXiv:2401.18059},
  year={2024}
}

@article{gutierrez2024hipporag,
  title={HippoRAG: Neurobiologically Inspired Long-Term Memory for Large Language Models},
  author={Guti{\'e}rrez, Bernal Jim{\'e}nez and Shu, Yiheng and Gu, Yu and Yasunaga, Michihiro and Su, Yu},
  journal={arXiv preprint arXiv:2405.14831},
  year={2024}
}

@misc{fastgraphrag,
    Author =  {FastGraphRAG},
    Title = {FastGraphRAG},
  year={2024},
    Howpublished = {\url{https://github.com/circlemind-ai/fast-graphrag}},
}

@article{multim3,
  title={M3-Embedding: Multi-Linguality, Multi-Functionality, Multi-Granularity Text Embeddings Through Self-Knowledge Distillation},
  author={Multi-Granularity, Multi-Linguality Multi-Functionality},
    year={2024}
}

@article{wang2023chatgpt,
  title={Is chatgpt a good nlg evaluator? a preliminary study},
  author={Wang, Jiaan and Liang, Yunlong and Meng, Fandong and Sun, Zengkui and Shi, Haoxiang and Li, Zhixu and Xu, Jinan and Qu, Jianfeng and Zhou, Jie},
  journal={arXiv preprint arXiv:2303.04048},
  year={2023}
}

@article{zheng2023judging,
  title={Judging llm-as-a-judge with mt-bench and chatbot arena},
  author={Zheng, Lianmin and Chiang, Wei-Lin and Sheng, Ying and Zhuang, Siyuan and Wu, Zhanghao and Zhuang, Yonghao and Lin, Zi and Li, Zhuohan and Li, Dacheng and Xing, Eric and others},
  journal={Advances in Neural Information Processing Systems},
  volume={36},
  pages={46595--46623},
  year={2023}
}

@article{schick2024toolformer,
  title={Toolformer: Language models can teach themselves to use tools},
  author={Schick, Timo and Dwivedi-Yu, Jane and Dess{\`\i}, Roberto and Raileanu, Roberta and Lomeli, Maria and Hambro, Eric and Zettlemoyer, Luke and Cancedda, Nicola and Scialom, Thomas},
  journal={Advances in Neural Information Processing Systems},
  volume={36},
  year={2024}
}

@article{mallen2022not,
  title={When not to trust language models: Investigating effectiveness of parametric and non-parametric memories},
  author={Mallen, Alex and Asai, Akari and Zhong, Victor and Das, Rajarshi and Khashabi, Daniel and Hajishirzi, Hannaneh},
  journal={arXiv preprint arXiv:2212.10511},
  year={2022}
}

@book{han2022data,
  title={Data mining: concepts and techniques},
  author={Han, Jiawei and Pei, Jian and Tong, Hanghang},
  year={2022},
  publisher={Morgan kaufmann}
}

@article{huang2023survey,
  title={A survey on hallucination in large language models: Principles, taxonomy, challenges, and open questions},
  author={Huang, Lei and Yu, Weijiang and Ma, Weitao and Zhong, Weihong and Feng, Zhangyin and Wang, Haotian and Chen, Qianglong and Peng, Weihua and Feng, Xiaocheng and Qin, Bing and others},
  journal={arXiv preprint arXiv:2311.05232},
  year={2023}
}

@article{gao2023retrieval,
  title={Retrieval-augmented generation for large language models: A survey},
  author={Gao, Yunfan and Xiong, Yun and Gao, Xinyu and Jia, Kangxiang and Pan, Jinliu and Bi, Yuxi and Dai, Yi and Sun, Jiawei and Wang, Haofen},
  journal={arXiv preprint arXiv:2312.10997},
  year={2023}
}

@article{devlin2018bert,
  title={Bert: Pre-training of deep bidirectional transformers for language understanding},
  author={Devlin, Jacob},
  journal={arXiv preprint arXiv:1810.04805},
  year={2018}
}

@inproceedings{khattab2020colbert,
  title={Colbert: Efficient and effective passage search via contextualized late interaction over bert},
  author={Khattab, Omar and Zaharia, Matei},
  booktitle={Proceedings of the 43rd International ACM SIGIR conference on research and development in Information Retrieval},
  pages={39--48},
  year={2020}
}

@article{traag2019louvain,
  title={From Louvain to Leiden: guaranteeing well-connected communities},
  author={Traag, Vincent A and Waltman, Ludo and Van Eck, Nees Jan},
  journal={Scientific reports},
  volume={9},
  number={1},
  pages={1--12},
  year={2019},
  publisher={Nature Publishing Group}
}

@article{tang2024multihop,
  title={Multihop-rag: Benchmarking retrieval-augmented generation for multi-hop queries},
  author={Tang, Yixuan and Yang, Yi},
  journal={arXiv preprint arXiv:2401.15391},
  year={2024}
}

@article{qian2024memorag,
  title={Memorag: Moving towards next-gen rag via memory-inspired knowledge discovery},
  author={Qian, Hongjin and Zhang, Peitian and Liu, Zheng and Mao, Kelong and Dou, Zhicheng},
  journal={arXiv preprint arXiv:2409.05591},
  year={2024}
}

@article{trivedi2022musique,
  title={MuSiQue: Multihop Questions via Single-hop Question Composition},
  author={Trivedi, Harsh and Balasubramanian, Niranjan and Khot, Tushar and Sabharwal, Ashish},
  journal={Transactions of the Association for Computational Linguistics},
  volume={10},
  pages={539--554},
  year={2022},
  publisher={MIT Press One Broadway, 12th Floor, Cambridge, Massachusetts 02142, USA~…}
}

@article{gao2023enabling,
  title={Enabling large language models to generate text with citations},
  author={Gao, Tianyu and Yen, Howard and Yu, Jiatong and Chen, Danqi},
  journal={arXiv preprint arXiv:2305.14627},
  year={2023}
}

@article{pang2021quality,
  title={QuALITY: Question answering with long input texts, yes!},
  author={Pang, Richard Yuanzhe and Parrish, Alicia and Joshi, Nitish and Nangia, Nikita and Phang, Jason and Chen, Angelica and Padmakumar, Vishakh and Ma, Johnny and Thompson, Jana and He, He and others},
  journal={arXiv preprint arXiv:2112.08608},
  year={2021}
}

@article{zhang2024raglab,
  title={Raglab: A modular and research-oriented unified framework for retrieval-augmented generation},
  author={Zhang, Xuanwang and Song, Yunze and Wang, Yidong and Tang, Shuyun and Li, Xinfeng and Zeng, Zhengran and Wu, Zhen and Ye, Wei and Xu, Wenyuan and Zhang, Yue and others},
  journal={arXiv preprint arXiv:2408.11381},
  year={2024}
}

@article{dubey2024llama,
  title={The llama 3 herd of models},
  author={Dubey, Abhimanyu and Jauhri, Abhinav and Pandey, Abhinav and Kadian, Abhishek and Al-Dahle, Ahmad and Letman, Aiesha and Mathur, Akhil and Schelten, Alan and Yang, Amy and Fan, Angela and others},
  journal={arXiv preprint arXiv:2407.21783},
  year={2024}
}

@article{yang2024qwen2,
  title={Qwen2. 5 Technical Report},
  author={Yang, An and Yang, Baosong and Zhang, Beichen and Hui, Binyuan and Zheng, Bo and Yu, Bowen and Li, Chengyuan and Liu, Dayiheng and Huang, Fei and Wei, Haoran and others},
  journal={arXiv preprint arXiv:2412.15115},
  year={2024}
}

@article{liu2024survey,
  title={A Survey on Medical Large Language Models: Technology, Application, Trustworthiness, and Future Directions},
  author={Liu, Lei and Yang, Xiaoyan and Lei, Junchi and Liu, Xiaoyang and Shen, Yue and Zhang, Zhiqiang and Wei, Peng and Gu, Jinjie and Chu, Zhixuan and Qin, Zhan and others},
  journal={arXiv preprint arXiv:2406.03712},
  year={2024}
}

@article{ghimire2024generative,
  title={Generative AI in Education: A Study of Educators' Awareness, Sentiments, and Influencing Factors},
  author={Ghimire, Aashish and Prather, James and Edwards, John},
  journal={arXiv preprint arXiv:2403.15586},
  year={2024}
}

@article{wang2024large,
  title={Large language models for education: A survey and outlook},
  author={Wang, Shen and Xu, Tianlong and Li, Hang and Zhang, Chaoli and Liang, Joleen and Tang, Jiliang and Yu, Philip S and Wen, Qingsong},
  journal={arXiv preprint arXiv:2403.18105},
  year={2024}
}

@article{zheng2024large,
  title={Large language models for medicine: a survey},
  author={Zheng, Yanxin and Gan, Wensheng and Chen, Zefeng and Qi, Zhenlian and Liang, Qian and Yu, Philip S},
  journal={International Journal of Machine Learning and Cybernetics},
  pages={1--26},
  year={2024},
  publisher={Springer}
}

@inproceedings{fan2024survey,
  title={A survey on rag meeting llms: Towards retrieval-augmented large language models},
  author={Fan, Wenqi and Ding, Yujuan and Ning, Liangbo and Wang, Shijie and Li, Hengyun and Yin, Dawei and Chua, Tat-Seng and Li, Qing},
  booktitle={Proceedings of the 30th ACM SIGKDD Conference on Knowledge Discovery and Data Mining},
  pages={6491--6501},
  year={2024}
}

@article{hu2024rag,
  title={Rag and rau: A survey on retrieval-augmented language model in natural language processing},
  author={Hu, Yucheng and Lu, Yuxing},
  journal={arXiv preprint arXiv:2404.19543},
  year={2024}
}

@article{li2024autodcworkflow,
  title={AutoDCWorkflow: LLM-based Data Cleaning Workflow Auto-Generation and Benchmark},
  author={Li, Lan and Fang, Liri and Torvik, Vetle I},
  journal={arXiv preprint arXiv:2412.06724},
  year={2024}
}

@article{li2024llm,
  title={LLM-R2: A Large Language Model Enhanced Rule-based Rewrite System for Boosting Query Efficiency},
  author={Li, Zhaodonghui and Yuan, Haitao and Wang, Huiming and Cong, Gao and Bing, Lidong},
  journal={arXiv preprint arXiv:2404.12872},
  year={2024}
}

@article{zheng2024revolutionizing,
  title={Revolutionizing Database Q\&A with Large Language Models: Comprehensive Benchmark and Evaluation},
  author={Zheng, Yihang and Li, Bo and Lin, Zhenghao and Luo, Yi and Zhou, Xuanhe and Lin, Chen and Su, Jinsong and Li, Guoliang and Li, Shifu},
  journal={arXiv preprint arXiv:2409.04475},
  year={2024}
}

@inproceedings{fan2024cost,
  title={Cost-effective in-context learning for entity resolution: A design space exploration},
  author={Fan, Meihao and Han, Xiaoyue and Fan, Ju and Chai, Chengliang and Tang, Nan and Li, Guoliang and Du, Xiaoyong},
  booktitle={2024 IEEE 40th International Conference on Data Engineering (ICDE)},
  pages={3696--3709},
  year={2024},
  organization={IEEE}
}

@article{chen2024automatic,
  title={Automatic Database Configuration Debugging using Retrieval-Augmented Language Models},
  author={Chen, Sibei and Fan, Ju and Wu, Bin and Tang, Nan and Deng, Chao and Wang, Pengyi and Li, Ye and Tan, Jian and Li, Feifei and Zhou, Jingren and others},
  journal={arXiv preprint arXiv:2412.07548},
  year={2024}
}

@article{li2024dawn,
  title={The Dawn of Natural Language to SQL: Are We Fully Ready?},
  author={Li, Boyan and Luo, Yuyu and Chai, Chengliang and Li, Guoliang and Tang, Nan},
  journal={arXiv preprint arXiv:2406.01265},
  year={2024}
}

@article{giannankouris2024lambda,
  title={$\{$$\backslash$lambda$\}$-Tune: Harnessing Large Language Models for Automated Database System Tuning},
  author={Giannankouris, Victor and Trummer, Immanuel},
  journal={arXiv preprint arXiv:2411.03500},
  year={2024}
}

@article{huang2024survey,
  title={A Survey on Retrieval-Augmented Text Generation for Large Language Models},
  author={Huang, Yizheng and Huang, Jimmy},
  journal={arXiv preprint arXiv:2404.10981},
  year={2024}
}

@article{wu2024medical,
  title={Medical graph rag: Towards safe medical large language model via graph retrieval-augmented generation},
  author={Wu, Junde and Zhu, Jiayuan and Qi, Yunli and Chen, Jingkun and Xu, Min and Menolascina, Filippo and Grau, Vicente},
  journal={arXiv preprint arXiv:2408.04187},
  year={2024}
}

@misc{appendix,
  author={The Technique Report},
  title = {In-depth Analysis of Graph-based RAG in a Unified Framework (technical report)},
  howpublished ={\url{https://github.com/JayLZhou/GraphRAG/blob/master/VLDB2025_GraphRAG.pdf}},
  year = 2025,
}

@article{wu2024retrieval,
  title={Retrieval-augmented generation for natural language processing: A survey},
  author={Wu, Shangyu and Xiong, Ying and Cui, Yufei and Wu, Haolun and Chen, Can and Yuan, Ye and Huang, Lianming and Liu, Xue and Kuo, Tei-Wei and Guan, Nan and others},
  journal={arXiv preprint arXiv:2407.13193},
  year={2024}
}

@article{yu2024evaluation,
  title={Evaluation of Retrieval-Augmented Generation: A Survey},
  author={Yu, Hao and Gan, Aoran and Zhang, Kai and Tong, Shiwei and Liu, Qi and Liu, Zhaofeng},
  journal={arXiv preprint arXiv:2405.07437},
  year={2024}
}

@article{zhao2024retrieval,
  title={Retrieval-augmented generation for ai-generated content: A survey},
  author={Zhao, Penghao and Zhang, Hailin and Yu, Qinhan and Wang, Zhengren and Geng, Yunteng and Fu, Fangcheng and Yang, Ling and Zhang, Wentao and Cui, Bin},
  journal={arXiv preprint arXiv:2402.19473},
  year={2024}
}

@article{liang2025revisiting,
  title={Revisiting Data Analysis with Pre-trained Foundation Models},
  author={Liang, Chen and Yang, Donghua and Liang, Zheng and Liang, Zhiyu and Zhang, Tianle and Xiao, Boyu and Yang, Yuqing and Wang, Wenqi and Wang, Hongzhi},
  journal={arXiv preprint arXiv:2501.01631},
  year={2025}
}

@article{liu2025palimpchat,
  title={PalimpChat: Declarative and Interactive AI analytics},
  author={Liu, Chunwei and Vitagliano, Gerardo and Rose, Brandon and Prinz, Matt and Samson, David Andrew and Cafarella, Michael},
  journal={arXiv preprint arXiv:2502.03368},
  year={2025}
}

@article{anderson2024design,
  title={The Design of an LLM-powered Unstructured Analytics System},
  author={Anderson, Eric and Fritz, Jonathan and Lee, Austin and Li, Bohou and Lindblad, Mark and Lindeman, Henry and Meyer, Alex and Parmar, Parth and Ranade, Tanvi and Shah, Mehul A and others},
  journal={arXiv preprint arXiv:2409.00847},
  year={2024}
}

@article{lin2025twix,
  title={TWIX: Automatically Reconstructing Structured Data from Templatized Documents},
  author={Lin, Yiming and Hasan, Mawil and Kosalge, Rohan and Cheung, Alvin and Parameswaran, Aditya G},
  journal={arXiv preprint arXiv:2501.06659},
  year={2025}
}

@article{lin2024towards,
  title={Towards Accurate and Efficient Document Analytics with Large Language Models},
  author={Lin, Yiming and Hulsebos, Madelon and Ma, Ruiying and Shankar, Shreya and Zeigham, Sepanta and Parameswaran, Aditya G and Wu, Eugene},
  journal={arXiv preprint arXiv:2405.04674},
  year={2024}
}

@article{patel2024lotus,
  title={Lotus: Enabling semantic queries with llms over tables of unstructured and structured data},
  author={Patel, Liana and Jha, Siddharth and Guestrin, Carlos and Zaharia, Matei},
  journal={arXiv preprint arXiv:2407.11418},
  year={2024}
}

@article{liu2024declarative,
  title={A Declarative System for Optimizing AI Workloads},
  author={Liu, Chunwei and Russo, Matthew and Cafarella, Michael and Cao, Lei and Chen, Peter Baille and Chen, Zui and Franklin, Michael and Kraska, Tim and Madden, Samuel and Vitagliano, Gerardo},
  journal={arXiv preprint arXiv:2405.14696},
  year={2024}
}

@article{chen2023seed,
  title={SEED: Domain-Specific Data Curation With Large Language Models},
  author={Chen, Zui and Cao, Lei and Madden, Sam and Kraska, Tim and Shang, Zeyuan and Fan, Ju and Tang, Nan and Gu, Zihui and Liu, Chunwei and Cafarella, Michael},
  journal={arXiv e-prints},
  pages={arXiv--2310},
  year={2023}
}

@article{khattab2023dspy,
  title={Dspy: Compiling declarative language model calls into self-improving pipelines},
  author={Khattab, Omar and Singhvi, Arnav and Maheshwari, Paridhi and Zhang, Zhiyuan and Santhanam, Keshav and Vardhamanan, Sri and Haq, Saiful and Sharma, Ashutosh and Joshi, Thomas T and Moazam, Hanna and others},
  journal={arXiv preprint arXiv:2310.03714},
  year={2023}
}

@article{malkov2018efficient,
  title={Efficient and robust approximate nearest neighbor search using hierarchical navigable small world graphs},
  author={Malkov, Yu A and Yashunin, Dmitry A},
  journal={IEEE transactions on pattern analysis and machine intelligence},
  volume={42},
  number={4},
  pages={824--836},
  year={2018},
  publisher={IEEE}
}

@article{wang2024survey,
  title={A survey on large language models from general purpose to medical applications: Datasets, methodologies, and evaluations},
  author={Wang, Jinqiang and Ning, Huansheng and Peng, Yi and Wei, Qikai and Tesfai, Daniel and Mao, Wenwei and Zhu, Tao and Huang, Runhe},
  journal={arXiv preprint arXiv:2406.10303},
  year={2024}
}

@article{nie2024survey,
  title={A survey of large language models for financial applications: Progress, prospects and challenges},
  author={Nie, Yuqi and Kong, Yaxuan and Dong, Xiaowen and Mulvey, John M and Poor, H Vincent and Wen, Qingsong and Zohren, Stefan},
  journal={arXiv preprint arXiv:2406.11903},
  year={2024}
}

@inproceedings{li2023large,
  title={Large language models in finance: A survey},
  author={Li, Yinheng and Wang, Shaofei and Ding, Han and Chen, Hang},
  booktitle={Proceedings of the fourth ACM international conference on AI in finance},
  pages={374--382},
  year={2023}
}

@article{huang2025ket,
  title={KET-RAG: A Cost-Efficient Multi-Granular Indexing Framework for Graph-RAG},
  author={Huang, Yiqian and Zhang, Shiqi and Xiao, Xiaokui},
  journal={arXiv preprint arXiv:2502.09304},
  year={2025}
}
